%% file: A.tex
\documentclass{article}

\usepackage[english]{babel}
\usepackage{graphicx}
\usepackage[letterpaper,top=2cm,bottom=2cm,left=3cm,right=3cm,marginparwidth=1.75cm]{geometry}
\usepackage{amsmath}
\usepackage[utf8]{inputenc}
\usepackage{amssymb}
\usepackage{booktabs}
\usepackage{float}
\usepackage{authblk}
\usepackage[numbers,sort&compress]{natbib}
\usepackage[colorlinks=true,allcolors=blue]{hyperref}

\newcommand{\keywords}[1]{%

  \par\noindent\textbf{Keywords:} #1

}

\title{Heavy Quarkonium Physics: An Experimental Review at the Tevatron, the LHC, and $e^+e^-$ Colliders}
\author{
Zhipeng Cui$^{1}$, Yilin Zhou$^{2}$, Gerry Bauer$^{1}$ and Kai Yi$^{1}$
}

\affil{
$^1$ Institute of Physics Frontiers and Interdisciplinary Sciences, School of Physics and Technology, Ministry of Education Key Laboratory of Numerical Simulation for Large Scale Complex Systems (NSLSCS), Nanjing Normal University, Nanjing 210023, China \\
$^2$ Institute of Modern Physics, Fudan University, 220 Handan Road, Shanghai, China
}

\date{}
\begin{document}
\setlength{\emergencystretch}{1em}
\maketitle

%%%%%%%%%%%%%%%%%%%%%%%%%%%%%%%%%%%%%%%%%%%%%%%%%%%%%%%%%%%%%%%%%%%%%%%%%%%%%%%%%%%%%%%
%%%%%%%%%%%%%%%%%%%%%%%%%%%%%%%%%%%%%%%%%%%%%%%%%%%%%%%%%%%%%%%%%%%%%%%%%%%%%%%%%%%%%%%
%%%%%%%%%%%%%%%%%%%%%%%%%%%%%%%%%%%%%%%%%%%%%%%%%%%%%%%%%%%%%%%%%%%%%%%%%%%%%%%%%%%%%%%

\begin{abstract}
We present a comprehensive review of recent experimental advances in heavy quarkonium physics, covering measurements from the Tevatron, $e^+e^-$ colliders (Belle, BaBar, Belle~II, BESIII, and LEP), and the LHC (ALICE, ATLAS, CMS, and LHCb) using data from Run~2 and early Run~3. Production studies include differential cross sections and polarization measurements for charmonium and bottomonium states, providing stringent tests of perturbative and non-perturbative QCD calculations within the Non-Relativistic QCD (NRQCD) framework and highlighting symmetry relations among spin and orbital configurations. At the Tevatron, the CDF and D0 experiments established key baselines for quarkonium cross sections and revealed important tensions in polarization, particularly the preference for longitudinal polarization at high transverse momentum, which challenged early NRQCD predictions. Complementary $e^+e^-$ collider experiments have delivered high-precision spectroscopy, decay branching fractions, radiative and hadronic transitions, and observations of exotic quarkonium-like states, offering clean probes of non-perturbative dynamics and serving as crucial inputs for theoretical models. At the LHC, unprecedented $pp$ luminosities and collision energies have enabled detailed production measurements across wide kinematic ranges, precise polarization studies, and groundbreaking spectroscopic discoveries, most notably the observation of fully-heavy tetraquark candidates and of the exotic $J/\psi J/\psi$ and $J/\psi\psi(2S)$ states reported by LHCb, alongside studies of toponium ($t\bar{t}$) quasi-bound states. In heavy-ion collisions, sequential suppression of bottomonium states in $\mathrm{Pb\text{--}Pb}$ collisions provides a powerful probe of the deconfined quark-gluon plasma (QGP), while comparisons with $\mathrm{p\text{--}Pb}$ data constrain cold nuclear matter effects. Multi-quarkonium production processes further explore multi-parton interactions and the underlying symmetry structure of exotic hadrons. These combined results from hadronic and leptonic colliders significantly advance our understanding of heavy quarkonium as both a QCD laboratory and a probe for physics beyond the Standard Model, while identifying remaining theoretical challenges. Future high-luminosity programs at the LHC, Belle~II, and BESIII promise even greater precision and new discoveries in quarkonium physics.

\end{abstract}


%%%%%%%%%%%%%%%%%%%%%%%%%%%%%%%%%%%%%%%%%%%%%%%%%%%%%%%%%%%%%%%%%%%%%%%%%%%%%%%%%%%%%%%
%%%%%%%%%%%%%%%%%%%%%%%%%%%%%%%%%%%%%%%%%%%%%%%%%%%%%%%%%%%%%%%%%%%%%%%%%%%%%%%%%%%%%%%
%%%%%%%%%%%%%%%%%%%%%%%%%%%%%%%%%%%%%%%%%%%%%%%%%%%%%%%%%%%%%%%%%%%%%%%%%%%%%%%%%%%%%%%

\include{B_Motivation}
\include{C_Tevatron}  
\include{D_EE_Colliders}
\include{E_LHC}
\include{F_Summary}
%%%%%%%%%%%%%%%%%%%%%%%%%%%%%%%%%%%%%%%%%%%%%%%%%%%%%%%%%%%%%%%%%%%%%%%%%%%%%%%%%%%%%%%
%%%%%%%%%%%%%%%%%%%%%%%%%%%%%%%%%%%%%%%%%%%%%%%%%%%%%%%%%%%%%%%%%%%%%%%%%%%%%%%%%%%%%%%
%%%%%%%%%%%%%%%%%%%%%%%%%%%%%%%%%%%%%%%%%%%%%%%%%%%%%%%%%%%%%%%%%%%%%%%%%%%%%%%%%%%%%%%

\bibliographystyle{unsrtnat}
\bibliography{References}

%%%%%%%%%%%%%%%%%%%%%%%%%%%%%%%%%%%%%%%%%%%%%%%%%%%%%%%%%%%%%%%%%%%%%%%%%%%%%%%%%%%%%%%
%%%%%%%%%%%%%%%%%%%%%%%%%%%%%%%%%%%%%%%%%%%%%%%%%%%%%%%%%%%%%%%%%%%%%%%%%%%%%%%%%%%%%%%
%%%%%%%%%%%%%%%%%%%%%%%%%%%%%%%%%%%%%%%%%%%%%%%%%%%%%%%%%%%%%%%%%%%%%%%%%%%%%%%%%%%%%%%

\end{document}

%% file: B_Motivation.tex
\section{Introduction and Theoretical Motivation}
\label{sec:introduction}

This review is organized around a single experimental question: how far can a
common QCD description of heavy-quarkonium production be tested across
different collision systems? The Tevatron established the reference
measurements that exposed the limitations of the simplest color-singlet
picture; $e^+e^-$ experiments provide complementary production,
spectroscopy, and decay measurements; and the LHC tests quarkonium
production models over a wide range of $p_T$, rapidity, collision energy,
and nuclear environment. Prompt and non-prompt yields, polarization,
feed-down, exotic-state spectroscopy, and heavy-ion suppression are
therefore treated here not as separate catalogues, but as complementary
observables that probe different aspects of short-distance production,
long-distance hadronization, and the properties of the QCD medium.

\subsection{Heavy Quarkonium as a Laboratory for QCD}

Heavy quarkonium -- the family of bound states composed of a heavy quark
and its antiquark, $Q\bar{Q}$ with $Q = c$ or $b$ -- occupies a unique
position in the landscape of strong-interaction physics. The discovery of
the $J/\psi$ in November 1974, simultaneously announced by the Brookhaven
E598 group studying $p+\mathrm{Be}\to e^+e^-+X$
\cite{Aubert:1974js} and by the SLAC-LBL collaboration studying
$e^+e^-$ annihilation at SPEAR \cite{Augustin:1974xw}, triggered what
became known as the "November Revolution" and sparked the decisive
experimental confirmation of the charm quark predicted by the
Glashow--Iliopoulos--Maiani mechanism. The $J/\psi$ on its own did not
settle the question: the charm interpretation was consolidated by the
subsequent observation of the $\psi(2S)$ and, decisively, by the discovery
of charmed $D$ mesons, which closed off the alternative explanations
that had been advanced at the time. Three years later, the even more
tightly bound $\Upsilon$ family was discovered at Fermilab
\cite{Herb:1977ek}, which initiated bottomonium as a comparatively clean
non-relativistic system and opened the era of systematic heavy-quarkonium
spectroscopy.

Fifty years on, quarkonium systems remain at the forefront of
strong-interaction physics, not because they are exotic, but precisely
because they offer a useful separation of scales. The large mass of the
heavy quark, $m_Q \gg \Lambda_{\rm QCD}$, introduces a hard scale that
makes heavy quarkonium one of the hadronic systems in which QCD
calculations can be confronted with high-precision data across a broad
range of collision environments. These include $e^+e^-$ annihilation,
$pp$, $p$Pb, and PbPb collisions. The breadth and depth of progress in
recent years is surveyed in the Quarkonium Working Group reviews
\cite{Brambilla:2004wh,Brambilla:2010cs}, which serve as the principal background references for the measurements discussed in this work. This
review does not attempt to duplicate those theory-oriented surveys. Its
aim is instead to follow the experimental side of the field in one place:
to collect, in a uniform notation and with consistent conventions, the
production, polarization, spectroscopy, and suppression measurements from
the Tevatron, $e^+e^-$, and LHC experiments, and to identify explicitly
those places where the data still resist a unified theoretical
description.

A recent review by Zhao et al. provides an overview of experimental
quarkonium measurements at the LHC, including production cross sections
and polarization, quarkonium suppression in heavy-ion collisions,
multiplicity-dependent production, ultraperipheral collisions, and
multi-quarkonium processes, based on results from ALICE, ATLAS, CMS, and
LHCb, primarily from Run 2 and early Run 3
\cite{Zhao:2025QuarkoniumReview}. The present review complements this
work by placing experimental quarkonium measurements from different
collider environments in a broader historical and phenomenological
context. It revisits the Tevatron measurements that established key
challenges for quarkonium production models, discusses complementary
spectroscopy, decay, and production measurements at $e^+e^-$ colliders,
and connects these results with measurements at the LHC. This
cross-collider perspective highlights the complementary constraints from
production cross sections, polarization, feed-down, spectroscopy, and
nuclear modification measurements on quarkonium production mechanisms,
non-perturbative QCD, and the properties of the quark-gluon plasma. While
some of the LHC results discussed here overlap with those covered in
Ref.~\cite{Zhao:2025QuarkoniumReview}, the broader collider coverage and
comparative synthesis of experimental results constitute the principal
distinction of the present review.

The physical picture underlying heavy quarkonium is that of a
non-relativistic two-body system, loosely analogous to positronium in
QED. The heavy-quark pair is characterized by a hierarchy of energy
scales,
\begin{equation}
m_Q \;\gg\; m_Q v \;\gg\; m_Q v^2 \;\gtrsim\; \Lambda_{\rm QCD},
\label{eq:hierarchy}
\end{equation}
where $v$ is the relative velocity of the quark and antiquark in the
quarkonium rest frame ($v^2 \sim 0.3$ for charmonium and $v^2 \sim 0.1$
for bottomonium). The scale $m_Q$ corresponds to the heavy-quark mass
(hard scale), $m_Q v$ to the typical relative momentum (soft scale), and
$m_Q v^2$ to the typical binding energy (ultrasoft scale). This
separation of scales motivates the effective-field-theory treatment of
quarkonium production and decay and provides the basis for systematically
separating perturbative short-distance physics from non-perturbative
long-distance dynamics \cite{Brambilla:2004jt}. The same hierarchy
organizes the lattice-QCD description of quarkonium, in which
non-perturbative quantities and the potentials entering the theoretical
description can be calculated from the QCD action
\cite{Brambilla:2004wh,Brambilla:2010cs}.

\subsection{From the Quark Model and Color-Singlet Model to QCD-Based Treatments}

\begin{figure}[htbp]
\centering
\includegraphics[width=0.45\textwidth]{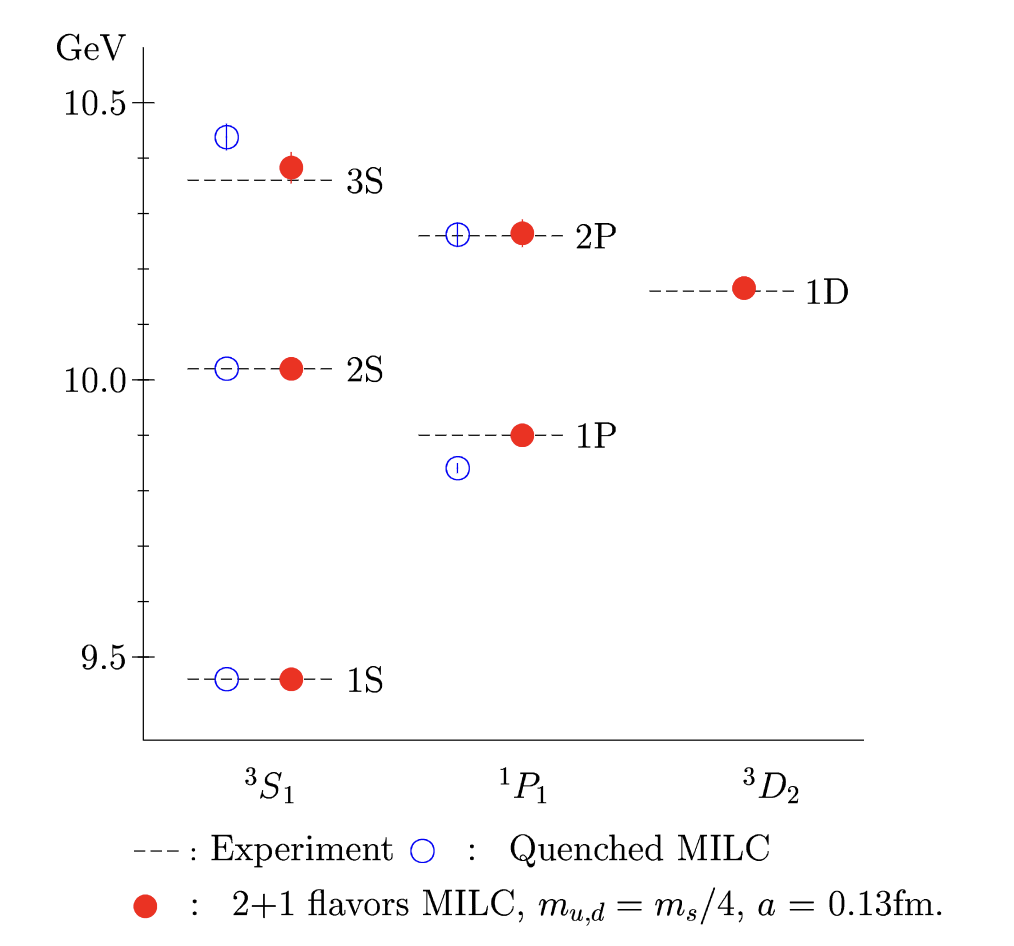}
\caption{Radial and orbital splittings in the $\Upsilon$ system computed
on the lattice in the quenched approximation (open circles) and including
dynamical $u,d,s$ sea quarks (filled circles), illustrating the progress
in first-principles calculations of bottomonium spectroscopy once the
lattice spacing is fixed by the $\Upsilon(2S)$--$\Upsilon(1S)$ splitting
and the $b$-quark mass is tuned to the $\Upsilon(1S)$ mass. Figure
reproduced from Ref.~\cite{Brambilla:2004jt}, Fig.~3.}
\label{fig:motivation_lattice_bottomonium}
\end{figure}

In the years immediately following the $J/\psi$ discovery, charmonium
spectroscopy was successfully described using non-relativistic potential
models, in which the $c\bar{c}$ pair was treated as moving in a static
potential of the Cornell form \cite{Eichten:1978tg},
\begin{equation}
V(r) = -\frac{\kappa}{r} + \frac{r}{a^2},
\label{eq:cornell}
\end{equation}
combining a short-distance Coulombic term generated by one-gluon exchange
with a long-distance linear confining term. While remarkably successful
in reproducing the charmonium and bottomonium mass spectra, such
potential models offered no first-principles description of how
quarkonium states are \emph{produced} in high-energy collisions. The
precision with which the charmonium and bottomonium level spacings are
reproduced in lattice QCD calculations
(Fig.~\ref{fig:motivation_lattice_bottomonium}) illustrates the
progress achievable in first-principles calculations of quarkonium
spectroscopy \cite{Brambilla:2004wh}. These calculations use lattice
QCD to determine the relevant non-perturbative quantities and provide
theoretical input for the description of quarkonium bound states.

The earliest hadroproduction framework, the Color-Singlet Model (CSM)
\cite{Chang:1979nn,Berger:1980ni}, assumed that the $Q\bar{Q}$ pair is
produced perturbatively directly in a color-singlet state with the same
quantum numbers as the observed quarkonium. It further assumed that the
transition probability to the physical bound state is governed solely by
the quarkonium wavefunction (or its derivatives) at the origin,
computable from potential models or extracted from leptonic decay
widths. While the CSM gave a reasonable account of quarkonium
photoproduction and some fixed-target hadroproduction data
\cite{Brambilla:2004wh,Brambilla:2010cs}, it failed to describe the
high-$p_T$ inclusive charmonium production rates observed at hadron
colliders within the leading-order calculations available at the time.
Measurements of inclusive $J/\psi$ and $\psi(2S)$ production by CDF at
the Tevatron in the early-to-mid 1990s revealed cross sections at high
transverse momentum exceeding leading-order CSM predictions by more than
an order of magnitude \cite{Abe:1992ww,Abe:1997yz}. This
``quarkonium $p_T$ anomaly'' constituted one of the central puzzles in
QCD phenomenology of the 1990s and motivated much of the subsequent
theoretical development discussed in this review. It is not, however, the
only long-standing theoretical challenge: the non-perturbative
hadronization of a colored $Q\bar{Q}$ pair, the treatment of feed-down,
and the nature of the exotic states remain open questions in their own
right, and are the subject of the following subsections.

\subsection{Non-Relativistic QCD and the Factorization Framework}

The resolution of the CSM crisis was pursued through the formulation of
Non-Relativistic QCD (NRQCD) as an effective field theory of QCD,
constructed by integrating out the hard scale $m_Q$ while retaining the
degrees of freedom relevant to the bound-state dynamics
\cite{Caswell:1985ui,Bodwin:1994jh}. NRQCD is organized as a double
expansion: an expansion in the strong coupling constant $\alpha_s(m_Q)$,
governing the perturbative matching coefficients, and an expansion in
powers of the heavy-quark relative velocity $v$, governing the relative
importance of different production and decay mechanisms.

The work of Bodwin, Braaten, and Lepage
\cite{Bodwin:1994jh} established the theoretical framework for separating
short-distance physics, which can be calculated perturbatively, from
long-distance, non-perturbative physics associated with the formation of
the physical quarkonium state. In the NRQCD factorization approach,
inclusive quarkonium production is conventionally expressed as a sum of
short-distance coefficients (SDCs) multiplied by long-distance matrix
elements (LDMEs). For an inclusive process producing a quarkonium state
$H$, the production cross section is written as
\begin{equation}
d\sigma(H+X) \;=\; \sum_{n}
d\hat{\sigma}_{Q\bar{Q}[n]}(\mu)\,
\langle \mathcal{O}^{H}_{n}(\mu) \rangle ,
\label{eq:nrqcd_factorization}
\end{equation}
where the sum runs over intermediate $Q\bar{Q}$ Fock states $n$, labeled
by their angular-momentum and color quantum numbers
$^{2S+1}L_J^{[1,8]}$ (singlet or octet). The short-distance
coefficients $d\hat{\sigma}_{Q\bar{Q}[n]}$ describe the perturbative
production of the $Q\bar{Q}$ pair in configuration $n$ at the
factorization scale $\mu$, and are calculable order-by-order in
$\alpha_s$ using perturbative QCD. The LDMEs
$\langle \mathcal{O}^{H}_n(\mu)\rangle$ encode the non-perturbative
transition of the intermediate $Q\bar{Q}[n]$ state into the physical
quarkonium $H$. The factorization framework hypothesizes that these
matrix elements are universal, i.e., independent of the production
process, and can therefore be used across different production
environments when the framework is applicable.

Although NRQCD factorization is theoretically motivated and has
achieved considerable phenomenological success, a general all-orders proof
of the factorization formula for inclusive quarkonium production has not
been established. NRQCD also provides an expansion for quarkonium decay rates, in which
the decay width is expressed in terms of perturbative short-distance
coefficients and decay matrix elements. Although production and decay
are described within the same effective-field-theory framework, the
corresponding observables involve different short-distance processes
and should not be directly identified. Experimentally measured
branching fractions are commonly used in the normalization of
quarkonium production measurements and in estimating feed-down
contributions. Certain decay observables, such as dilepton decay
widths, can also constrain specific matrix elements, particularly
color-singlet contributions, within the NRQCD framework. However,
branching-fraction measurements do not generally provide a direct or
unique determination of the color-octet production LDMEs, which are
primarily extracted from production data through phenomenological
analyses. The theoretical status of factorization has been
investigated in specific processes and perturbative settings, but its
general validity remains an open question. Accordingly, the
factorization formula in Eq.~(\ref{eq:nrqcd_factorization}) is used
here as a theoretical framework whose applicability and predictive power
are tested against experimental measurements, rather than as a theorem
whose validity has been established for all inclusive production
processes.

Each LDME scales with a definite power of $v$ according to the
velocity-scaling rules of NRQCD \cite{Lepage:1992tx}, so that
Eq.~(\ref{eq:nrqcd_factorization}) can be organized as an expansion in
$v^2$ and truncated at a given order. For S-wave charmonium and
bottomonium states, the leading color-singlet contribution is
supplemented by color-octet contributions entering at relative order
$v^4$,
\begin{equation}
\begin{split}
d\sigma(\psi+X) \;=\;
& d\hat{\sigma}_{[^3S_1^{[1]}]}
\langle \mathcal{O}^{\psi}[^3S_1^{[1]}]\rangle
\;+\; d\hat{\sigma}_{[^1S_0^{[8]}]}
\langle \mathcal{O}^{\psi}[^1S_0^{[8]}]\rangle \\
&+\; d\hat{\sigma}_{[^3S_1^{[8]}]}
\langle \mathcal{O}^{\psi}[^3S_1^{[8]}]\rangle
\;+\; d\hat{\sigma}_{[^3P_J^{[8]}]}
\langle \mathcal{O}^{\psi}[^3P_J^{[8]}]\rangle
\;+\; \cdots
\end{split}
\label{eq:nrqcd_jpsi}
\end{equation}
It is precisely the inclusion of these color-octet channels -- in
which the $Q\bar{Q}$ pair is produced in a colored configuration and
subsequently neutralizes its color through non-perturbative dynamics --
that enabled NRQCD-based calculations to address the large
hadroproduction cross sections measured at the Tevatron, beyond the
leading-order color-singlet description. Braaten and Fleming argued
\cite{Braaten:1994vv} that at high $p_T$ the dominant mechanism can be
gluon fragmentation into a $c\bar{c}$ pair in a color-octet
$^3S_1^{[8]}$ state, enhanced relative to the color-singlet contribution
by a short-distance factor of $\sim 1/\alpha_s^2$ that compensates the
$v^4$ velocity suppression of the corresponding LDME.
Systematic numerical implementations of the color-octet framework were
carried out by Cho and Leibovich \cite{Cho:1995ce,Cho:1995vh}.
Their fits to CDF data represented important phenomenological support
for the NRQCD production framework and its predictive potential,
although agreement with the data used in the fits does not independently
establish LDME universality (reviewed in Section~\ref{sec:tevatron}).

More recently, fragmentation-based approaches have also been extended
to pseudoscalar heavy quarkonia, with the NRFF1.0 framework providing
publicly available fragmentation functions for $\eta_c$ and $\eta_b$
production across all partonic channels, incorporating NLO NRQCD
initial conditions and DGLAP evolution with consistent heavy-flavor
thresholds~\cite{Celiberto:2025NRFF}.

\subsection{The Polarization Puzzle: NRQCD's Central Theoretical Challenge}

Cross-section measurements alone, however, provide limited
discriminating power among different combinations of LDMEs, since
multiple parameter sets can be tuned to reproduce similar $p_T$ spectra.
A more stringent test of the production mechanism is provided by the
\emph{polarization} of the produced quarkonium, since the angular
distribution of the subsequent leptonic decay is sensitive to the spin
state of the $Q\bar{Q}$ pair and to the production dynamics.

\begin{figure}[htbp]
\centering
\includegraphics[width=0.45\textwidth]{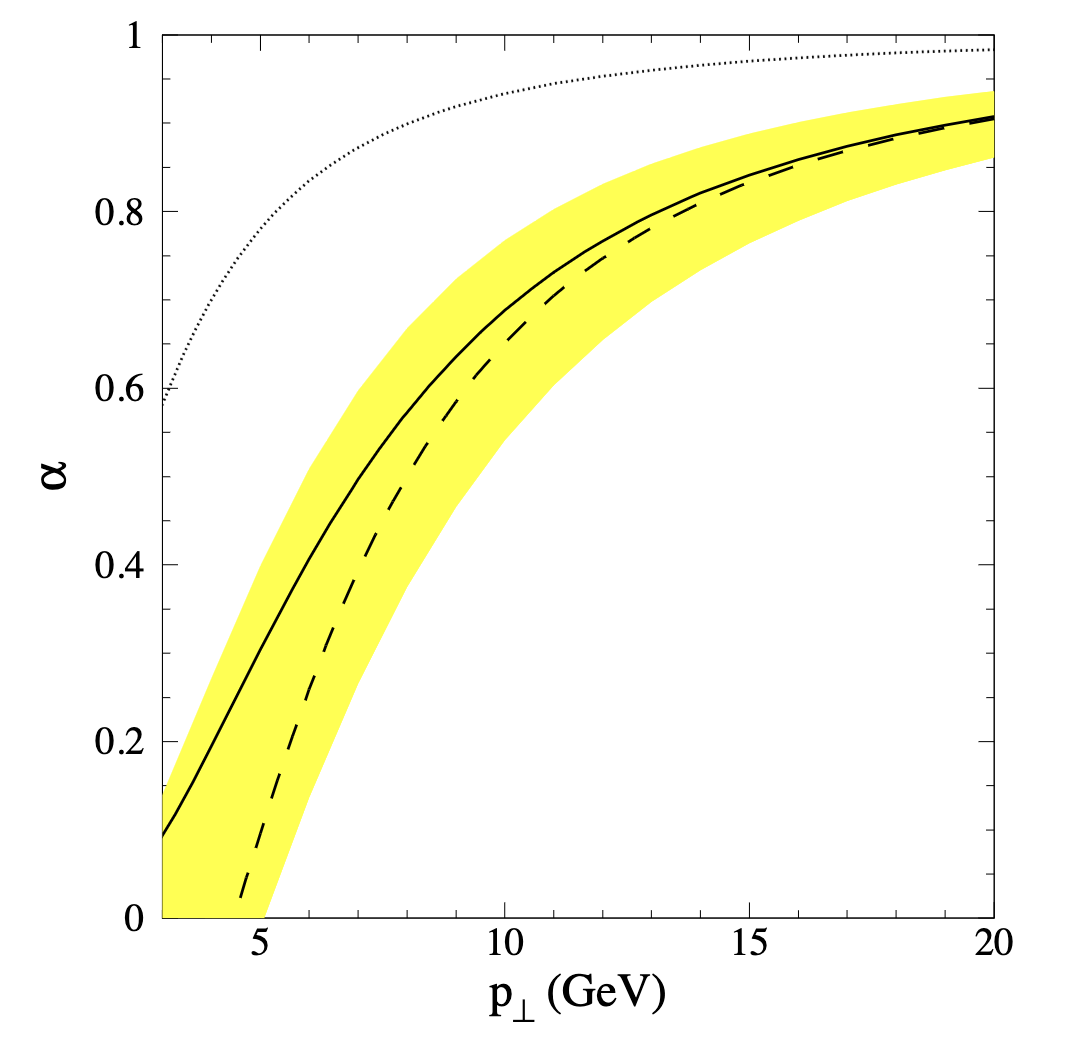}
\caption{Original NRQCD theoretical prediction for the polarization
parameter $\alpha$ of prompt $\psi(2S)$ as a function of $p_T$ at the
Tevatron, computed from color-octet $^3P_J^{[8]}$ and
$^1S_0^{[8]}$ contributions (solid and dashed curves, shaded band
reflecting LDME uncertainties) together with the leading-order
color-singlet contribution (dotted curve). The calculation predicts a
transition from near-zero polarization at low $p_T$ to fully transverse
polarization ($\alpha\to+1$) at high $p_T$, in direct tension with the
Tevatron measurements discussed in Section~\ref{sec:tevatron}. Figure
reproduced from Ref.~\cite{Leibovich:1996pa}, Fig.~1.}
\label{fig:motivation_nrqcd_psiprime_polarization_prediction}
\end{figure}

In the helicity frame, the normalized angular distribution of the decay
leptons from $J/\psi \to \mu^+\mu^-$ (or
$\Upsilon \to \mu^+\mu^-$) takes the general form
\begin{equation}
\frac{1}{N}\frac{dN}{d\Omega} \;\propto\;
1 + \lambda_\theta \cos^2\theta^*
+ \lambda_\phi \sin^2\theta^*\cos 2\phi^*
+ \lambda_{\theta\phi}\sin 2\theta^*\cos\phi^* ,
\label{eq:angular_distribution}
\end{equation}
which, when integrated over the azimuthal angle $\phi^*$, reduces to
the familiar one-dimensional form
\begin{equation}
\frac{dN}{d\cos\theta^*} \;\propto\;
1 + \alpha\cos^2\theta^*, \qquad
\alpha \equiv \lambda_\theta ,
\label{eq:polar_alpha}
\end{equation}
with $\alpha = +1$ corresponding to fully transverse polarization and
$\alpha = -1$ to fully longitudinal polarization. At leading order in
$\alpha_s$ and in the limit of large $p_T$, gluon fragmentation into a
$^3S_1^{[8]}$ color-octet pair is a dominant channel in the original
NRQCD picture. It preserves the gluon's transverse polarization,
leading to the characteristic leading-order expectation that prompt
$J/\psi$ and $\Upsilon$ production should become increasingly
\emph{transversely} polarized ($\alpha \to +1$) as $p_T \to \infty$
\cite{Cho:1995ce,Beneke:1998re,Leibovich:1996pa}.
Figure~\ref{fig:motivation_nrqcd_psiprime_polarization_prediction}
illustrates this prediction quantitatively for the $\psi(2S)$.

This prediction was challenged by CDF's polarization measurements:
analyses of prompt $J/\psi$ and $\psi(2S)$ production found $\alpha$
values consistent with zero or mildly \emph{negative} (longitudinal)
across the measured $p_T$ range, with no indication of the predicted
rise toward $+1$ \cite{CDF:2000polarization}. The measured $p_T$
dependence of $\alpha$, shown together with the original NRQCD
prediction in Fig.~\ref{fig:tevatron_pol_alpha}, is discussed in detail
in Section~\ref{sec:tevatron}. This conflict between the original
leading-order polarization prediction and experimental observation --
commonly referred to as the ``quarkonium polarization puzzle'' --
represents one of the persistent open problems in the application of
NRQCD-based calculations to hadroproduction.

Proposed resolutions have included the incorporation of higher-order QCD
corrections (which can substantially soften or even reverse the
polarization at moderate $p_T$) \cite{Butenschoen:2012px}, a more
careful treatment of feed-down contributions from higher-mass resonances
(e.g.\ $\psi(2S)\to J/\psi$, $\chi_{cJ}\to J/\psi\,\gamma$,
$\Upsilon(nS)\to\Upsilon(1S)+X$), and global refits of the color-octet
LDMEs that attempt to describe cross-section and polarization data
simultaneously. The consistency of such fits with the available
polarization measurements remains an important challenge, motivating
continued precision measurements across a broad kinematic range and
across different production environments
\cite{Brambilla:2010cs}.

\subsection{Long-Distance Matrix Elements, Universality, and Global Fits}

A central hypothesis of the NRQCD factorization framework in
Eq.~(\ref{eq:nrqcd_factorization}) is the universality of the LDMEs:
the same set of matrix elements $\langle \mathcal{O}^H_n(\mu)\rangle$
is expected to describe quarkonium production in $e^+e^-$ annihilation,
in $ep$ deep-inelastic and photoproduction, and in $pp$
hadroproduction, provided that the factorization framework applies and
the appropriate short-distance coefficients for each process are used.
This universality is a hypothesis to be tested rather than an
experimentally established property. Applications to nuclear collisions
add further assumptions concerning initial-state nuclear effects and
possible final-state interactions, which may affect the interpretation
of production measurements and the extraction of matrix elements.
Testing universality therefore requires comparisons among independent
data sets, different production processes, and observables -- precisely
the complementary measurements from the Tevatron, $e^+e^-$, and LHC
experiments discussed in this review.

Global NLO fits of the color-octet LDMEs to Tevatron and LHC production
data describe many individual $p_T$ spectra reasonably well
\cite{Butenschoen:2011yh}. However, agreement with the data used in a
fit does not by itself provide an independent test of universality.
Different global analyses arrive at LDME sets that differ not only in
magnitude but occasionally in relative sign. These differences depend
on the choice of data sets, the treatment of feed-down, and the order
in $\alpha_s$ at which the short-distance coefficients are computed,
particularly for the $^1S_0^{[8]}$ and $^3P_J^{[8]}$ channels. These
channels have strongly correlated short-distance coefficients at the
level of total cross sections but can in principle be disentangled
through their distinct $p_T$ shapes and polarization signatures.
Consequently, the non-uniqueness of LDME extractions and the
differences among phenomenological analyses remain important issues
throughout this review and motivate high-precision measurements over a
wide $p_T$ range. A particularly relevant example is the CMS
$\Upsilon(nS)$ measurement at $\sqrt{s}=13.6$ TeV, which reaches
$p_T$ up to 200 GeV \cite{CMS-BPH-24-004} and provides additional
experimental leverage through the shapes of the
$\Upsilon(2S)/\Upsilon(1S)$ and $\Upsilon(3S)/\Upsilon(1S)$
cross-section ratios.

\subsection{Bottomonium and Potential-NRQCD}

Bottomonium offers a complementary laboratory for testing NRQCD, since
the larger $b$-quark mass implies a smaller characteristic velocity
($v_b^2 \sim 0.1$ versus $v_c^2 \sim 0.3$), which is expected to
improve the convergence of the velocity expansion and reduce the relative
importance of relativistic corrections. The $\Upsilon(nS)$ family
additionally benefits from a comparatively simple feed-down structure,
although contributions from
$\chi_{bJ}(nP)\to\Upsilon(mS)\,\gamma$ transitions remain a
non-negligible effect that must be considered in interpreting prompt
production measurements and in extracting production matrix elements
\cite{Digal:2001ue}.

The validity and practical usefulness of the heavy-quark expansion can
also be studied through potential-NRQCD (pNRQCD), a lower-energy
effective field theory obtained by further integrating out the soft
scale $m_Qv$ from NRQCD \cite{Brambilla:2004jt,Brambilla:1999xf}.
In this framework, the soft modes are encoded in quantum-mechanical
potentials. In the strongly-coupled regime relevant to ground-state and
low-lying quarkonia, pNRQCD provides a description in terms of a
Schr\"{o}dinger-like equation governed by potentials whose properties
can be related to QCD,
\begin{equation}
\left[ \frac{p^2}{m_Q} + V^{(0)}(r) + \frac{V^{(1)}(r)}{m_Q}
+ \frac{V^{(2)}(r)}{m_Q^2} + \cdots \right]\psi(\mathbf{r}) = E\,\psi(\mathbf{r}),
\label{eq:pnrqcd_schrodinger}
\end{equation}
providing a theoretical connection between phenomenological
Cornell-type potentials of Eq.~(\ref{eq:cornell}) and QCD. The
systematic tower of effective field theories, NRQCD $\to$ pNRQCD,
underlies many quantitative theoretical descriptions of quarkonium
production, decay, and spectroscopy discussed in this review
\cite{Brambilla:2004jt}.

So far this introduction has dealt primarily with quarkonium in vacuum.
The next subsection turns to quarkonium produced in the hot, deconfined
medium created in heavy-ion collisions, where the theoretical
description is modified: the potentials and dynamical properties of
quarkonium become temperature dependent, and dissociation and
regeneration become central questions.

\subsection{Quarkonium Suppression as a Probe of the Quark--Gluon Plasma}

\begin{figure}[htbp]
\centering
\includegraphics[width=0.55\textwidth]{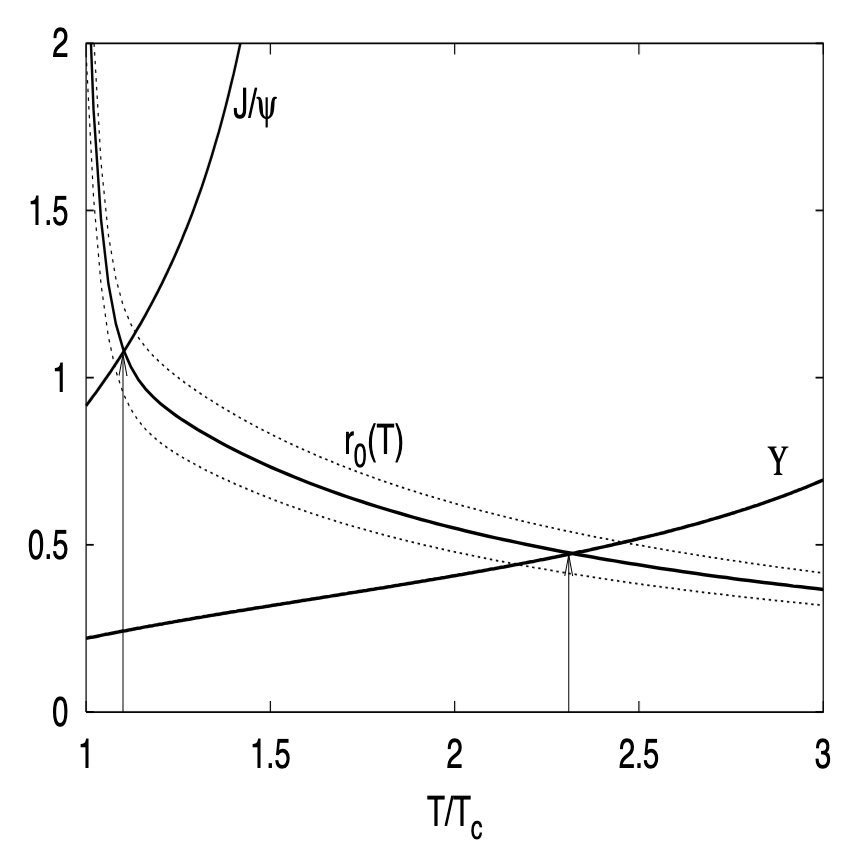}
\caption{Radii of the $J/\psi$ and $\Upsilon$ ground states (vertical
axis, in fm), obtained by solving the Schr\"{o}dinger equation with the
lattice-determined, temperature-dependent color-singlet potential
$V_1(T,r)$, compared with the limiting screening radius $r_0(T)$
beyond which the $Q\bar Q$ potential becomes flat (color-screened).
The intersection of each curve with $r_0(T)$ is used in this model to
estimate the dissociation temperature of that state: $T_d\simeq1.1\,T_c$
for the $J/\psi$ and $T_d\simeq2.5\,T_c$ for the $\Upsilon(1S)$, where
$T_c$ denotes the QCD critical temperature. Figure reproduced from
Ref.~\cite{Digal:2001ue}, Fig.~9.}
\label{fig:motivation_dissociation_radii}
\end{figure}

A theoretically distinct, but historically and conceptually closely
related, motivation for quarkonium physics arises in the context of
heavy-ion collisions. Matsui and Satz proposed in 1986
\cite{Matsui:1986dk} that the formation of a deconfined quark--gluon
plasma (QGP) in ultra-relativistic heavy-ion collisions should lead to
the dissolution, or ``melting,'' of quarkonium bound states, because the
color charges of the heavy-quark pair would be screened by the
surrounding medium of free color charges once the screening length
$\lambda_D(T)$ becomes comparable to or smaller than the quarkonium
binding radius. This is conventionally illustrated by the
temperature-dependent dissociation condition,
\begin{equation}
\lambda_D(T) \;\lesssim\; r_{Q\bar{Q}},
\label{eq:debye_screening}
\end{equation}
where $r_{Q\bar Q}$ is the characteristic radius of the quarkonium
state. Because different quarkonium states have different binding
energies and radii, the tightly bound $\Upsilon(1S)$ (binding energy
$\sim 1.1$ GeV, radius $\sim 0.14$ fm) is expected to be more robust
against screening than the more weakly bound $\Upsilon(3S)$ or
$\chi_{c1}$. This motivates a hierarchical, or
\emph{sequential}, pattern of suppression
\cite{Digal:2001ue,Karsch:1987pv}. Figure~\ref{fig:motivation_dissociation_radii}
illustrates how the intersection of the state radius with the screening
radius $r_0(T)$ can be used in a model to estimate dissociation
temperatures.

\begin{figure}[htbp]
\centering
\includegraphics[width=0.45\textwidth]{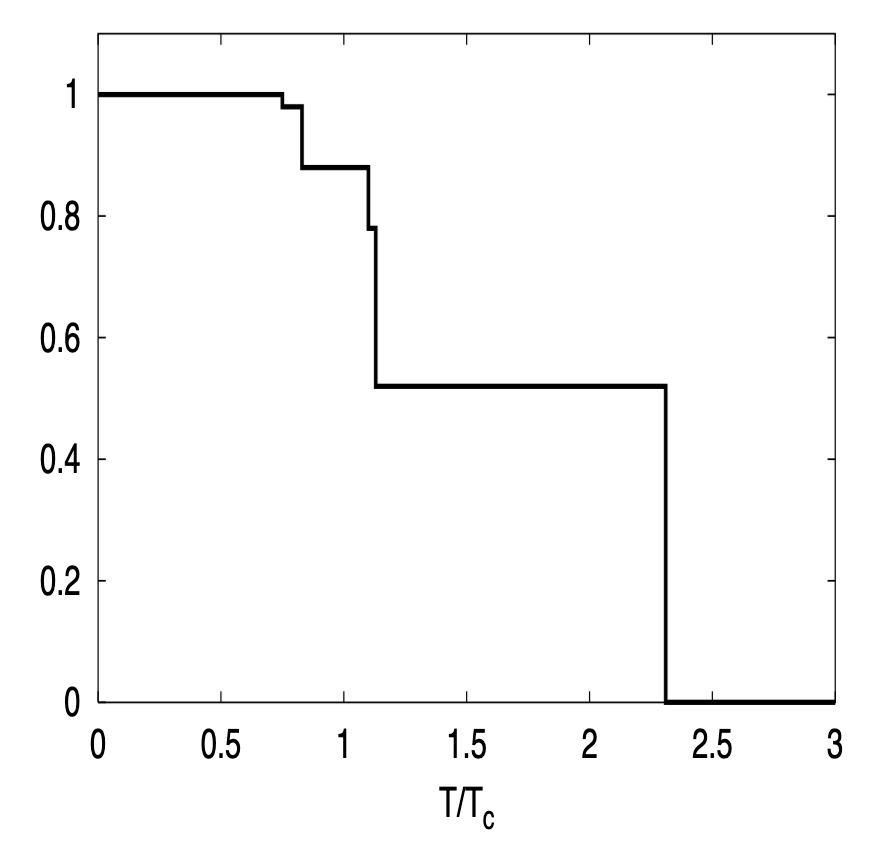}
\caption{Predicted sequential suppression pattern for inclusive
$\Upsilon(1S)$ yield as a function of temperature, obtained by combining
the dissociation temperatures of the $\Upsilon(1S)$, $\chi_b(1P)$,
$\Upsilon(2S)$, $\Upsilon(3S)$, and $\chi_b(2P)$ states with their
respective feed-down fractions to the observed $\Upsilon(1S)$ yield.
Each step corresponds to the suppression of one feed-down component,
illustrating a qualitative sequential-suppression scenario. Figure
reproduced from Ref.~\cite{Digal:2001ue}, Fig.~11.}
\label{fig:motivation_sequential_suppression}
\end{figure}

When dissociation temperatures are combined with estimates of the
feed-down fractions from higher excitations -- approximately 30\% of
the observed $J/\psi$ yield originates from $\chi_c$ and $\psi(2S)$
decays, while nearly 50\% of the $\Upsilon(1S)$ yield arises from
$\chi_b$ and $\Upsilon(2S,3S)$ feed-down in the cited phenomenological
description \cite{Digal:2001ue} -- the result is a characteristic
staircase suppression pattern, illustrated in
Fig.~\ref{fig:motivation_sequential_suppression}. Each step in this
pattern corresponds to the suppression of an additional feed-down
component as the medium temperature increases. This is a qualitative
sequential-suppression picture rather than a model-independent
prediction: its detailed shape depends on the temperature-dependent
potentials, feed-down fractions, and the treatment of regeneration and
other medium effects.

The modern theoretical understanding of quarkonium suppression has been
substantially refined since the original Matsui--Satz proposal through
finite-temperature extensions of the pNRQCD framework. In these
calculations, quarkonium binding and dissociation may be described using
a complex, temperature-dependent potential,
$V(r,T)=V_{\rm Re}(r,T)+iV_{\rm Im}(r,T)$. The imaginary part is
associated with thermal processes that generate a finite in-medium
width, including gluo-dissociation and inelastic parton scattering
\cite{Brambilla:2004jt}.

In heavy-ion collisions, suppression mechanisms compete with regeneration.
At LHC energies, the number of independently produced $c\bar{c}$ pairs
per PbPb event can be large, so uncorrelated $c$ and $\bar c$ quarks
may recombine into $J/\psi$ mesons during hadronization
\cite{BraunMunzinger:2000px}. Regeneration can partially restore the
observed yield and substantially complicates the interpretation of
charmonium nuclear modification factors relative to the comparatively
cleaner bottomonium sector. Disentangling suppression and regeneration,
as well as these effects from cold nuclear matter effects such as
nuclear parton shadowing, requires systematic comparisons of PbPb,
$p$Pb, and $pp$ reference data across multiple quarkonium states --
precisely the program pursued at the LHC and discussed in
Section~\ref{sec:lhc}.

With this, the survey of quarkonium as a probe of the medium is
complete, and the remainder of this introduction changes subject once
more. The suppression picture just described -- like the production
discussion before it -- presumes that quarkonium can be described as a
non-relativistic $Q\bar{Q}$ bound state with the properties of a
conventional quark-model meson. The final subsection asks what happens
when that presumption is relaxed.

\subsection{Exotic States and the Limits of the Conventional Picture}

The conventional $Q\bar{Q}$ picture of quarkonium, while
extraordinarily successful, has been challenged over the past two
decades by the discovery -- primarily at $e^+e^-$ $B$-factories
(Belle, BaBar) and later studied extensively at BESIII and the LHC
experiments -- of numerous exotic quarkonium-like states (the so-called
$XYZ$ states). These include charged states such as
$Z_c(3900)^+$ \cite{Ablikim:2013mio} and
$Z_b(10610)^+/Z_b(10650)^+$ \cite{Bondar:2011xe} that cannot be
accommodated within a simple $c\bar c$ or $b\bar b$ assignment since
they carry non-zero electric charge while decaying to quarkonium plus a
pion, as well as neutral states such as the $X(3872)$
\cite{Choi:2003ue} whose quantum numbers and radiative decay pattern
continue to challenge a simple charmonium interpretation. These
observations have motivated alternative interpretations including
compact tetraquarks, hadronic molecules bound by pion exchange,
hadro-quarkonium configurations, gluon-hybrid states, and
coupled-channel or threshold-induced dynamical effects
\cite{Brambilla:2010cs}. Such states matter for a production review for
two concrete reasons: their line shapes and production rates probe the
boundary between perturbative quark-pair creation and non-perturbative
hadronization, while their decays supply feed-down and normalization
inputs needed for conventional charmonium and bottomonium measurements.
Section~\ref{sec:eecolliders} therefore retains the spectroscopy
program as a focused test of those interfaces, not as an unrelated
inventory of exotic candidates.

\subsection{Outline and Scope of This Review}

Motivated by the theoretical framework outlined above, this review is
organized to track the experimental confrontation of NRQCD-based
production approaches, the Matsui--Satz suppression picture, and related
theoretical frameworks across three complementary classes of collider
experiments. Section~\ref{sec:tevatron} reviews quarkonium production
and polarization measurements at the Tevatron, where the original
color-singlet-model discrepancy and the subsequent polarization puzzle
were first identified. Section~\ref{sec:eecolliders} discusses the
complementary spectroscopy, decay, and production program carried out
at $e^+e^-$ colliders (Belle, Belle~II, and BESIII), which provides
important experimental inputs, including branching fractions,
transition rates, and resonance parameters, for cross-section
normalization, feed-down studies, and spectroscopy. These measurements
also provide complementary information for testing theoretical
descriptions of quarkonium production and decay, but are not assumed to
directly determine the production LDMEs in general. Section~\ref{sec:lhc}
presents the current state of the art at the LHC, where the broad
kinematic reach in $pp$ collisions provides stringent tests of
quarkonium production models, while PbPb and $p$Pb data provide
complementary information on suppression, regeneration, and nuclear
effects in the quark--gluon plasma. Together, these results illustrate
the phenomenological successes and open challenges of using quarkonium
as a precision QCD laboratory.

%% file: C_Tevatron.tex
\section{Quarkonium Production at the Tevatron}
\label{sec:tevatron}

This section uses representative CDF and D0 measurements to establish the
empirical baseline against which later NRQCD fits and LHC measurements are
benchmarked. It addresses two questions in turn: whether a common set of
non-perturbative matrix elements can describe inclusive and differential
quarkonium yields, and whether the measured spin alignment and multi-parton
final states are consistent with that same production picture. The discussion
is ordered conceptually rather than chronologically: cross-section
measurements first, then polarization, followed by double-parton scattering and
the CDF spectroscopy results that test the limits of the conventional
quarkonium assignment. Tables~\ref{tab:tevatron_summary} and
\ref{tab:tevatron_d0_summary} summarize the principal CDF and D0 inputs;
``Run~I'' and ``Run~II'' below refer to the Tevatron runs, and the relevant CDF
or D0 analysis is named explicitly whenever a result is introduced.

\begin{table}[htbp]
\centering
\caption{Summary of the main CDF Tevatron measurements discussed in this section.}
\label{tab:tevatron_summary}
\footnotesize                    % 或 \small
\setlength{\tabcolsep}{4pt} 
\begin{tabular}{p{2.6cm}p{1.6cm}p{1.8cm}p{2.0cm}p{5.4cm}}
\hline\hline
Reference & $\sqrt{s}$ & $\int\!\mathcal{L}\,dt$ & States & Principal result \\
\hline
\cite{Acosta:2004yw} & 1.96 TeV & 39.7~pb$^{-1}$ & $J/\psi$, $H_b$ &
First $0<p_T<20$~GeV/$c$ inclusive $J/\psi$ measurement; $\sigma(J/\psi)\cdot
\mathrm{Br}=4.08\,\mu$b; total single $b$-hadron cross section $17.6\,\mu$b, in agreement
with FONLL \\
\cite{Aaltonen:2009tha} & 1.96 TeV & 1.1~fb$^{-1}$ & $\psi(2S)$ &
Feed-down-free cross section to $p_T=30$~GeV/$c$; $\sigma(\psi(2S))\cdot
\mathrm{Br}=3.29$~nb for $p_T>2$~GeV/$c$ \\
\cite{Abulencia:2007us} & 1.96 TeV & 800~pb$^{-1}$ & $J/\psi$, $\psi(2S)$ &
Polarization parameter $\alpha$ becomes increasingly longitudinal for
$p_T>10$~GeV/$c$, in conflict with leading-order NRQCD \\
\cite{Acosta:2001gc} & 1.8 TeV & 77~pb$^{-1}$ & $\Upsilon(1S,2S,3S)$ &
First hadron-collider $\Upsilon(1S)$ polarization measurement; consistent with
unpolarized production over $0<p_T<20$~GeV/$c$ \\
\hline\hline
\end{tabular}
\end{table}

\begin{table}[htbp]
\centering
\caption{Summary of the main D0 Tevatron measurements discussed in this section.}
\label{tab:tevatron_d0_summary}
\footnotesize                    % 或 \small
\setlength{\tabcolsep}{4pt} 
\begin{tabular}{p{2.6cm}p{1.6cm}p{1.8cm}p{2.0cm}p{5.4cm}}
\hline\hline
Reference & $\sqrt{s}$ & $\int\!\mathcal{L}\,dt$ & States & Principal result \\
\hline
\cite{Abachi:1996fu} & 1.8 TeV & 6.6--13~pb$^{-1}$ & $J/\psi$ &
First D0 inclusive $J/\psi$ cross section using isolation/impact-parameter separation;
$b$-fraction $f_b=0.35\pm0.09\pm0.10$; $\chi_c$ feed-down fraction
$f_\chi=0.32\pm0.07\pm0.07$, smaller than direct-charmonium/fragmentation
predictions \\
\cite{Abbott:1998wf} & 1.8 TeV & 9.8~pb$^{-1}$ & $J/\psi$ &
First $J/\psi$ cross section in the forward region $2.5\leq|\eta|\leq3.7$ using SAMUS;
described by color-octet+color-singlet+$b$-decay model across the full rapidity range \\
\cite{Abazov:2008za} & 1.96 TeV & 1.3~fb$^{-1}$ & $\Upsilon(1S,2S)$ &
Significant $p_T$-dependent \emph{longitudinal} polarization observed for $\Upsilon(1S)$,
in tension with both NRQCD and the earlier CDF $\Upsilon(1S)$ result \\
\cite{Abazov:2014qba} & 1.96 TeV & 8.1~fb$^{-1}$ & $J/\psi J/\psi$ &
First observation of double $J/\psi$ production at the Tevatron; SP/DP separation via
$|\Delta\eta(J/\psi,J/\psi)|$; $\sigma_{\rm eff}=4.8\pm0.5\pm2.5$~mb \\
\cite{Abazov:2015fbq} & 1.96 TeV & 8.1~fb$^{-1}$ & $J/\psi+\Upsilon$ &
First evidence (3.2$\sigma$) for simultaneous $J/\psi+\Upsilon$ production;
$\sigma_{\rm eff}=2.2\pm0.7\pm0.9$~mb, indicating a very compact gluonic core \\
\hline\hline
\end{tabular}
\end{table}

\subsection{Inclusive $J/\psi$ and Associated $b$-Hadron Production}
\label{sec:tevatron_jpsi_xsec}
\begin{figure}[htbp]
\centering
% Source: Acosta et al. 2004 (hep-ex/0412071), Fig. 1, p. 15
\includegraphics[width=0.55\textwidth]{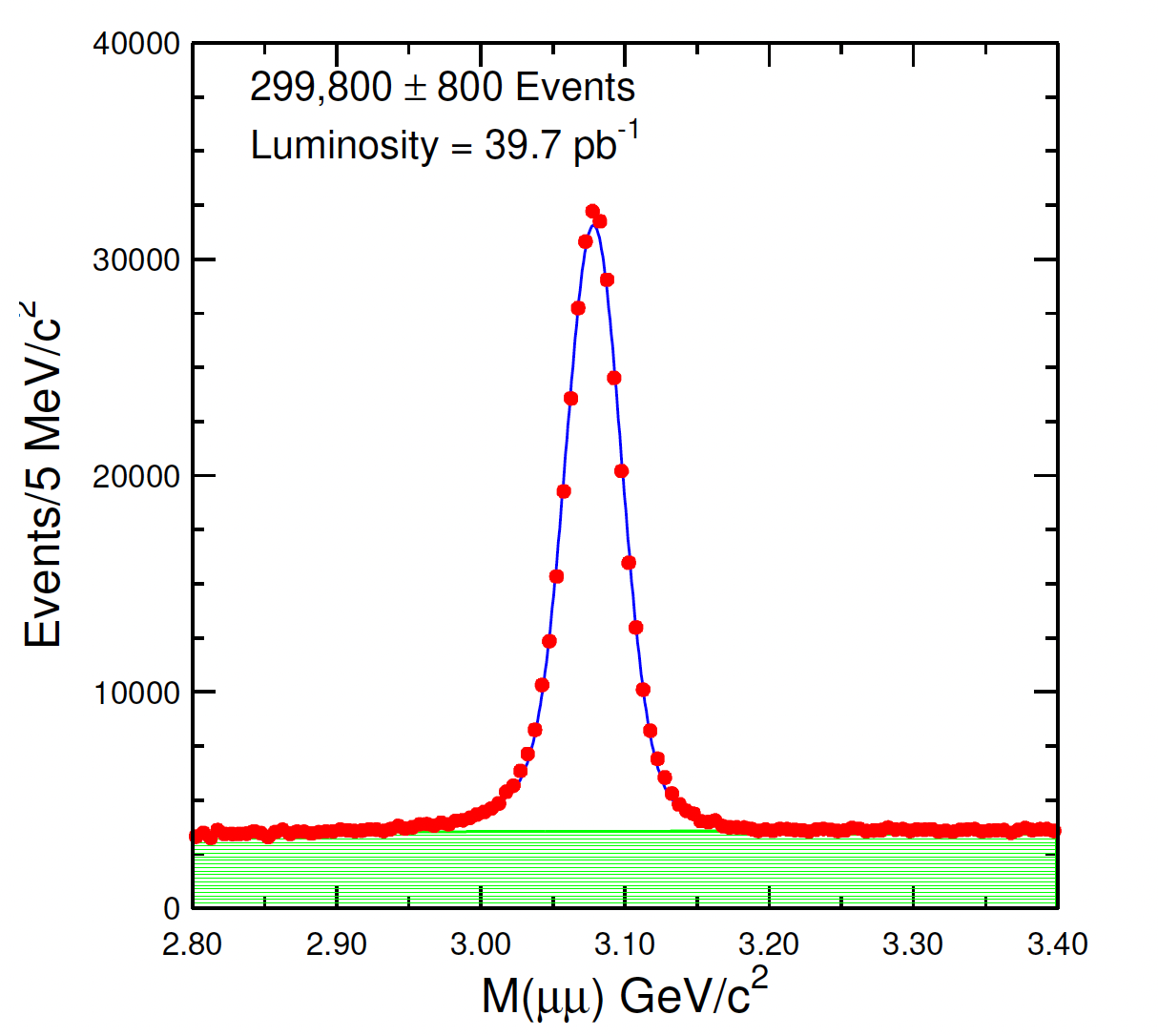}
\caption{Invariant-mass distribution of reconstructed $J/\psi\to\mu^+\mu^-$ candidates in
the full CDF Run~II sample, showing $299{,}800\pm800$ signal events over a linear
background. Figure reproduced from Ref.~\cite{Acosta:2004yw}, Fig.~1.}
\label{fig:tevatron_jpsi_mass}
\end{figure}

The CDF Tevatron Run~II measurement \cite{Acosta:2004yw} used $39.7\pm2.3$~pb$^{-1}$
of $p\bar{p}$ collisions at $\sqrt{s}=1960$~GeV, collected with an upgraded di-muon trigger
capable of reconstructing $J/\psi\to\mu^+\mu^-$ candidates with transverse momentum down to
zero. This represented a qualitative advance over earlier Tevatron measurements, which had
been restricted to $p_T(J/\psi)>5$~GeV/$c$ and covered only 10--13\% of the inclusive
$p_T$ spectrum. Using the central sample at $|y|<0.6$
(Fig.~\ref{fig:tevatron_jpsi_mass}), the integrated inclusive cross section
was determined to be
\begin{equation}
\sigma(p\bar{p}\to J/\psi X,\,|y(J/\psi)|<0.6)\,
\mathcal{B}(J/\psi\to\mu^+\mu^-)
= 4.08\pm0.02\,(\mathrm{stat.})^{+0.36}_{-0.33}\,(\mathrm{syst.})\;\mu\mathrm{b}.
\label{eq:cdf_jpsi_incl_xsec}
\end{equation}

For measurements extrapolated beyond the fiducial phase space defined by
the daughter-muon selection, the differential production cross section is
obtained from the fitted signal yield after correcting for detector
acceptance and reconstruction efficiency:
\begin{equation}
\mathcal{B}\frac{d^2\sigma}{dp_T\,dy}
=
\frac{N(p_T,y)}
{\mathcal{L}\,\Delta p_T\,\Delta y\,
\epsilon(p_T,y)\,
A(p_T,y;\lambda_\theta,\lambda_\phi,\lambda_{\theta\phi})}.
\label{eq:differential_cross_section}
\end{equation}
Here, $N$ is the fitted signal yield, $\mathcal{L}$ is the integrated
luminosity, and $\epsilon$ denotes the reconstruction and trigger
efficiency for events within the detector acceptance. The acceptance
$A$ accounts for the fraction of quarkonium decays satisfying the
daughter-muon kinematic and geometric requirements. In general, it
depends on the assumed polarization of the quarkonium state, since
the decay angular distribution determines the directions of the
daughter muons. Consequently, an unpolarized simulation corresponds
to a particular acceptance assumption rather than a polarization-
independent correction. When polarization is not measured
simultaneously, the associated acceptance variation can be evaluated
for alternative polarization scenarios and included as a systematic
uncertainty.

A different observable is the fiducial cross section, which is defined
within an explicitly specified phase space of the daughter muons.
For this observable, no extrapolation beyond the fiducial muon
kinematic requirements is applied, and the cross section is obtained
by correcting the observed yield for the efficiency within that
fiducial region:
\begin{equation}
\mathcal{B}\frac{d^2\sigma_{\mathrm{fid}}}{dp_T\,dy}
=
\frac{N_{\mathrm{fid}}(p_T,y)}
{\mathcal{L}\,\Delta p_T\,\Delta y\,
\epsilon_{\mathrm{fid}}(p_T,y)}.
\label{eq:fiducial_differential_cross_section}
\end{equation}
The fiducial definition must be accompanied by the corresponding
daughter-muon kinematic requirements. Unlike a cross section
extrapolated to the full decay phase space, a fiducial cross section
does not require an acceptance correction for the phase space outside
those requirements. It therefore avoids the corresponding
extrapolation and its dependence on the assumed polarization,
although the detector efficiency within the fiducial region may
still depend on the polarization.

The CDF inclusive $J/\psi$ result was a central experimental driver of the
so-called color-singlet-model (CSM) crisis. Early Run~I measurements showed
that the prompt $J/\psi$ rate at moderate-to-high $p_T$ exceeded the
leading-order CSM prediction by factors of $30$--$50$; color-singlet
fragmentation reduced, but did not remove, the discrepancy
\cite{CDF:1997psi2S,Abe:1997yz}. Several explanations were advanced, including
higher-order QCD corrections, soft-color interactions, the color-evaporation
model, and the color-octet mechanism of NRQCD
\cite{Bodwin:1994jh,Braaten:1994vv}. In NRQCD, production of a color-octet
$c\bar{c}$ pair followed by a non-perturbative transition to the observed
$J/\psi$ can accommodate the Tevatron rates once the long-distance matrix
elements are fitted.

Next-to-leading-order calculations and global fits to HERA, RHIC, Tevatron, and
early LHC data describe the measured $p_T$ spectra reasonably well
\cite{Butenschoen:2010rq,Ma:2010jj,Gong:2012ru}. That success makes
polarization the sharper discriminator: different matrix-element
combinations can reproduce a $p_T$ spectrum, whereas the angular distribution
tests the spin state directly. The CDF and D0 polarization measurements below
are therefore the central Tevatron test of the production mechanism, rather
than a separate historical narrative.

\begin{figure}[htbp]
\centering
% Source: Acosta et al. 2004 (hep-ex/0412071), Fig. 8, p. 33
\includegraphics[width=0.6\textwidth]{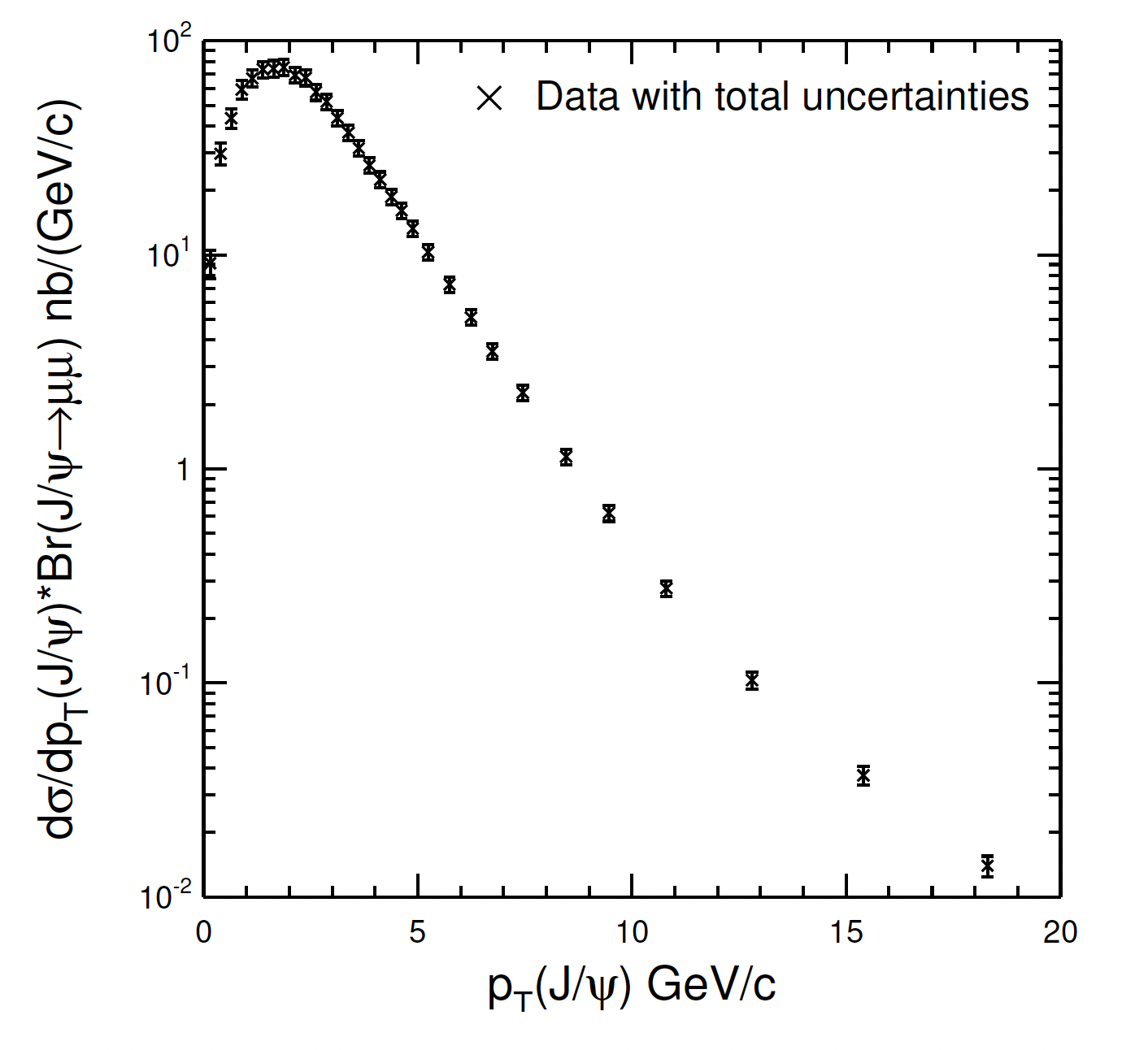}
\caption{Inclusive differential $J/\psi$ cross section $d\sigma/dp_T\cdot
\mathrm{Br}(J/\psi\to\mu\mu)$ as a function of $p_T(J/\psi)$, integrated over $|y|<0.6$,
measured over the full range $0<p_T<20$~GeV/$c$ for the first time at the Tevatron.
Figure reproduced from \cite{Acosta:2004yw}, Fig.~8.}
\label{fig:tevatron_jpsi_diffxsec}
\end{figure}

\begin{figure}[htbp]
\centering
% Source: Acosta et al. 2004 (hep-ex/0412071), Fig. 13, p. 43
\includegraphics[width=0.55\textwidth]{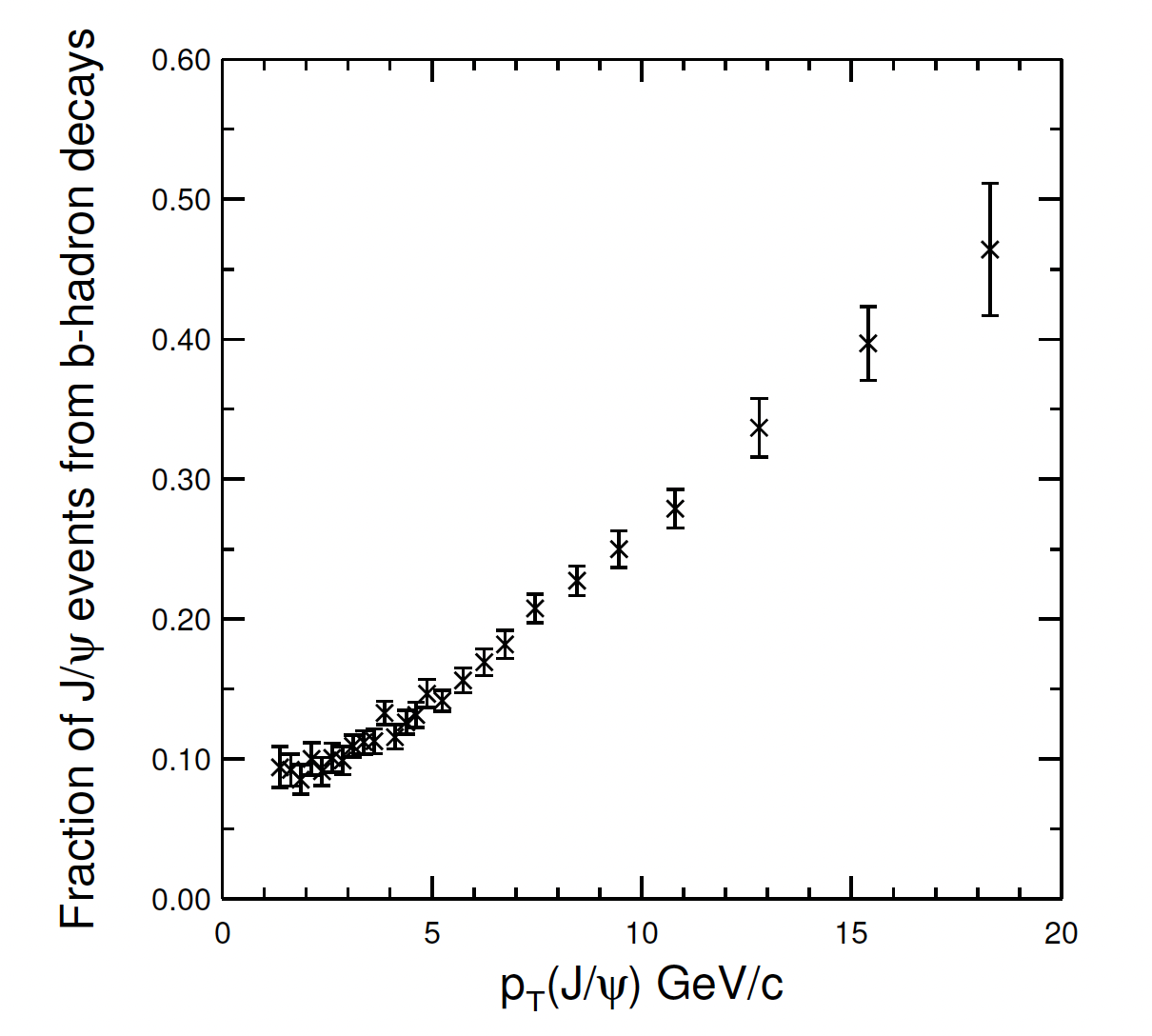}
\caption{Fraction $f_b$ of prompt and non-prompt $J/\psi$ candidates originating
from $b$-hadron decays in the CDF Tevatron Run~II $p\bar p$ sample, shown as a
function of $J/\psi$ transverse momentum and extracted from the pseudo-proper
decay-length fit. Figure
reproduced from \cite{Acosta:2004yw}, Fig.~13.}
\label{fig:tevatron_bfraction}
\end{figure}

\begin{figure}[htbp]
\centering
% Source: Acosta et al. 2004 (hep-ex/0412071), Fig. 16, p. 52
\includegraphics[width=0.6\textwidth]{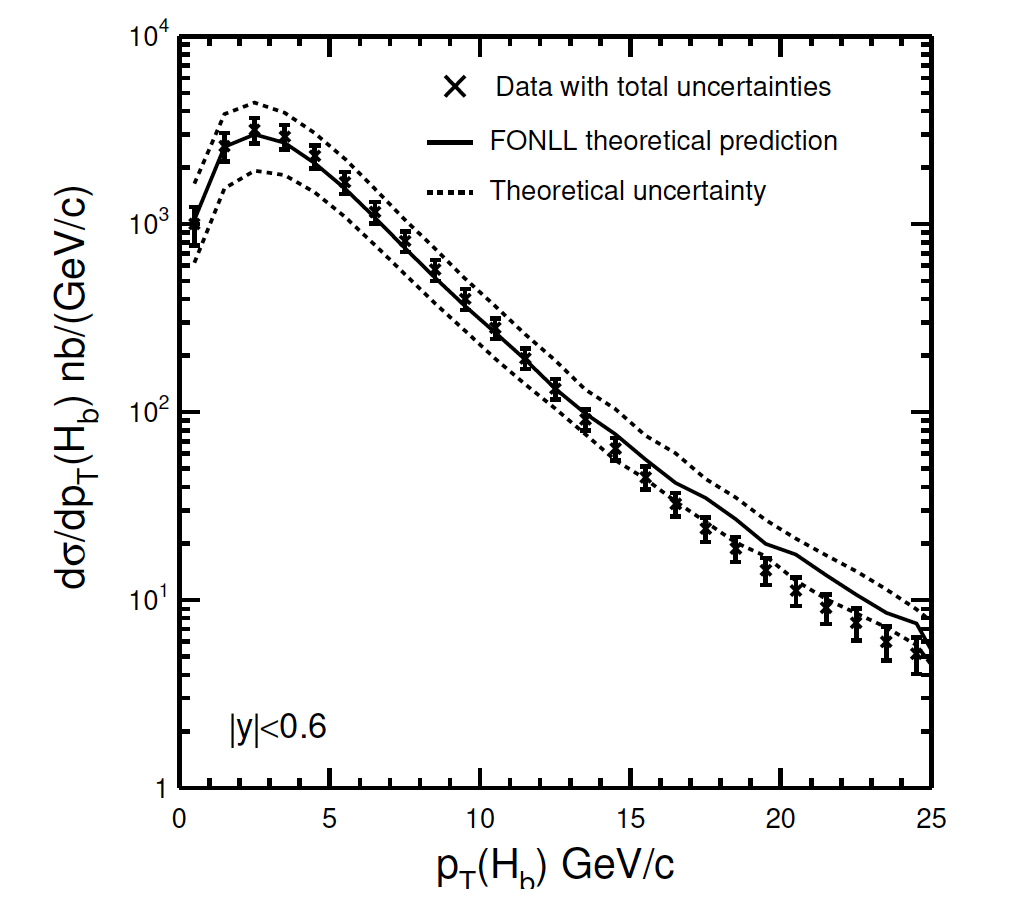}
\caption{CDF measurement in Tevatron Run~II $p\bar p$ collisions of the
differential $b$-hadron production cross section as a function of
$p_T(H_b)$, compared to the FONLL theoretical prediction (solid line) and its
uncertainty band (dotted lines). Figure reproduced from
\cite{Acosta:2004yw}, Fig.~16.}
\label{fig:tevatron_hb_diffxsec}
\end{figure}

\begin{figure}[htbp]
\centering
% Source: Acosta et al. 2004 (hep-ex/0412071), Fig. 18, p. 55
\includegraphics[width=0.6\textwidth]{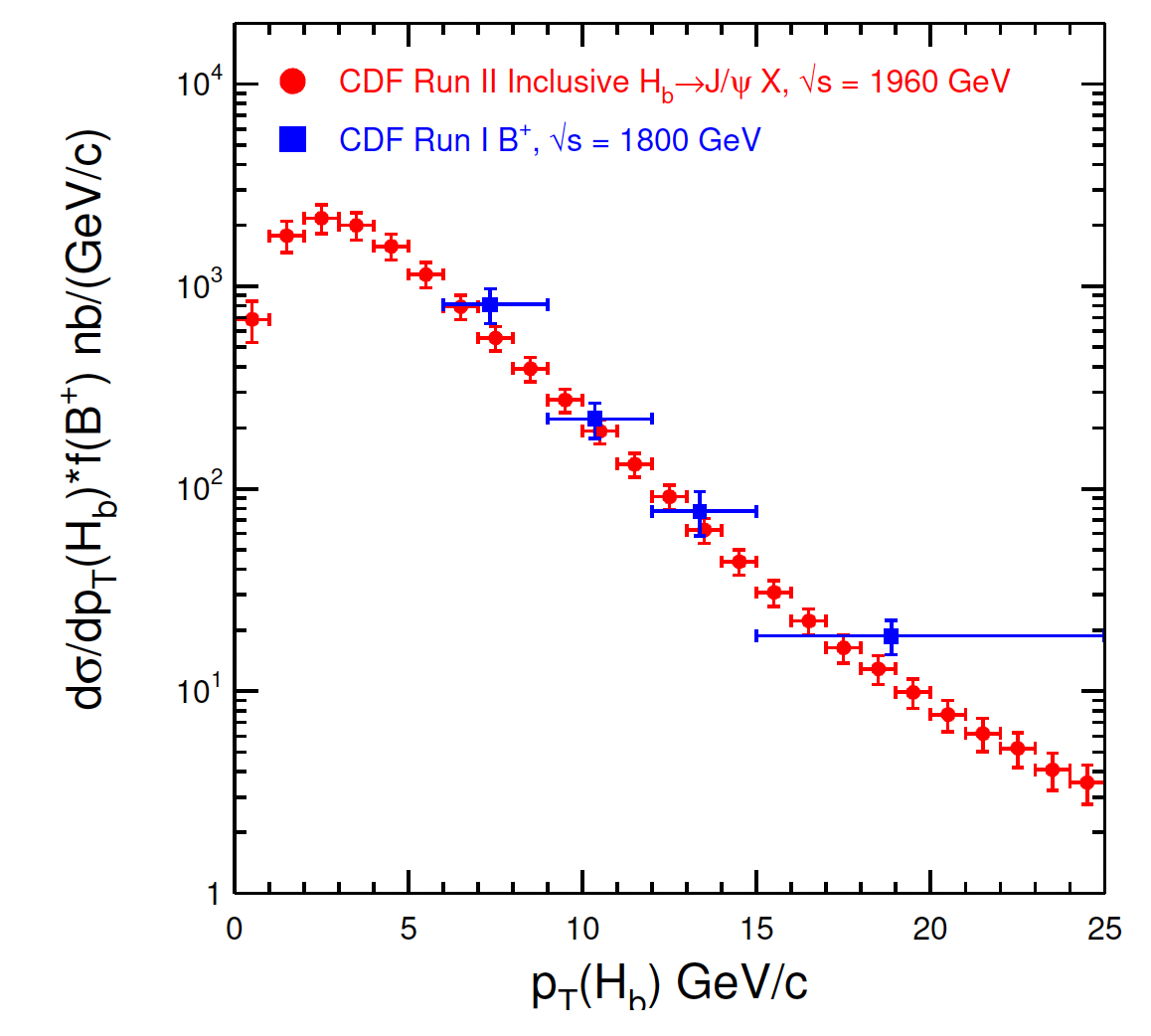}
\caption{Comparison of the CDF Tevatron Run~II inclusive $H_b\to J/\psi X$ differential cross section
(scaled by the $B^+$ fragmentation fraction) with the direct Run~I $B^+$ cross-section
measurement, both integrated over $|y(H_b)|<1.0$. Figure reproduced from \cite{Acosta:2004yw}, Fig.~18.}
\label{fig:tevatron_run1_run2}
\end{figure}

Separation of prompt $J/\psi$ production from the decays of long-lived $b$-hadrons was
achieved in the CDF analysis~\cite{Acosta:2004yw} using the pseudo-proper decay length,
$x=L_{xy}(J/\psi)M(J/\psi)/p_T(J/\psi)$, where $L_{xy}$ is the
transverse distance from the primary $p\bar p$ vertex to the $J/\psi$
decay vertex. A joint fit to the mass and pseudo-proper-decay-time
distribution in $p_T$ bins extracted the fraction originating from
$H_b\to J/\psi X$ decays. The fraction rises smoothly with $p_T$
(Fig.~\ref{fig:tevatron_bfraction}) because the $b$-hadron decay products
have a harder intrinsic $p_T$ spectrum than prompt charmonium.
Using this $b$-fraction together with a Monte Carlo model of the $H_b\to J/\psi X$ decay
kinematics, the differential and integrated $H_b\to J/\psi X$ cross sections were extracted
(Fig.~\ref{fig:tevatron_jpsi_diffxsec} and Fig.~\ref{fig:tevatron_hb_diffxsec}), and, for the
first time at a hadron collider, unfolded down to zero $b$-hadron transverse momentum to
obtain the total inclusive single $b$-hadron cross section,
\begin{equation}
\sigma(p\bar{p}\to H_bX,\,|y|<0.6) = 17.6\pm0.4\,(\mathrm{stat.})^{+2.5}_{-2.3}\,(\mathrm{syst.})\;\mu\mathrm{b}.
\label{eq:cdf_hb_xsec}
\end{equation}
This result was found to be in good agreement with the fixed-order-plus-next-to-leading-log
(FONLL) QCD calculation \cite{Cacciari:2012fonll},
$\sigma^{\rm FONLL}_{(|y|<0.6)}=16.8^{+7.0}_{-5.0}\,\mu$b, resolving
the factor-of-two-to-three excess over NLO predictions that had persisted from the Run~I
measurements at $\sqrt{s}=1800$~GeV. A direct comparison of the extracted Run~II $b$-hadron
spectrum with the Run~I $B^+$ measurement (Fig.~\ref{fig:tevatron_run1_run2}) showed
consistent shapes across the full measured $p_T$ range, confirming the expected mild
energy-scaling behavior between $\sqrt{s}=1.8$ and $1.96$~TeV.

\subsection{D0 Run~I Inclusive $J/\psi$ Production and Charmonium Production Mechanisms}
\label{sec:tevatron_d0_jpsi_run1}

\begin{figure}[htbp]
\centering
% Source: Abachi et al. 1996 (FERMILAB-Pub-96/003-E), Fig. 2, p. 5
\includegraphics[width=0.55\textwidth]{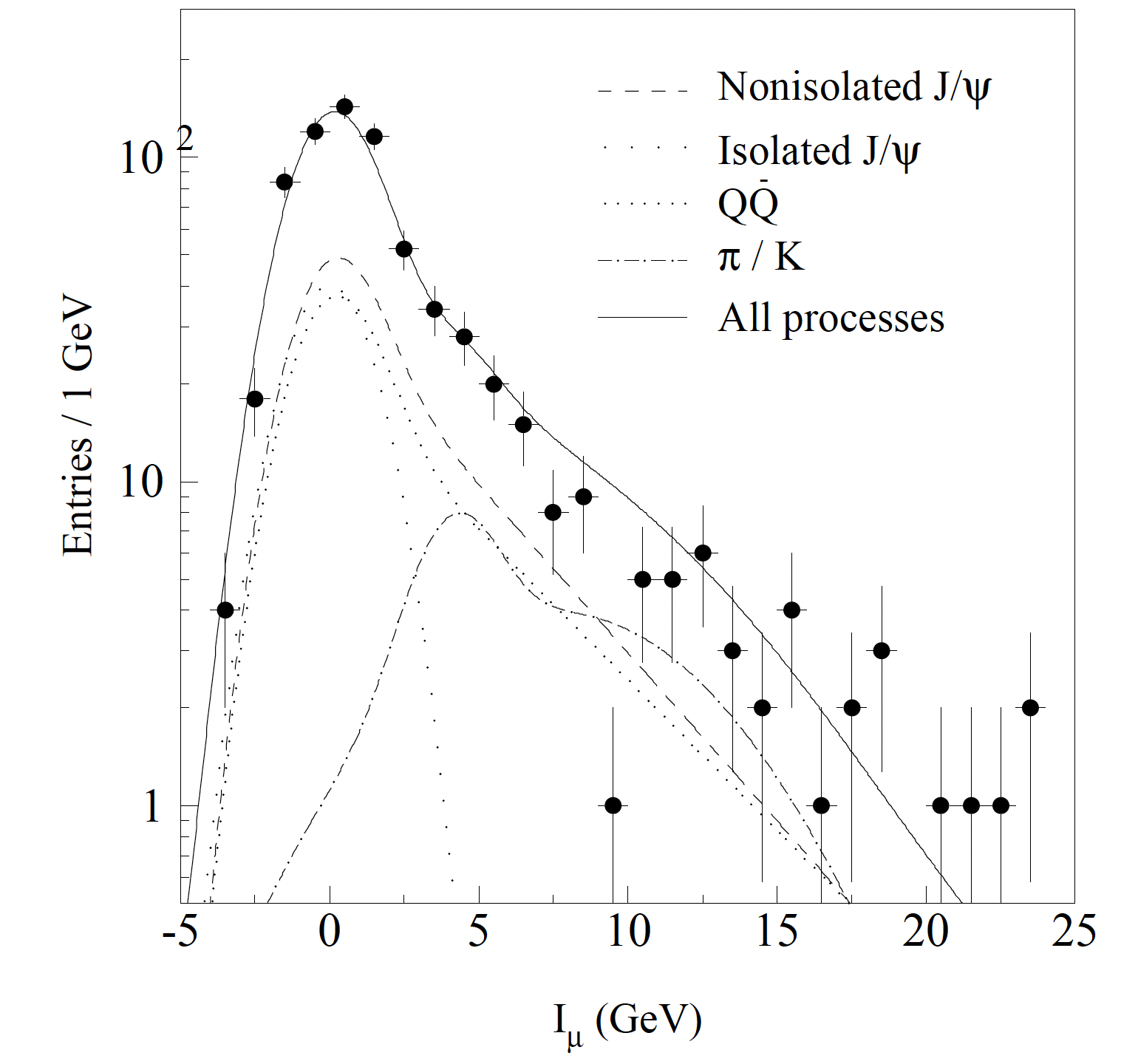}
\caption{Muon isolation energy $I_\mu$ distribution for opposite-sign dimuon events with
$2<M_{\mu\mu}<4.4$~GeV/$c^2$, showing the fitted sum of the $J/\psi$ signal and the
nonisolated, isolated, $Q\bar Q$, and $\pi/K$ background contributions used to determine
$f_{\rm nonisol}$. Figure reproduced from Ref.~\cite{Abachi:1996fu},
Fig.~2.}
\label{fig:tevatron_d0_1996_isolation}
\end{figure}

Almost a decade before the CDF Run~II measurement of Section~\ref{sec:tevatron_jpsi_xsec},
the D0 Collaboration reported the first D0 measurement of inclusive $J/\psi$ production in
$p\bar{p}$ collisions at $\sqrt{s}=1.8$~TeV~\cite{Abachi:1996fu}, based on $6.6$--$13$~pb$^{-1}$
of Run~I data. Rather than relying on the pseudo-proper decay length alone, the D0 analysis
used a complementary source-separation strategy. Opposite-sign dimuon events with
$M_{\mu\mu}<6$~GeV/$c^2$ were statistically decomposed into six physical sources --
$B\to J/\psi X$, direct charmonium
production, $b\bar b/c\bar c$ semileptonic (``$Q\bar Q$'') decays, associated $\pi^\pm/K^\pm$
decays, Drell--Yan production, and light-meson ($\rho,\phi,\eta$) decays -- using a combined
likelihood fit to the dimuon mass $M_{\mu\mu}$, the calorimeter isolation of the leading muon
$I_\mu$, and the momentum of the dimuon system transverse to any nearby jet axis,
$p_{T\,\mathrm{rel}}^{*}$ (Fig.~\ref{fig:tevatron_d0_1996_isolation}). This isolation-based
approach treated $B\to J/\psi X$ decays as a template for all ``nonisolated'' (jet-embedded)
$J/\psi$ production, including gluon-fragmentation contributions for which no simulation was
then available, while direct charmonium production served as the template for ``isolated''
sources.

\begin{figure}[htbp]
\centering
% Source: Abachi et al. 1996 (FERMILAB-Pub-96/003-E), Fig. 4, p. 7
\includegraphics[width=0.6\textwidth]{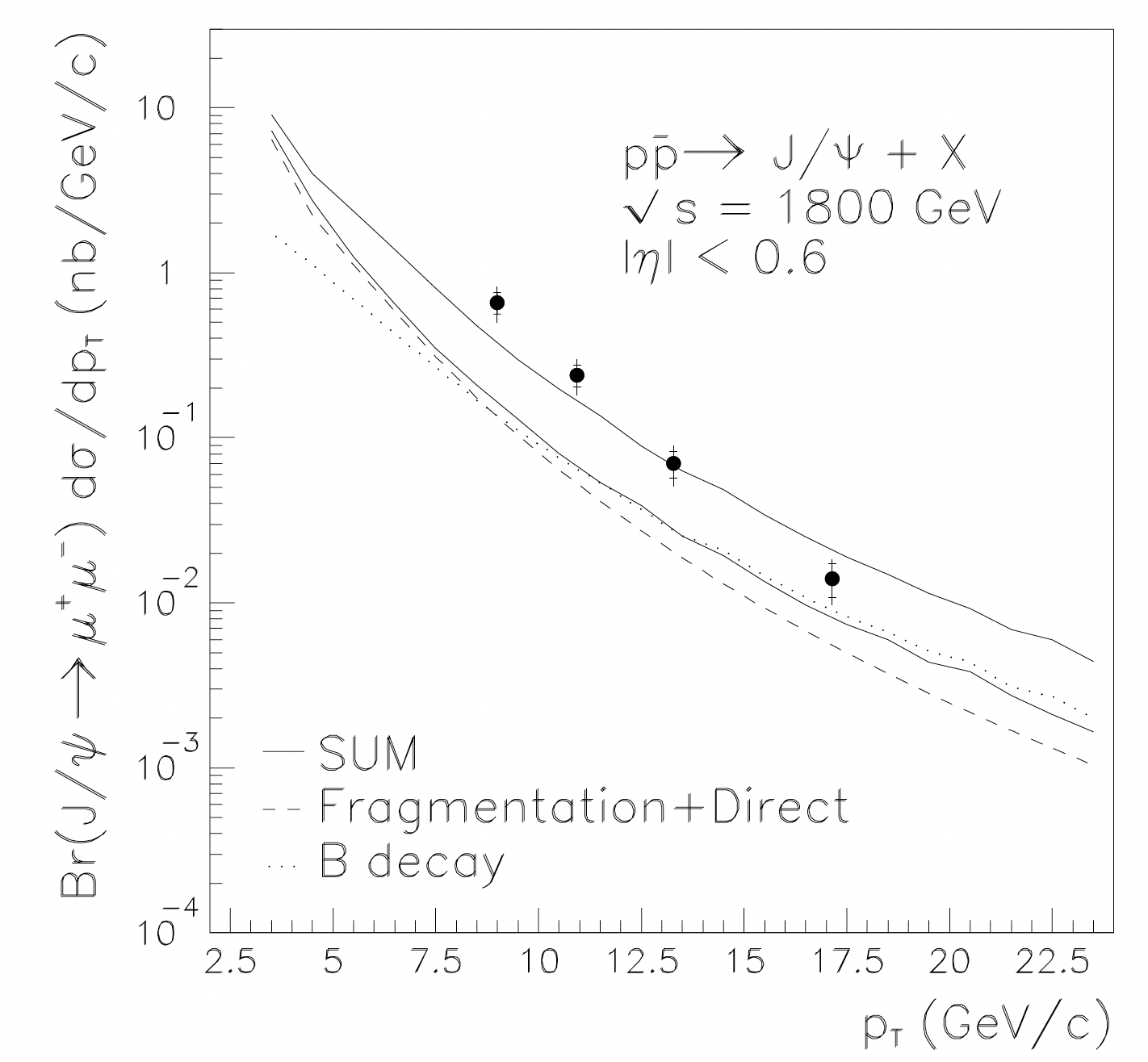}
\caption{D0 measurement of
$d\sigma/dp_T\cdot\mathcal{B}(J/\psi\to\mu^+\mu^-)$ for inclusive
$J/\psi$ production in $p\bar p$ collisions at $\sqrt{s}=1.8$~TeV. The dotted
curve is the $B\to J/\psi$ contribution, the dashed curve is prompt $J/\psi$
production, and the solid lines delimit the theoretical sum with its
uncertainty. Figure reproduced from \cite{Abachi:1996fu}, Fig.~4.}
\label{fig:tevatron_d0_1996_diffxsec}
\end{figure}

\begin{figure}[htbp]
\centering
% Source: Abachi et al. 1996 (FERMILAB-Pub-96/003-E), Fig. 5, p. 8
\includegraphics[width=0.55\textwidth]{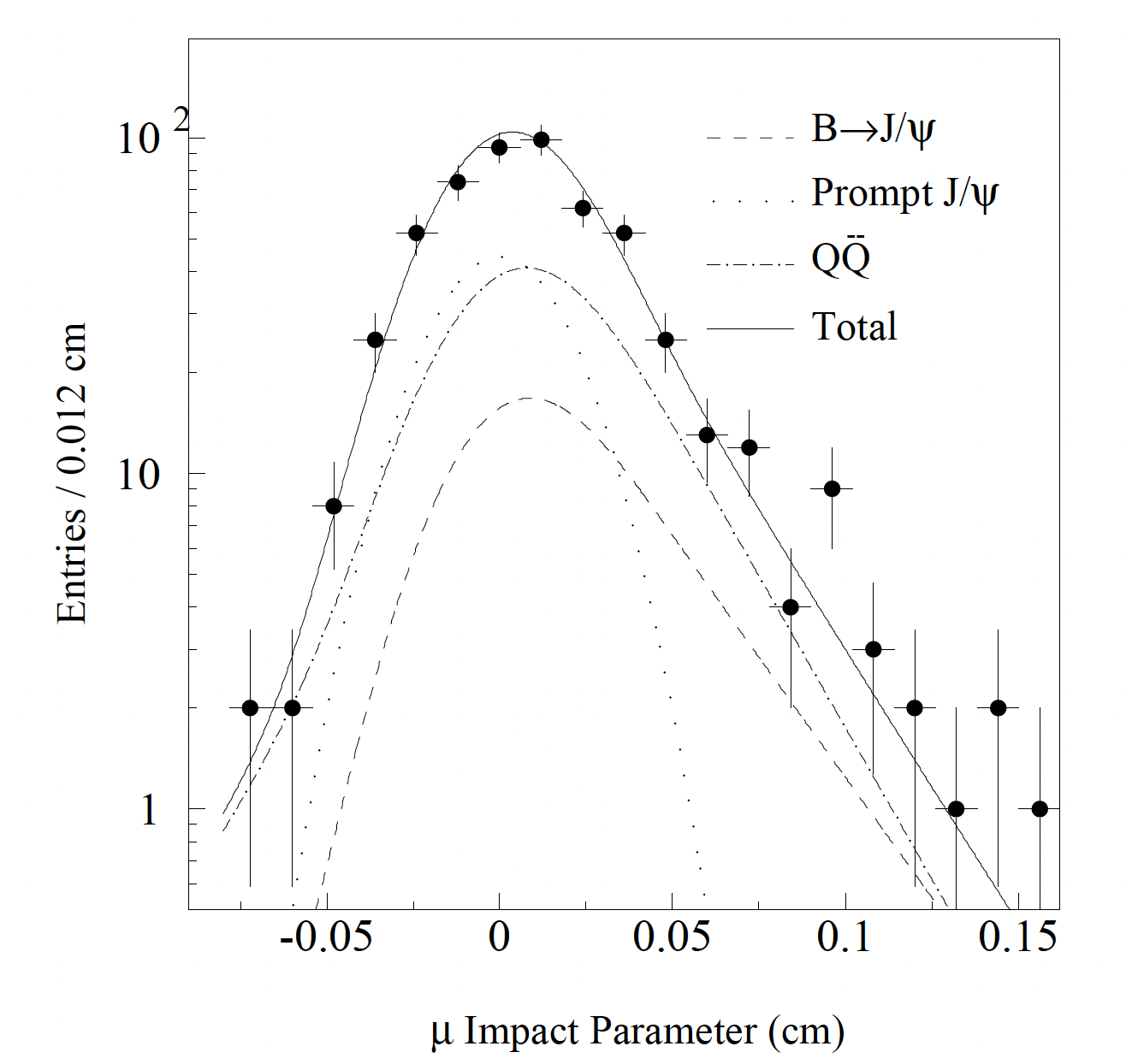}
\caption{Distribution of the muon impact parameter with respect to the primary
event vertex in the transverse plane for muons in the $J/\psi$ region of the
D0 $p\bar p$ data at $\sqrt{s}=1.8$~TeV, with the fitted contributions from
prompt $J/\psi$, $B$-produced $J/\psi$, and the $Q\bar Q$ continuum. Figure reproduced from
\cite{Abachi:1996fu}, Fig.~5.}
\label{fig:tevatron_d0_1996_impactparam}
\end{figure}

The fit yielded $407\pm28\,(\mathrm{stat.})\pm55\,(\mathrm{syst.})$ $J/\psi$
events in the 6.6~pb$^{-1}$ subsample. It also determined a nonisolated
fraction
$f_{\rm nonisol}=0.64\pm0.08\,(\mathrm{stat.})\pm0.06\,(\mathrm{syst.})$.
The unfolded and acceptance-corrected differential cross section
(Fig.~\ref{fig:tevatron_d0_1996_diffxsec}), integrated over $p_T(J/\psi)>8$~GeV/$c$ and
$|\eta(J/\psi)|<0.6$, gives
\begin{equation}
\small
\sigma(p\bar{p}\to J/\psi X,\,p_T>8~\mathrm{GeV}/c,\,|\eta|<0.6)
\cdot\mathcal{B}(J/\psi\to\mu^+\mu^-)
= 2.08\pm0.17\,(\mathrm{stat.})\pm0.46\,(\mathrm{syst.})\;\mathrm{nb},
\label{eq:d0_jpsi_1996_xsec}
\end{equation}

in agreement with the contemporaneous UA1 and CDF measurements. Using the transverse muon
impact parameter relative to the event vertex (Fig.~\ref{fig:tevatron_d0_1996_impactparam}),
a simultaneous mass-and-impact-parameter fit to the larger 13~pb$^{-1}$ sample determined the
fraction of $J/\psi$ mesons from $B$-hadron decay, $f_b=0.35\pm0.09\,(\mathrm{stat.})\pm0.10\,(\mathrm{syst.})$.
From this fraction the inclusive $b$-quark cross section was inferred. It lies at the upper
edge of the contemporary NLO QCD uncertainty band, in excellent agreement with the independent
D0 single-muon measurement.

\begin{figure}[htbp]
\centering
% Source: Abachi et al. 1996 (FERMILAB-Pub-96/003-E), Fig. 7, p. 10
\includegraphics[width=0.5\textwidth]{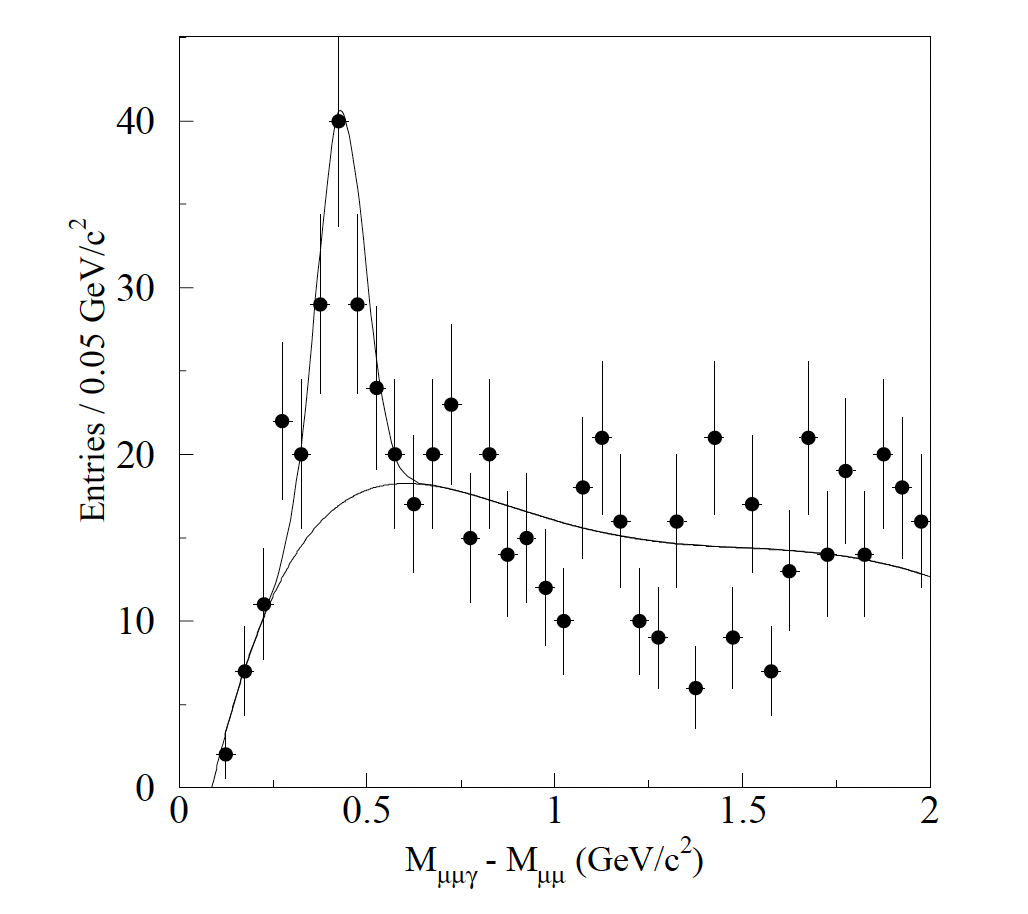}
\caption{Distribution of $\Delta M=M_{\mu\mu\gamma}-M_{\mu\mu}$ for
D0 dimuon-plus-photon candidates in the $J/\psi$ region, showing the
$\chi_c\to J/\psi\,\gamma$ signal near $\Delta M\simeq0.4$~GeV. The fitted
signal yield determines $f_\chi$, the fraction of all $J/\psi$ mesons that
originate from radiative $\chi_c$ decay. Figure reproduced from
\cite{Abachi:1996fu}, Fig.~7.}
\label{fig:tevatron_d0_1996_chic}
\end{figure}

A distinctive feature of this analysis was a direct search for the radiative decay
$\chi_c\to J/\psi\,\gamma$ using full reconstruction of the $J/\psi+\gamma$ decay chain
(Fig.~\ref{fig:tevatron_d0_1996_chic}), motivated by the theoretical expectation -- from both
the direct color-singlet model and gluon/charm-fragmentation
calculations~\cite{Braaten:1994vv} -- that $P$-wave $\chi_c$ states should dominate prompt $J/\psi$
production at the Tevatron. The measured feed-down fraction,
$f_\chi=0.32\pm0.07\,(\mathrm{stat.})\pm0.07\,(\mathrm{syst.})$, is smaller than the
$f_\chi=0.45\pm0.05\pm0.15$ reported by CDF and, combined with $f_b$ and the known
$B\to\chi_c\to J/\psi$ branching chain, implies that a large fraction,
$1-f_b-f_\chi+0.08=0.41\pm0.17$, of all $J/\psi$ mesons originate from neither $B$-hadron
decay nor $\chi_c$ radiative decay. This ``missing'' direct-$J/\psi$ component, in direct
tension with the $\chi_c$-dominated predictions of the contemporary direct-production and
fragmentation models, is among the earliest pieces of Tevatron evidence pointing toward the
need for an additional short-distance production mechanism -- later understood within NRQCD
as color-octet ${}^1S_0^{[8]}$ and ${}^3S_1^{[8]}$ production -- that does not proceed through
an intermediate $\chi_c$ state.

\subsection{D0 Forward-Rapidity (``Small-Angle'') $J/\psi$ Production}
\label{sec:tevatron_d0_smallangle}

A second, complementary D0 Run~I measurement extended the kinematic reach of charmonium
hadroproduction studies into a rapidity region entirely inaccessible to CDF. Using the Small
Angle MUon Spectrometer (SAMUS), which provides muon coverage over $2.2<|\eta_\mu|<3.3$,
The D0 Collaboration \cite{Abbott:1998wf} reported the first measurement of inclusive
$J/\psi$ production in the forward pseudorapidity interval $2.5\leq|\eta(J/\psi)|\leq3.7$, in
$9.8\pm0.5$~pb$^{-1}$ of $\sqrt{s}=1.8$~TeV data. 

\begin{figure}[htbp]
\centering
% Source: Abbott et al. 1998 (hep-ex/9807029), Fig. 2, p. 4
\includegraphics[width=0.6\textwidth]{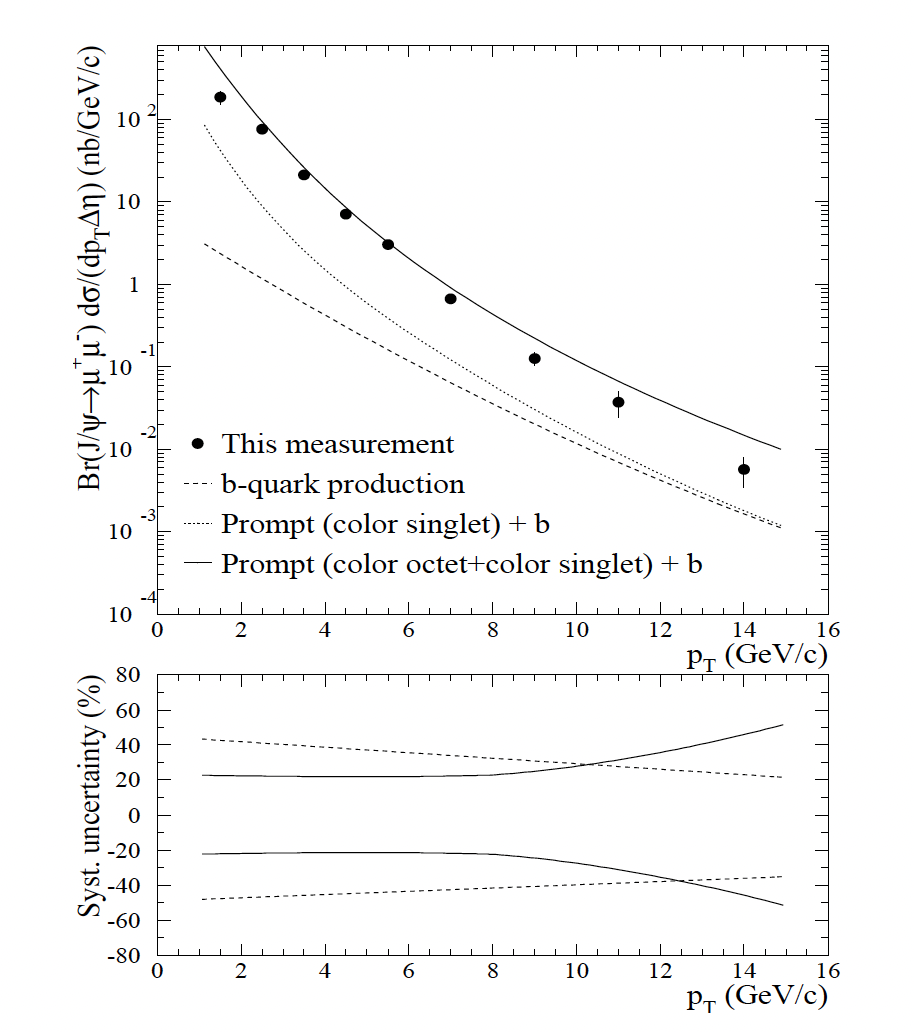}
\caption{$p_T$ dependence of the forward-rapidity $J/\psi$ differential cross section
(points) compared with the $b$-quark-only, color-singlet-plus-$b$, and
color-octet-plus-color-singlet-plus-$b$ predictions (upper panel), and the corresponding
total systematic uncertainty, including the polarization-dependent acceptance band (lower
panel). Figure reproduced from \cite{Abbott:1998wf}, Fig.~2.}
\label{fig:tevatron_d0_1998_diffxsec}
\end{figure}

\begin{figure}[htbp]
\centering
% Source: Abbott et al. 1998 (hep-ex/9807029), Fig. 3, p. 5 -- KEY FIGURE
\includegraphics[width=0.55\textwidth]{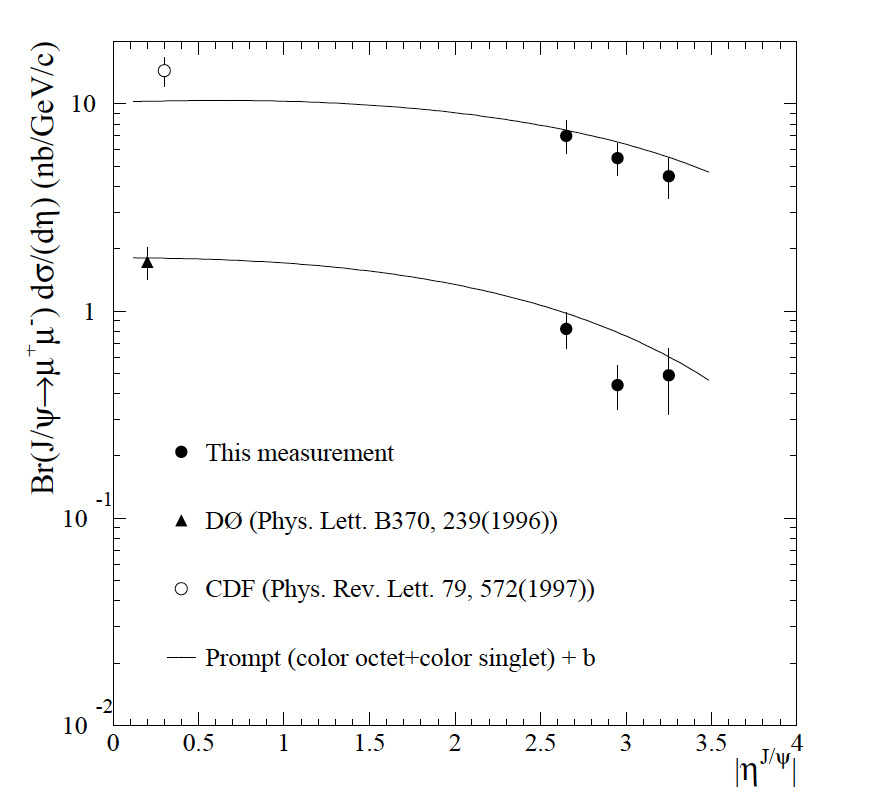}
\caption{Pseudorapidity dependence of the $J/\psi$ production cross section for
$p_T>5$~GeV/$c$ and $p_T>8$~GeV/$c$, combining this forward measurement with the earlier
central-rapidity D0 and CDF results and compared to the color-octet-plus-color-singlet-plus-$b$
prediction. Figure reproduced from \cite{Abbott:1998wf}, Fig.~3.}
\label{fig:tevatron_d0_1998_etadep}
\end{figure}

The resulting inclusive $p_T$ spectrum, unfolded for momentum and angular
smearing and averaged over the fixed interval $2.5\leq|\eta(J/\psi)|\leq3.7$ (Fig.~\ref{fig:tevatron_d0_1998_diffxsec} and Table~\ref{tab:tevatron_d0_1998_diffxsec}),
falls by roughly four orders of magnitude between $p_T=1.5$ and $14$~GeV/$c$ and is well
described, within the sizable ($\sim20$--$47$\%) systematic uncertainties, by the sum of the
NLO $b$-quark contribution and the color-octet-plus-color-singlet direct-charmonium
prediction of Cho and Leibovich~\cite{Cho:1995vh}, increased by 12\% to account for
$\psi(2S)$ feed-down; the color-singlet-only prediction falls well short of the data, exactly
as observed at central rapidity. Comparing the $\eta$-dependence of the cross section for
$p_T>5$ and $p_T>8$~GeV/$c$ with the earlier central-rapidity D0 and CDF measurements
(Fig.~\ref{fig:tevatron_d0_1998_etadep}) shows that the same color-octet-plus-$b$-decay model
consistently describes the full measured pseudorapidity range, $0\leq|\eta(J/\psi)|\lesssim3.7$,
providing an important cross-check of the universality of the color-octet mechanism away from
central rapidity and additional leverage on the low-$x$ gluon distribution probed by forward
charmonium production.

\begin{table}[htbp]
\centering
\caption{D0 forward-rapidity inclusive $J/\psi$ cross section
$\mathcal{B}(J/\psi\to\mu^+\mu^-)\,d\sigma/dp_T$ (nb/GeV/$c$), averaged over
$2.5\leq|\eta^{J/\psi}|\leq3.7$~\cite{Abbott:1998wf}.}
\label{tab:tevatron_d0_1998_diffxsec}
\begin{tabular}{cc}
\hline\hline
$p_T^{J/\psi}$ (GeV/$c$) & $\mathcal{B}\,d\sigma/dp_T$ (nb/GeV/$c$) \\
\hline
1.5 & $187\pm34$ \\
3.5 & $21.3\pm2.1$ \\
5.5 & $3.03\pm0.36$ \\
7.0 & $0.667\pm0.075$ \\
9.0 & $0.126\pm0.023$ \\
11.0 & $0.037\pm0.013$ \\
14.0 & $0.0057\pm0.0023$ \\
\hline\hline
\end{tabular}
\end{table}

\subsection{$\psi(2S)$ Hadroproduction: A Feed-Down-Free Charmonium Probe}
\label{sec:tevatron_psi2s_xsec}

Because no charmonium state with a significant hadronic production cross section decays
into the $\psi(2S)$, this state offers an essentially feed-down-free probe of charmonium
hadroproduction, in contrast to the $J/\psi$, whose inclusive rate contains contributions
from $\chi_{cJ}$ and $\psi(2S)$ decays in addition to direct production -- precisely the
mixture of sources that the D0 analysis of Section~\ref{sec:tevatron_d0_jpsi_run1} attempted
to disentangle statistically. Exploiting this feature, the CDF Collaboration \cite{Aaltonen:2009tha}
used $1.1$~fb$^{-1}$ of CDF~II data to measure the differential $\psi(2S)\to\mu^+\mu^-$ cross
section over the transverse-momentum range $2<p_T<30$~GeV/$c$ and rapidity $|y|<0.6$,
extending the reach of the earlier Run~I measurement by an order of magnitude in statistics.

\begin{figure}[htbp]
\centering
% Source: Aaltonen et al. 2009 (arXiv:0905.1982), Fig. 2, p. 9
\includegraphics[width=0.85\textwidth]{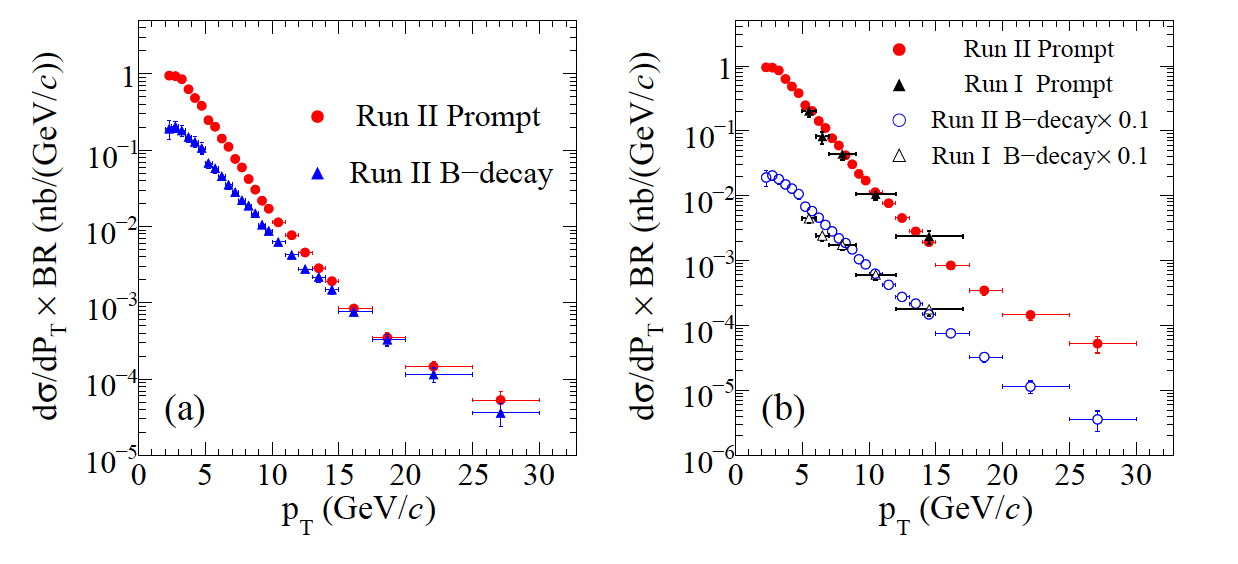}
\caption{(a) Prompt and $B$-decay $\psi(2S)$ differential cross sections as a function of
$p_T$; (b) the same data compared to the corresponding Run~I measurements. Figure
reproduced from \cite{Aaltonen:2009tha}, Fig.~2.}
\label{fig:tevatron_psi2s_diffxsec}
\end{figure}

A joint unbinned maximum-likelihood fit to the dimuon invariant mass and proper decay
length simultaneously separated the signal from combinatorial background and the prompt
component from $H_b\to\psi(2S)X$ decays. The
prompt fraction was found to decrease approximately linearly with $p_T$, from $\sim$87\%
near $p_T=2.3$~GeV/$c$ to $\sim$55--61\% for $p_T>17$~GeV/$c$, reflecting the same relative
hardening of the $b$-hadron decay spectrum observed in the $J/\psi$ analysis. Integrating
the resulting differential cross sections (Fig.~\ref{fig:tevatron_psi2s_diffxsec}) yields
\begin{equation}
\small
\sigma(p\bar{p}\to\psi(2S)X,\,p_T>2~\mathrm{GeV}/c,\,|y|<0.6)\cdot\mathrm{Br}(\psi(2S)\to\mu^+\mu^-)
= 3.29\pm0.04\,(\mathrm{stat.})\pm0.32\,(\mathrm{syst.})\;\mathrm{nb}.
\label{eq:cdf_psi2s_xsec}
\end{equation}
Restricting to $p_T>5$~GeV/$c$ for comparison with the Run~I measurement at
$\sqrt{s}=1800$~GeV, the cross section increases by $(21\pm19)\%$ between $\sqrt{s}=1.8$
and $1.96$~TeV, consistent with the $(14\pm8)\%$ increase predicted from the change in
parton luminosities at the higher collision energy.

\begin{figure}[htbp]
\centering
% Source: Aaltonen et al. 2009 (arXiv:0905.1982), Fig. 3, p. 10
\includegraphics[width=0.85\textwidth]{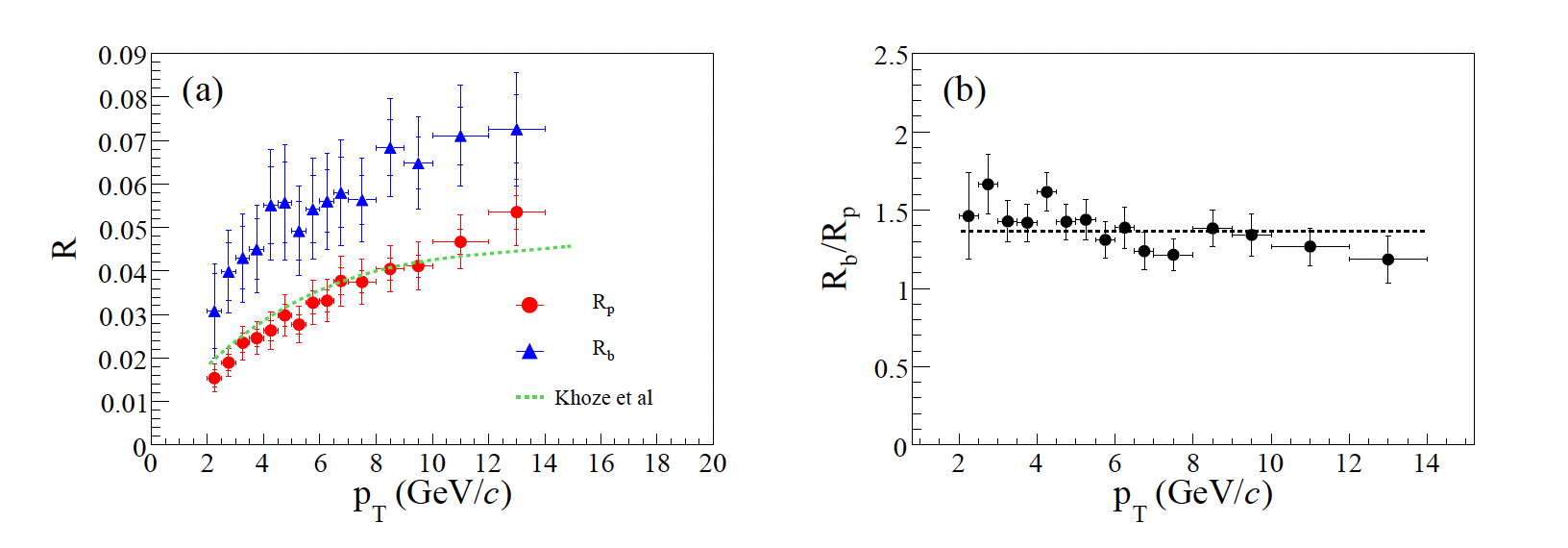}
\caption{(a) Ratio of the $\psi(2S)$ to $J/\psi$ differential cross sections as a function
of $p_T$ for prompt ($R_p$) and $B$-decay ($R_b$) events, compared to the gluon-tower
model prediction of Khoze \textit{et al.}; (b) the ratio $R_b/R_p$, consistent with a
constant. Figure reproduced from \cite{Aaltonen:2009tha}, Fig.~3.}
\label{fig:tevatron_psi2s_ratio}
\end{figure}

A particularly informative observable constructed from this measurement is the ratio of
the prompt $\psi(2S)$ to prompt $J/\psi$ differential cross sections as a function of
$p_T$ (Fig.~\ref{fig:tevatron_psi2s_ratio}). Because the $\psi(2S)$ spectrum is
intrinsically harder than that of the (feed-down-contaminated) $J/\psi$, this ratio
increases steadily with $p_T$; the gluon-tower model of Khoze \textit{et al.}
\cite{Khoze:2005inelastic}
qualitatively reproduces this trend once normalized to the data, while an analogous ratio
constructed from the $H_b$-decay components of both spectra is found to track the prompt
ratio with an approximately constant proportionality factor $R_b/R_p$, for which no
theoretical explanation currently exists.

The physics payoff is twofold. Because the prompt $\psi(2S)$ sample is not
dominated by feed-down from a higher charmonium state, its absolute rate tests
the short-distance production calculation more directly than the inclusive
$J/\psi$ yield. Its rising ratio to the prompt $J/\psi$ further shows that the
two states do not share a single $p_T$ dependence; any successful LDME fit must
reproduce both normalizations and this relative shape.

\subsection{Charmonium Polarization: Confronting the NRQCD Prediction}
\label{sec:tevatron_charmonium_pol}

\begin{figure}[htbp]
\centering
% Source: Abulencia et al. 2007 (arXiv:0704.0638), Fig. 1, p. 5
\includegraphics[width=0.75\textwidth]{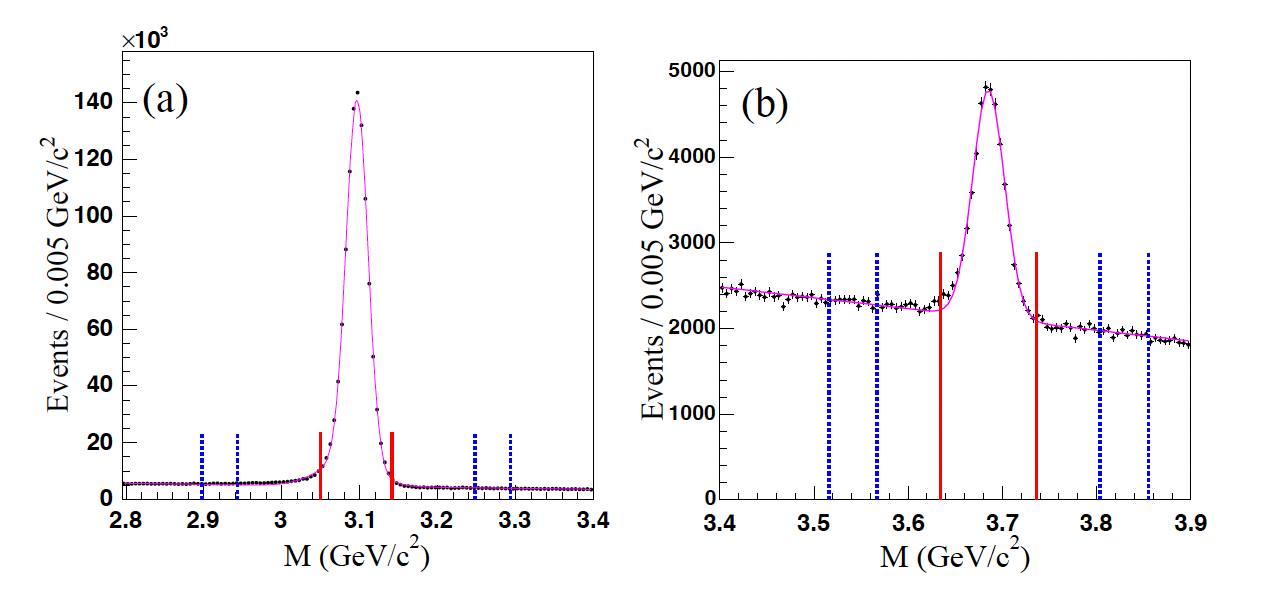}
\caption{Invariant-mass distributions for (a) $J/\psi$ and (b) $\psi(2S)$ candidates used
in the polarization analysis, with the fitted signal (solid) and sideband (dashed) regions
indicated. Figure reproduced from Ref.~\cite{Abulencia:2007us}, Fig.~1.}
\label{fig:tevatron_pol_massmc}
\end{figure}

\begin{figure}[htbp]
\centering
% Source: Abulencia et al. 2007 (arXiv:0704.0638), Fig. 3, p. 6
\includegraphics[width=0.55\textwidth]{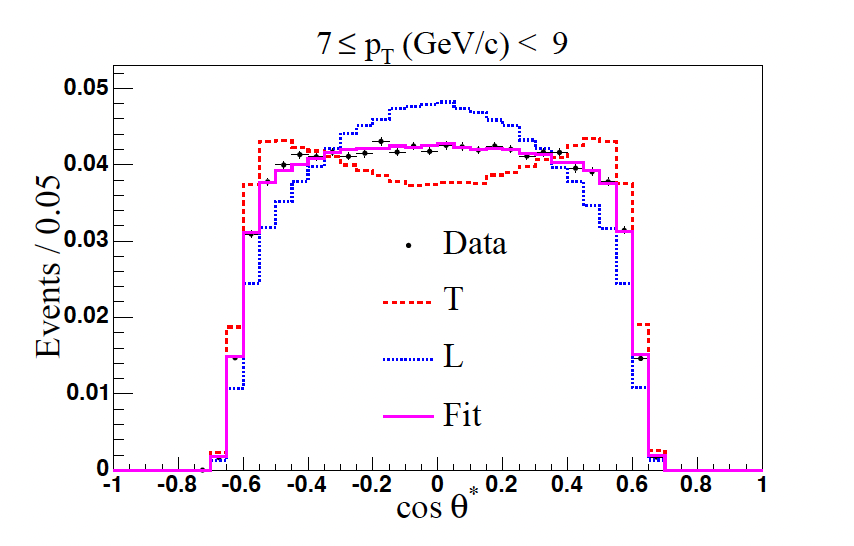}
\caption{$\cos\theta^*$ distribution of data (points) for the worst-fit $J/\psi$ $p_T$
bin ($7\leq p_T<9$~GeV/$c$), together with the fully longitudinal (dotted) and fully
transverse (dashed) templates and the resulting fit (solid). Figure reproduced from
\cite{Abulencia:2007us}, Fig.~3.}
\label{fig:tevatron_pol_costheta}
\end{figure}

The most theoretically incisive Tevatron measurement discussed in this section is the CDF
Tevatron Run~II polarization analysis \cite{Abulencia:2007us}, which
used $\sim$800~pb$^{-1}$ of data to measure the polarization parameter $\alpha$
[Eq.~\eqref{eq:polar_alpha}] of prompt $J/\psi$ and $\psi(2S)$ mesons as a function of
$p_T$ over the range $5\leq p_T\leq30$~GeV/$c$ in the central rapidity region $|y|<0.6$.
Prompt candidates were isolated from $B$-hadron decay background using the impact-parameter
significance of the two muon tracks, and the decay-angle distribution
$dN/d\cos\theta^*\propto1+\alpha\cos^2\theta^*$ was extracted in six $p_T$ bins for $J/\psi$
and three $p_T$ bins for $\psi(2S)$ via a template fit to fully transverse and fully
longitudinal Monte Carlo distributions (Fig.~\ref{fig:tevatron_pol_massmc} and
Fig.~\ref{fig:tevatron_pol_costheta}).

\begin{figure}[htbp]
\centering
% Source: Abulencia et al. 2007 (arXiv:0704.0638), Fig. 4, p. 7 -- KEY FIGURE
\includegraphics[width=0.9\textwidth]{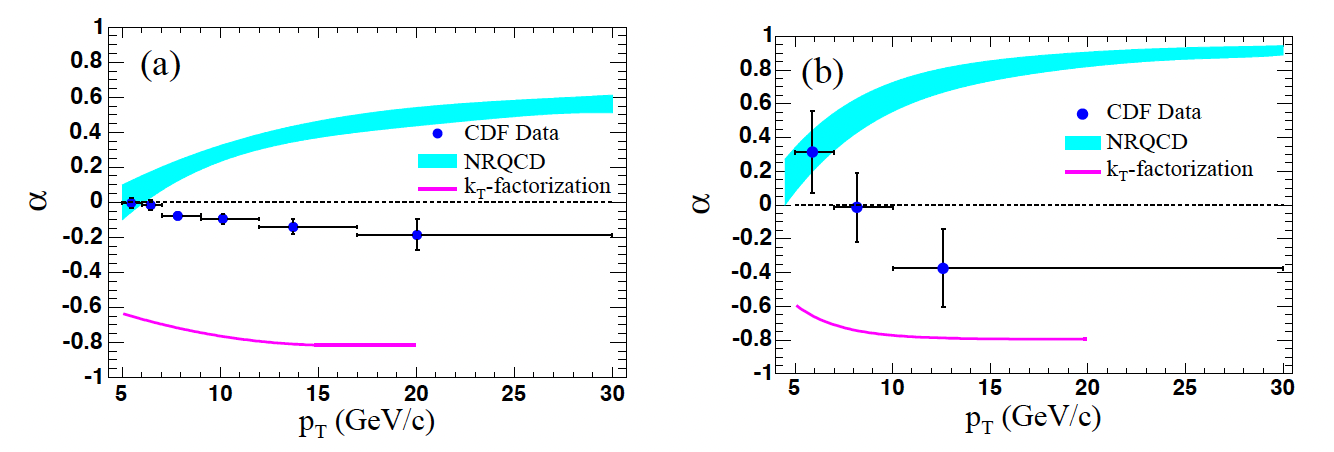}
\caption{Prompt polarization parameter $\alpha$ as a function of $p_T$ for (a) $J/\psi$
and (b) $\psi(2S)$, compared to the NRQCD prediction band and the $k_T$-factorization
model. The cyan band is the NRQCD prediction including its uncertainty, while
the magenta curve is the $k_T$-factorization result. Both charmonium states become increasingly longitudinally polarized as $p_T$
increases beyond 10~GeV/$c$, in clear disagreement with the NRQCD expectation of
increasing transverse polarization. Figure reproduced from \cite{Abulencia:2007us}, Fig.~4.}
\label{fig:tevatron_pol_alpha}
\end{figure}

For both charmonium states, $\alpha$ is consistent with zero at $p_T\simeq5$~GeV/$c$
and the central values become increasingly negative -- that is, increasingly
longitudinal -- as $p_T$ rises toward 30~GeV/$c$
(Fig.~\ref{fig:tevatron_pol_alpha}). The numerical values are collected in
Table~\ref{tab:tevatron_pol_values}; the $\psi(2S)$ points carry substantially
larger uncertainties than those for the $J/\psi$. The central trend is in
conflict with the leading-order NRQCD expectation of increasingly \emph{transverse} polarization at
high $p_T$ [cf.\ Fig.~\ref{fig:motivation_nrqcd_psiprime_polarization_prediction}], driven
by gluon fragmentation into the $^3S_1^{[8]}$ color-octet channel. The $k_T$-factorization
model of Baranov \cite{Baranov:2011kt}, while predicting longitudinal polarization qualitatively in the correct
direction, overshoots the observed degree of longitudinal polarization at all measured
$p_T$ (Fig.~\ref{fig:tevatron_pol_alpha}).

The analysis also reports the effective polarization of $J/\psi$ and $\psi(2S)$ mesons
produced in $B$-hadron decays, finding $\alpha_{\rm eff}=-0.106\pm0.033\,(\mathrm{stat.})
\pm0.007\,(\mathrm{syst.})$ for $J/\psi$ -- consistent with, and more precise than, the
value inferred from the BaBar measurement of $B\to J/\psi X$ decays -- and reports the
first measurement of the analogous quantity for $\psi(2S)$,
$\alpha_{\rm eff}=0.36\pm0.25\,(\mathrm{stat.})\pm0.03\,(\mathrm{syst.})$. The authors
attribute the disagreement between this measurement and the earlier CDF Run~I polarization
result (based on 110~pb$^{-1}$, consistent with $\alpha\approx0$ across all $p_T$; cf.\
Section~\ref{sec:introduction}) to improved modeling of the muon trigger and reconstruction
efficiencies as a smooth function of $p_T$, in contrast to the discontinuous efficiency
corrections used in the earlier analysis. Together with the cross-section measurements of
Sections~\ref{sec:tevatron_jpsi_xsec}--\ref{sec:tevatron_d0_smallangle} and~\ref{sec:tevatron_psi2s_xsec},
this result crystallizes the central tension identified in Section~\ref{sec:introduction}: NRQCD
successfully reproduces the magnitude and $p_T$ dependence of the charmonium production
cross sections, yet fails qualitatively to describe their polarization.

\begin{table}[htbp]
\centering
\caption{CDF measurement of the prompt $J/\psi$ and $\psi(2S)$ polarization
parameter $\alpha$ in Tevatron Run~II $p\bar p$ collisions as a function of
$p_T$~\cite{Abulencia:2007us}. Uncertainties are statistical followed by
systematic.}
\label{tab:tevatron_pol_values}
\begin{tabular}{ccc}
\hline\hline
State & $\langle p_T\rangle$ (GeV/$c$) & $\alpha$ \\
\hline
$J/\psi$ & 5.5  & $-0.004\pm0.029\pm0.009$ \\
$J/\psi$ & 6.5  & $-0.015\pm0.028\pm0.010$ \\
$J/\psi$ & 7.8  & $-0.077\pm0.023\pm0.013$ \\
$J/\psi$ & 10.1  & $-0.094\pm0.028\pm0.007$ \\
$J/\psi$ & 13.7 & $-0.140\pm0.043\pm0.007$ \\
$J/\psi$ & 20.0 & $-0.187\pm0.090\pm0.007$ \\
$\psi(2S)$ & 5.9  & $+0.314\pm0.242\pm0.028$ \\
$\psi(2S)$ & 8.2  & $-0.013\pm0.201\pm0.035$ \\
$\psi(2S)$ & 12.6 & $-0.374\pm0.222\pm0.062$ \\
\hline\hline
\end{tabular}
\end{table}

\subsection{Bottomonium Production and Polarization at the Tevatron}
\label{sec:tevatron_upsilon}

Bottomonium provides another test of the same production mechanisms. Its smaller
characteristic velocity, $v_b^2\sim0.1$, improves the convergence of the
NRQCD velocity expansion, and its feed-down structure is relatively
simple~(Section~\ref{sec:introduction}). Using $77\pm3$~pb$^{-1}$ of
Tevatron Run~I
data at $\sqrt{s}=1.8$~TeV, CDF reconstructed $\Upsilon(1S)$, $\Upsilon(2S)$, and
$\Upsilon(3S)\to\mu^+\mu^-$ decays in the rapidity range $|y|<0.4$ and measured the differential cross sections
$(d^2\sigma/dp_Tdy)\cdot\mathrm{Br}(\Upsilon\to\mu^+\mu^-)$ for all three states over
$0<p_T<20$~GeV/$c$~\cite{Acosta:2001gc}.

\begin{figure}[htbp]
\centering
% Source: Acosta et al. 2001/2002 (FERMILAB-Pub-01/355-E), Figure 3, p. 21 -- KEY FIGURE
\includegraphics[width=0.45\textwidth]{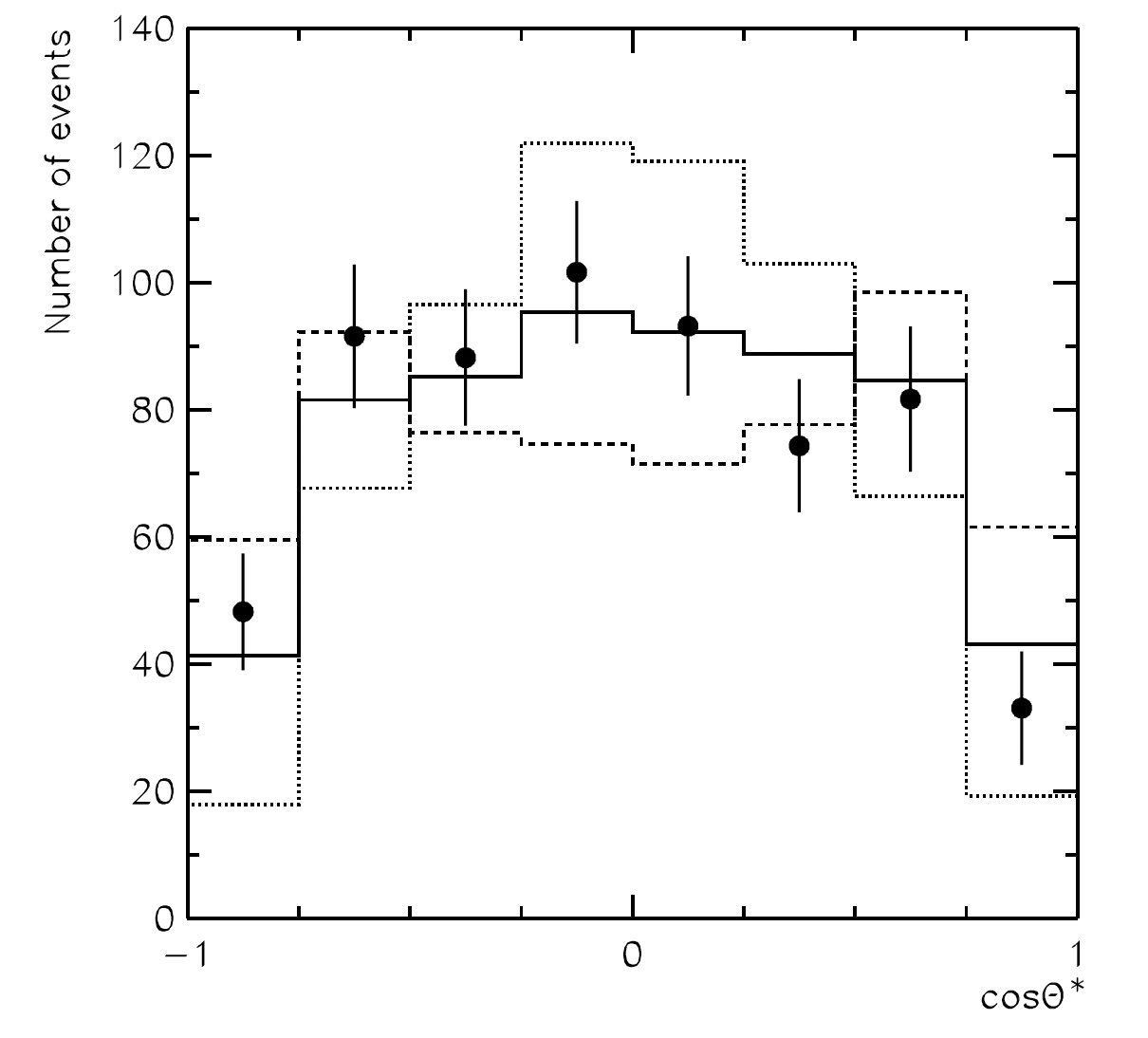}
\caption{Uncorrected $\cos\theta$ distribution for $\Upsilon(1S)$ candidates with
$8<p_T<20$~GeV/$c$ and $|y|<0.4$ (points), together with the combined fit (solid) and the
individually normalized longitudinal (dotted) and transverse (dashed) Monte Carlo
templates. Figure reproduced from \cite{Acosta:2001gc}, Fig.~3.}
\label{fig:tevatron_upsilon_costheta}
\end{figure}

The same data set was used to perform the first measurement, at a hadron collider, of the
$\Upsilon(1S)$ polarization. Fitting the $\cos\theta$ distribution of the decay muons in
the $\Upsilon(1S)$ rest frame to longitudinally and transversely polarized Monte Carlo
templates (Fig.~\ref{fig:tevatron_upsilon_costheta}) in four $p_T$ bins yields polarization
parameters consistent with unpolarized production across the full measured range: $\alpha
=+0.05\pm0.14$, $+0.01\pm0.14$, $+0.10\pm0.17$, and $-0.12\pm0.22$ for the bins
$0$--3, 3--5, 5--8, and 8--20~GeV/$c$, respectively (Table~\ref{tab:tevatron_upsilon_pol}).
The highest-$p_T$ result was found to agree with the NRQCD-based calculation of Braaten
and Lee current at the time. Combined with the charmonium results of
Section~\ref{sec:tevatron_charmonium_pol}, this measurement showed that a rising
transverse polarization was not required by the bottomonium data either. The
limitation was statistical: the CDF result was compatible with zero throughout
the measured range rather than establishing a definite non-zero polarization.
The higher-statistics D0 Run~II measurement discussed next therefore provided a
more decisive test of the same question.

\begin{figure}[htbp]
\centering
% Source: Acosta et al. 2001/2002 (FERMILAB-Pub-01/355-E), Figure 2, p. 20 -- KEY FIGURE
\includegraphics[width=0.65\textwidth]{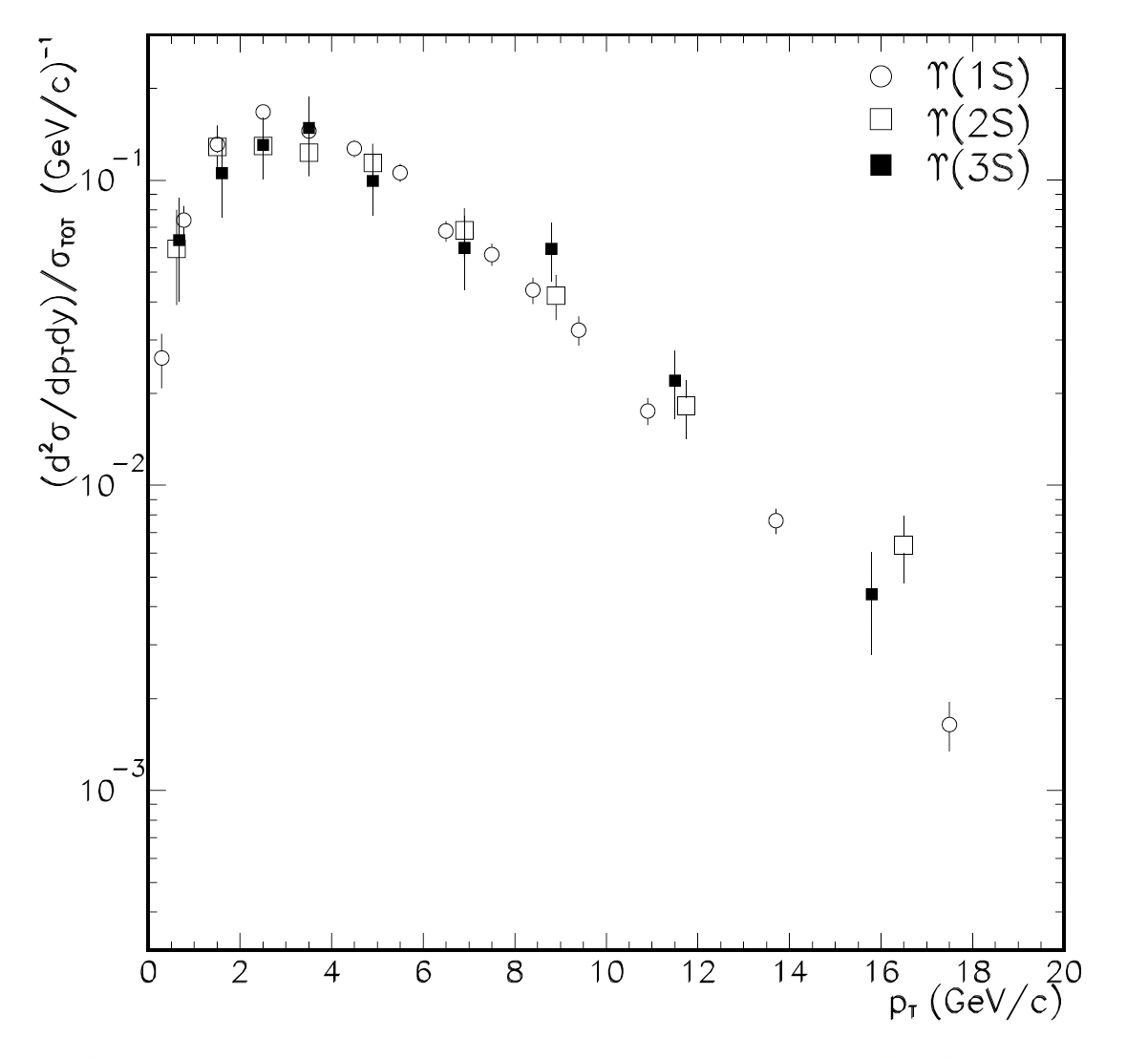}
\caption{Normalized differential cross sections
$(1/\sigma_{\rm TOT})(d^2\sigma/dp_Tdy)_{|y|<0.4}$ for $\Upsilon(1S)$, $\Upsilon(2S)$, and
$\Upsilon(3S)\to\mu^+\mu^-$ as a function of $p_T$, showing a common shape across all
three states. Figure reproduced from \cite{Acosta:2001gc}, Fig.~2.}
\label{fig:tevatron_upsilon_diffxsec}
\end{figure}

\begin{table}[htbp]
\centering
\caption{CDF measurement of the $\Upsilon(1S)$ polarization parameter $\alpha$
for inclusive $\Upsilon(1S)\to\mu^+\mu^-$ production in $p\bar p$ collisions
at $\sqrt{s}=1.8$~TeV and $|y|<0.4$~\cite{Acosta:2001gc}.}
\label{tab:tevatron_upsilon_pol}
\begin{tabular}{cc}
\hline\hline
$p_T$ range (GeV/$c$) & $\alpha$ \\
\hline
$0.0$--$3.0$ & $+0.05\pm0.14$ \\
$3.0$--$5.0$ & $+0.01\pm0.14$ \\
$5.0$--$8.0$ & $+0.10\pm0.17$ \\
$8.0$--$20.0$ & $-0.12\pm0.22$ \\
\hline\hline
\end{tabular}
\end{table}

\begin{figure}[htbp]
    \centering
    \includegraphics[width=0.40\textwidth]{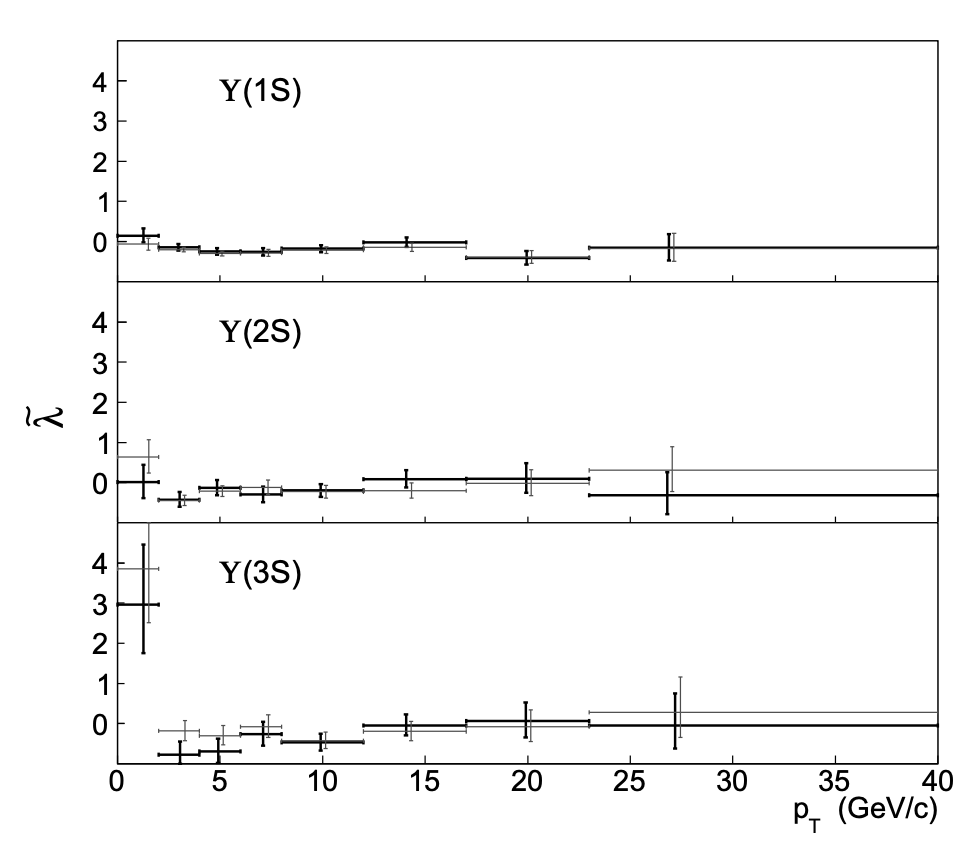}
    \includegraphics[width=0.50\textwidth]{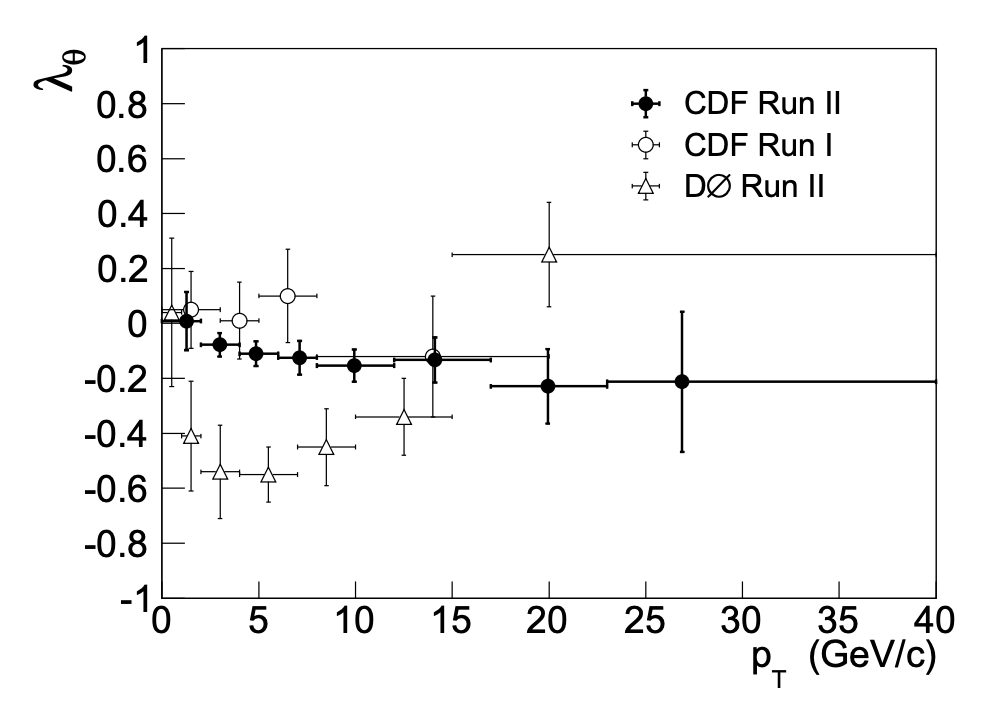}

    \caption{Left: The frame-invariant polarization parameter
    $\widetilde{\lambda}$ measured as a function of $p_T$ for the
    $\Upsilon(1S)$, $\Upsilon(2S)$, and $\Upsilon(3S)$ states in the CDF
    Run II analysis. Right: Comparison of the polar anisotropy parameter
    $\lambda_\theta$ for $\Upsilon(1S)$ measured by CDF Run II with the
    previous CDF Run I and D0 Run II measurements in the s-channel
    helicity frame. Figure reproduced from
    Ref.~\cite{CDF:2012UpsilonAngularDistributions}.}

    \label{fig:cdf_run2_upsilon_polarization}
\end{figure}

A more comprehensive bottomonium polarization measurement was subsequently performed by CDF using $6.7~\mathrm{fb}^{-1}$ of Run II data in $p\bar p$ collisions at $\sqrt{s}=1.96$ TeV \cite{CDF:2012UpsilonAngularDistributions}. The angular distributions of $\Upsilon(1S)$, $\Upsilon(2S)$, and $\Upsilon(3S)$ decays were measured as functions of $p_T$ up to $40~\mathrm{GeV}/c$ in the central rapidity region $|y|<0.6$, with the full angular parameters determined in both the Collins--Soper and s-channel helicity frames. The measured angular distributions are generally consistent with being nearly isotropic, and the analysis provided the first measurement of the spin alignment of $\Upsilon(3S)$ mesons at a hadron collider. The $\Upsilon(1S)$ polarization results are statistically consistent with the earlier CDF Run I measurement but differ from the D0 Run II result, highlighting the unresolved experimental discrepancy in bottomonium polarization measurements at the Tevatron \cite{CDF:2012UpsilonAngularDistributions}. Figure~\ref{fig:cdf_run2_upsilon_polarization} summarizes the measured $p_T$ dependence of the frame-invariant parameter $\widetilde{\lambda}$ for all three $\Upsilon$ states, alongside a comparison of the $\Upsilon(1S)$ polar anisotropy parameter $\lambda_\theta$ with the earlier CDF Run I and D0 Run II measurements.

\paragraph{D0 Run~II measurement.}
\label{sec:tevatron_d0_upsilon_pol}

\begin{figure}[htbp]
\centering
% Source: Abazov et al. 2008 (arXiv:0804.2799), Fig. 1, p. 5
\includegraphics[width=0.80\textwidth]{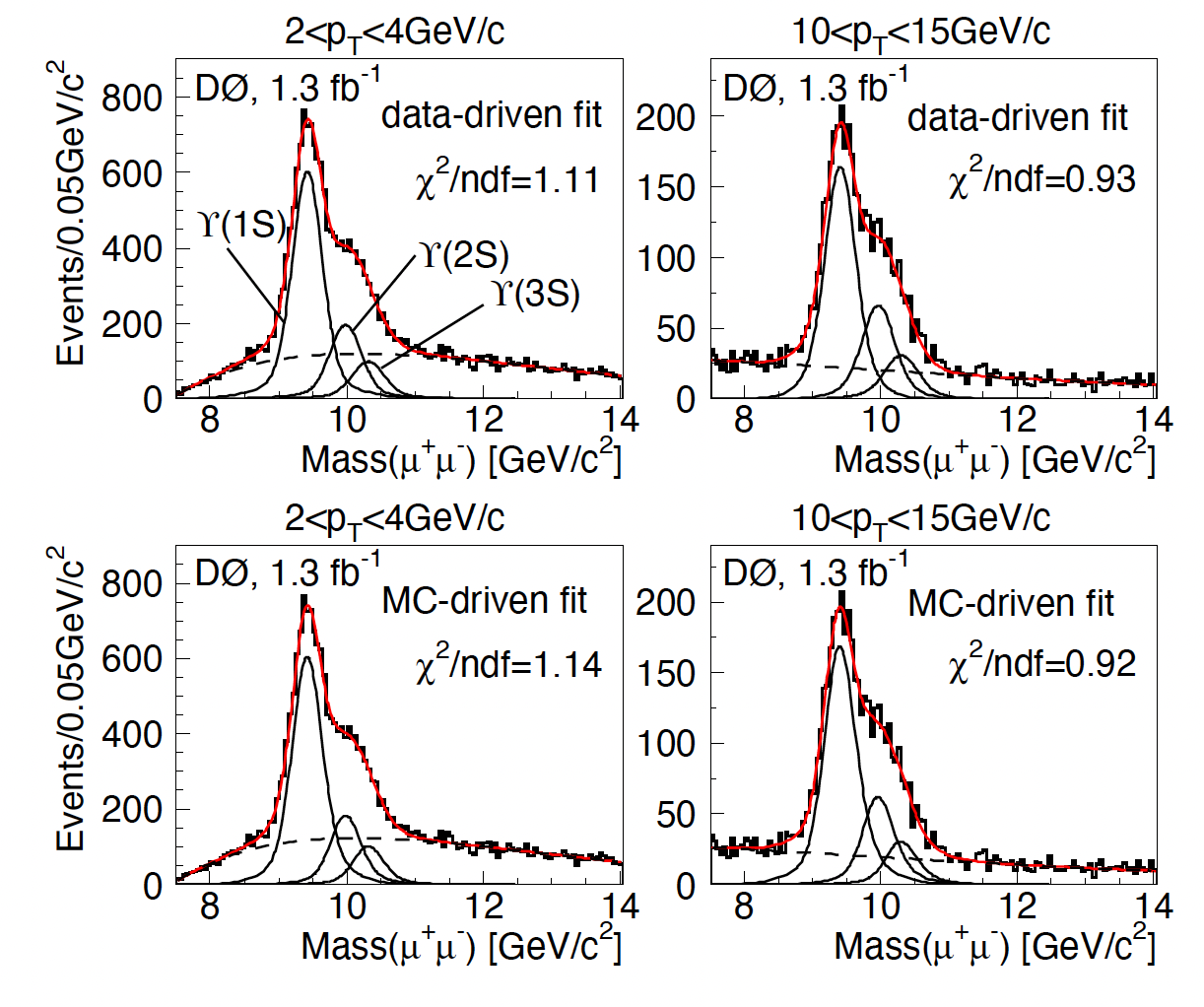}
\caption{D0 dimuon invariant-mass fits for inclusive $\Upsilon\to\mu^+\mu^-$
production in $p\bar p$ collisions at $\sqrt{s}=1.96$~TeV, in the
$0.4<|\cos\theta^*|<0.5$ region for (a,c)
$2<p_T^\Upsilon<4$~GeV/$c$ and (b,d) $10<p_T^\Upsilon<15$~GeV/$c$, using the data-driven
(top) and MC-driven (bottom) signal parameterizations. Figure reproduced from
Ref.~\cite{Abazov:2008za}, Fig.~1.}
\label{fig:tevatron_d0_2008_massfit}
\end{figure}

\begin{figure}[htbp]
\centering
% Source: Abazov et al. 2008 (arXiv:0804.2799), Fig. 2, p. 5
\includegraphics[width=0.7\textwidth]{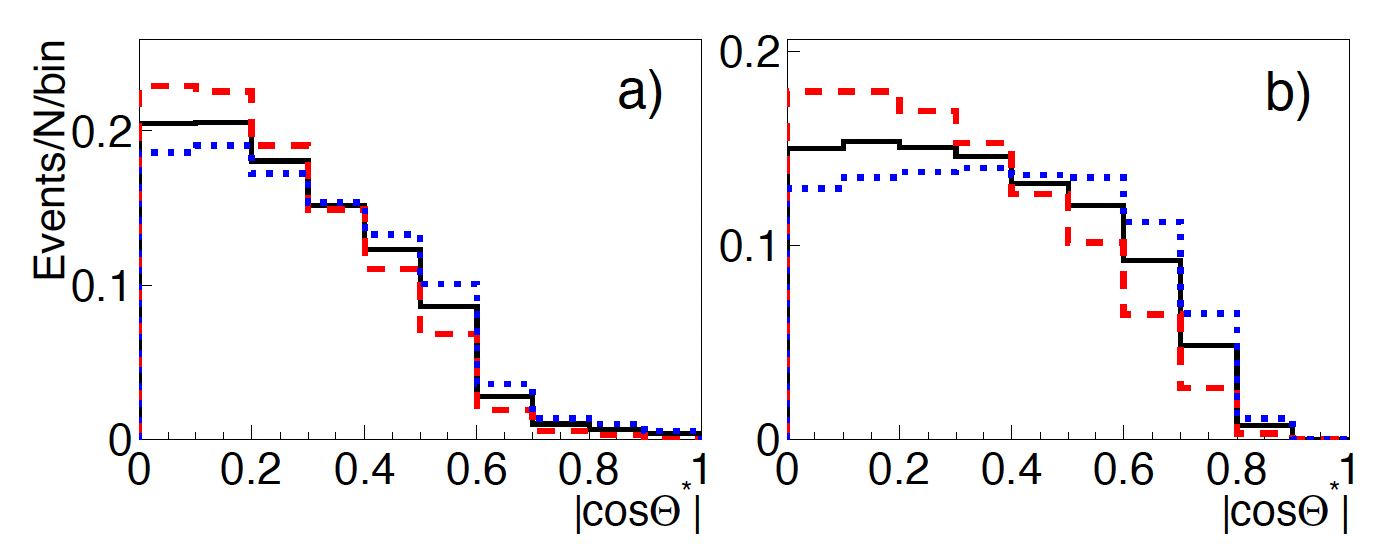}
\caption{D0 Monte Carlo $|\cos\theta^*|$ distributions for $\alpha=-1$ (red
dashed), $\alpha=0$ (black solid), and $\alpha=+1$ (blue dotted) in the lowest
(a) and highest (b) $p_T^\Upsilon$ intervals. The comparisons illustrate the
detector sensitivity to $\Upsilon(1S)$ polarization in inclusive
$\Upsilon\to\mu^+\mu^-$ events from $p\bar p$ collisions at
$\sqrt{s}=1.96$~TeV. Figure reproduced
from \cite{Abazov:2008za}, Fig.~2.}
\label{fig:tevatron_d0_2008_mcshapes}
\end{figure}

The highest-statistics bottomonium polarization measurement at the Tevatron was performed
by D0 using $1.3$~fb$^{-1}$ of Tevatron Run~II data collected between 2002 and 2006, in which
approximately $260{,}000$ $\Upsilon(nS)\to\mu^+\mu^-$ decays were reconstructed over the
wider rapidity range $|y^\Upsilon|<1.8$~\cite{Abazov:2008za}. Two independent
parameterizations of the dimuon invariant-mass line shape (``data-driven,'' using a
double-Gaussian signal function, and ``MC-driven,'' incorporating a Monte-Carlo-derived
non-Gaussian tail) were fit simultaneously in bins of $p_T^\Upsilon$ and $|\cos\theta^*|$
(Fig.~\ref{fig:tevatron_d0_2008_massfit}), and the resulting $|\cos\theta^*|$ distributions
were compared to \textsc{pythia} Monte Carlo templates spanning
$\alpha=-1,0,+1$. The generated events were passed through the D0 detector
simulation and weighted to the measured kinematic distributions before the
angular fit (Fig.~\ref{fig:tevatron_d0_2008_mcshapes}) to extract the polarization parameter of
Eq.~\eqref{eq:polar_alpha} for the $\Upsilon(1S)$ and $\Upsilon(2S)$ states separately;
insufficient $\Upsilon(3S)$ statistics precluded an angular analysis for that state.

\begin{figure}[htbp]
\centering
% Source: Abazov et al. 2008 (arXiv:0804.2799), Fig. 4, p. 6 -- KEY FIGURE
\includegraphics[width=0.6\textwidth]{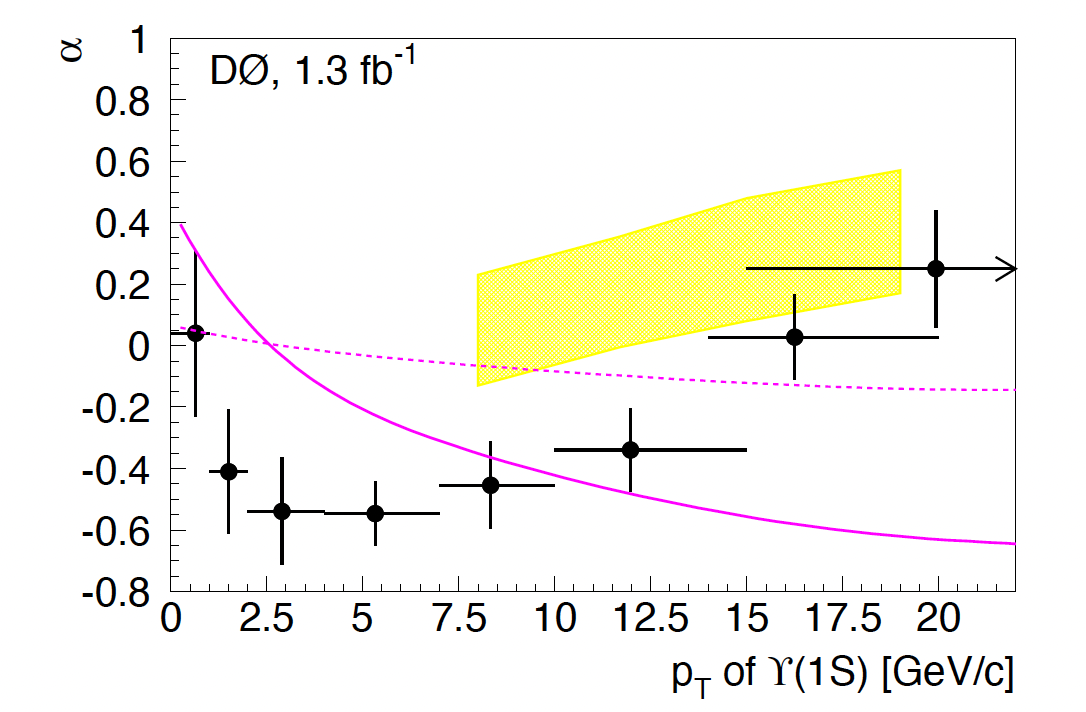}
\caption{D0 measurement of the polarization parameter $\alpha$ for inclusive
$\Upsilon(1S)\to\mu^+\mu^-$ candidates (points) in $p\bar p$ collisions at
$\sqrt{s}=1.96$~TeV, compared to the NRQCD prediction band and the two
limiting cases of the $k_T$-factorization model. The yellow shaded region is
the NRQCD prediction and its uncertainty; the magenta solid and dotted curves
show the two limiting $k_T$-factorization cases. Significant longitudinal polarization is
observed for $2\lesssim p_T^\Upsilon\lesssim10$~GeV/$c$, in tension with both NRQCD and the
earlier CDF result. Figure reproduced from \cite{Abazov:2008za}, Fig.~4.}
\label{fig:tevatron_d0_2008_alpha_upsilon1s}
\end{figure}

\begin{figure}[htbp]
\centering
% Source: Abazov et al. 2008 (arXiv:0804.2799), Fig. 5, p. 6
\includegraphics[width=0.6\textwidth]{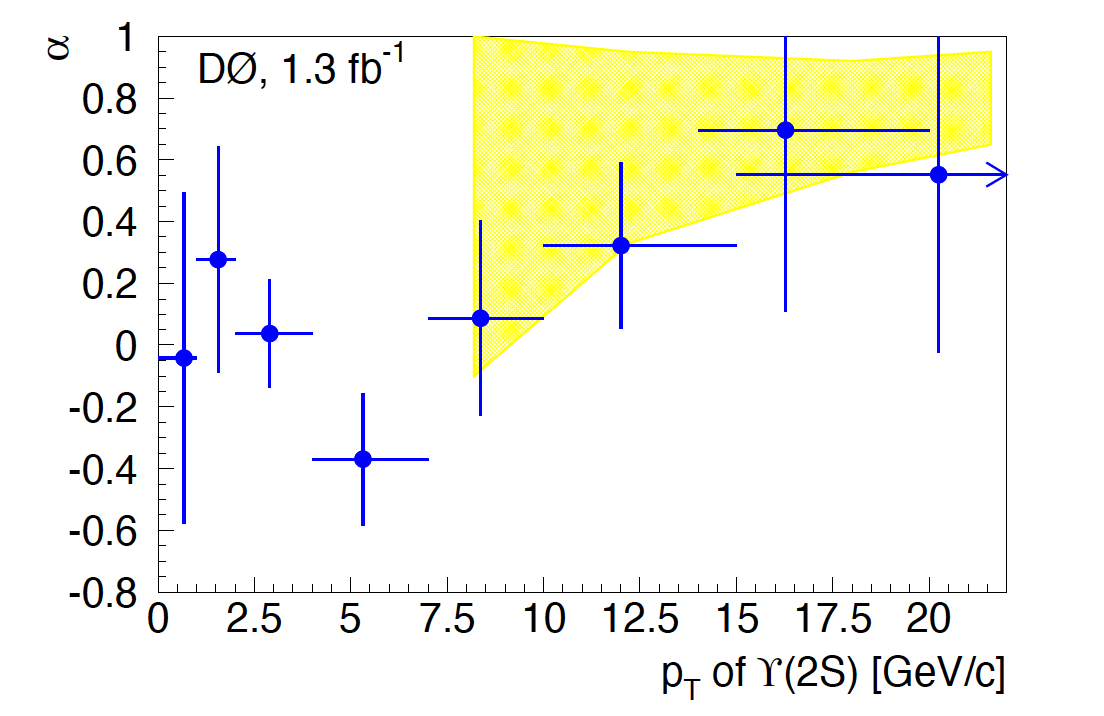}
\caption{D0 measurement of $\alpha$ for inclusive
$\Upsilon(2S)\to\mu^+\mu^-$ candidates in $p\bar p$ collisions at
$\sqrt{s}=1.96$~TeV. The yellow band is the NRQCD prediction and its
uncertainty. Figure reproduced
from \cite{Abazov:2008za}, Fig.~5.}
\label{fig:tevatron_d0_2008_alpha_upsilon2s}
\end{figure}

The central result is a statistically significant \emph{longitudinal}
polarization of the $\Upsilon(1S)$ for
$2\lesssim p_T^\Upsilon\lesssim10$~GeV/$c$, with the central values becoming
less negative again at the highest measured $p_T$
(Fig.~\ref{fig:tevatron_d0_2008_alpha_upsilon1s} and
Table~\ref{tab:tevatron_d0_2008_alpha}). This trend is in clear disagreement with the
Braaten--Lee NRQCD prediction of dominant, steadily increasing transverse polarization for
$p_T^\Upsilon\gtrsim10$~GeV/$c$, and lies below even the more longitudinal of the two
limiting $k_T$-factorization curves of Baranov and Zotov \cite{Baranov:2011kt}. Remarkably, this result is also in
direct tension with the earlier CDF $\Upsilon(1S)$ measurement of
Section~\ref{sec:tevatron_upsilon}, which found $\alpha$ consistent with zero over a
similar $p_T$ range but a narrower rapidity acceptance ($|y^\Upsilon|<0.4$).
D0 did not identify an explanation for the numerical difference between the
two results. This is best described as one unresolved bottomonium-versus-NRQCD
puzzle with an experiment-to-experiment discrepancy in its details, rather
than as two independent theoretical failures. For the
$\Upsilon(2S)$ (Fig.~\ref{fig:tevatron_d0_2008_alpha_upsilon2s}), the measured polarization
is systematically less longitudinal than that of the $\Upsilon(1S)$ at the same $p_T$, in
qualitative agreement with the NRQCD expectation that higher radial excitations should be
more transversely polarized.

The Tevatron therefore presents a consistent challenge to the simplest NRQCD
picture, but not a single clean contradiction: NRQCD can describe the yield
spectra, while the angular observables favor zero or longitudinal polarization
where the leading color-octet expectation tends toward transverse
polarization. The CDF and D0 differences in charmonium and bottomonium add an
experimental dimension that must be resolved before the remaining discrepancy
can be attributed uniquely to a short-distance production mechanism
\cite{Abulencia:2007us,Abazov:2008za,Butenschoen:2010rq,Gong:2012ru}.

\begin{table}[htbp]
\centering
\caption{D0 measurements of $\alpha$ for inclusive $\Upsilon(1S)$ and
$\Upsilon(2S)$ production in $p\bar p$ collisions at
$\sqrt{s}=1.96$~TeV, for statistically independent $p_T^\Upsilon$
intervals~\cite{Abazov:2008za}. Uncertainties
are statistical and systematic added in quadrature.}
\label{tab:tevatron_d0_2008_alpha}
\begin{tabular}{ccc}
\hline\hline
$p_T^\Upsilon$ (GeV/$c$) & $\alpha[\Upsilon(1S)]$ & $\alpha[\Upsilon(2S)]$ \\
\hline
$0$--$1$   & $0.04\pm0.27$  & $-0.04\pm0.54$ \\
$1$--$2$   & $-0.41\pm0.20$ & $0.28\pm0.37$ \\
$2$--$4$   & $-0.54\pm0.17$ & $0.04\pm0.18$ \\
$4$--$7$   & $-0.55\pm0.10$ & $-0.37\pm0.21$ \\
$7$--$10$  & $-0.45\pm0.14$ & $0.09\pm0.32$ \\
$10$--$15$ & $-0.34\pm0.14$ & $0.32\pm0.27$ \\
$>15$      & $0.25\pm0.19$  & $0.55\pm0.58$ \\
\hline\hline
\end{tabular}
\end{table}

\subsection{Double-Parton Scattering in Double $J/\psi$ Production}
\label{sec:tevatron_d0_doublejpsi}

Beyond single-quarkonium production and polarization, the D0 experiment exploited its full
$8.1\pm0.5$~fb$^{-1}$ Run~II data set to measure a qualitatively different class of
quarkonium observable: the production of \emph{two} $J/\psi$ mesons in the same $p\bar p$
collision~\cite{Abazov:2014qba}. Because the dominant $J/\psi$ hadroproduction mechanism is
$gg$ fusion (Section~\ref{sec:introduction}), simultaneous double-$J/\psi$ production
can proceed either through a single hard $2\to4$ parton scattering (SP) or through two
independent, uncorrelated $2\to2$ parton scatterings occurring in the same $p\bar p$
collision -- a double parton (DP) interaction. Separating these two mechanisms provides a
sensitive probe of the transverse spatial distribution of gluons within the proton under
the standard DP-factorization assumption, complementary to the cross-section and polarization observables
discussed above~\cite{Diehl:2011yj,Blok:2013bpa}.

\begin{figure}[htbp]
\centering
% Source: Abazov et al. 2014 (arXiv:1406.2380), Fig. 1, p. 5
\includegraphics[width=0.55\textwidth]{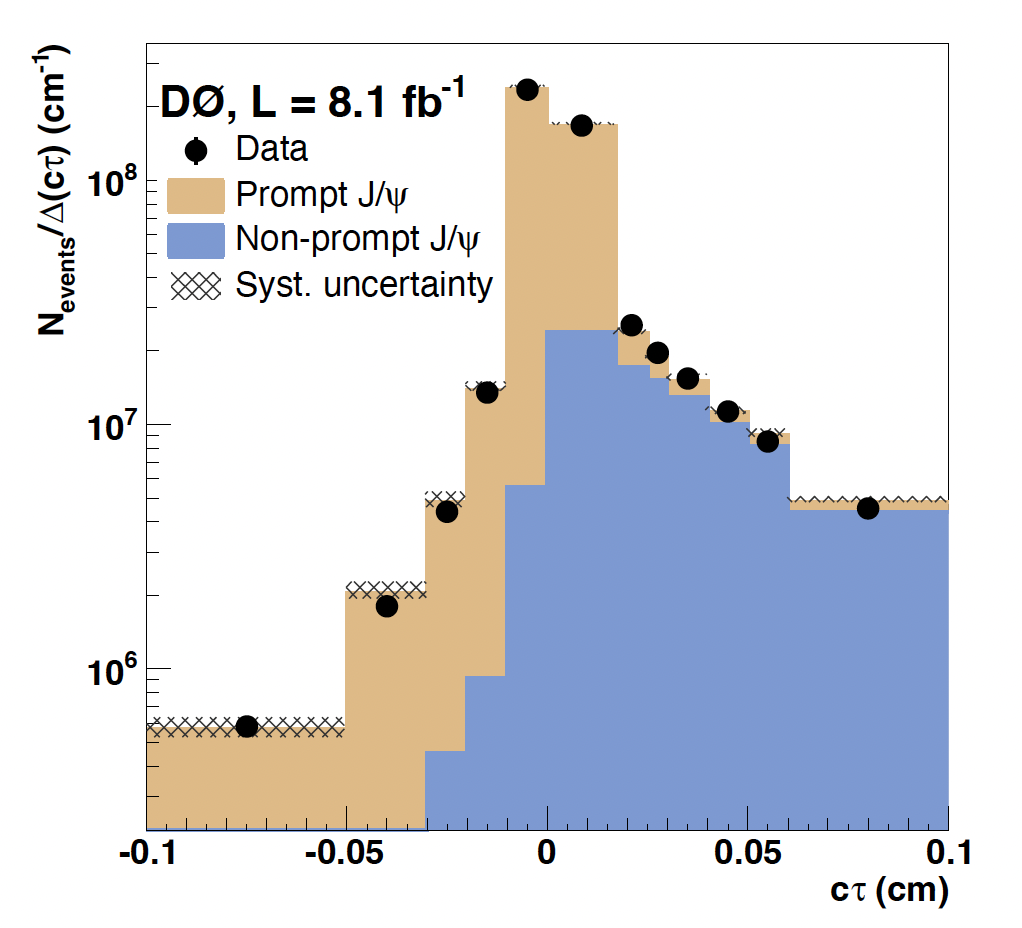}
\caption{Background-subtracted $c\tau$ distribution of single $J/\psi$ events, with the
prompt and non-prompt (b-hadron decay) template fractions used to extract the prompt
fraction. Figure reproduced from Ref.~\cite{Abazov:2014qba}, Fig.~1.}
\label{fig:tevatron_d0_2014_ctau}
\end{figure}

\begin{figure}[htbp]
\centering
% Source: Abazov et al. 2014 (arXiv:1406.2380), Fig. 2, p. 5
\includegraphics[width=0.55\textwidth]{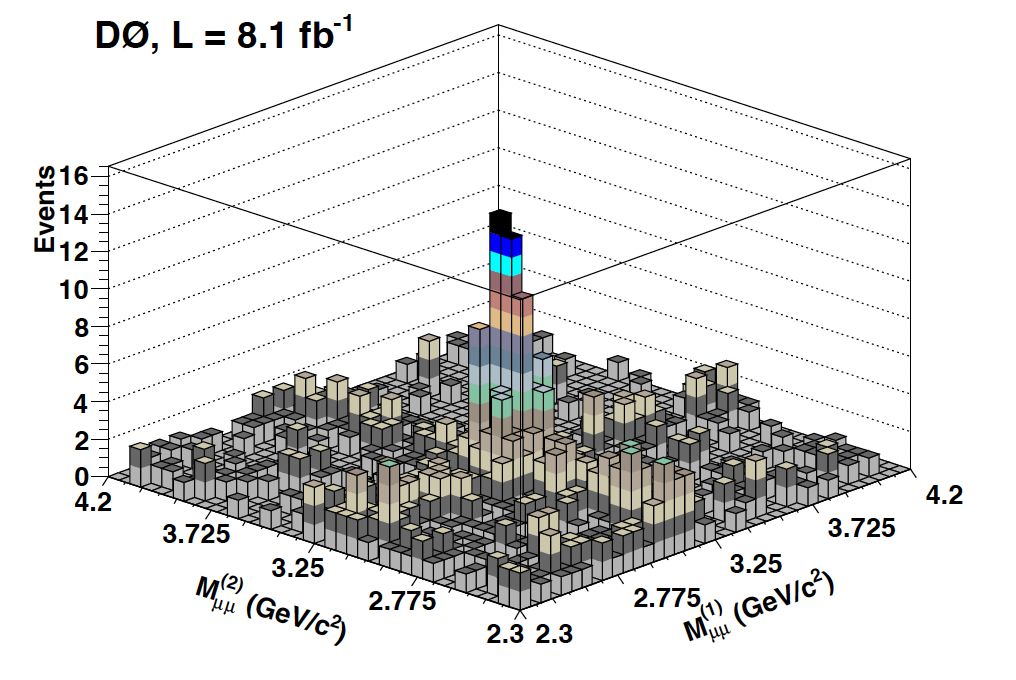}
\caption{Two-dimensional dimuon invariant-mass distribution, $M^{(1)}_{\mu\mu}$ versus
$M^{(2)}_{\mu\mu}$, for double $J/\psi$ candidates after selection, used together with the
two-dimensional fit surface to estimate accidental and $J2\mu$ backgrounds. Figure
reproduced from \cite{Abazov:2014qba}, Fig.~2.}
\label{fig:tevatron_d0_2014_massplane}
\end{figure}

\begin{figure}[htbp]
\centering
% Source: Abazov et al. 2014 (arXiv:1406.2380), Fig. 4, p. 7 -- KEY FIGURE
\includegraphics[width=0.6\textwidth]{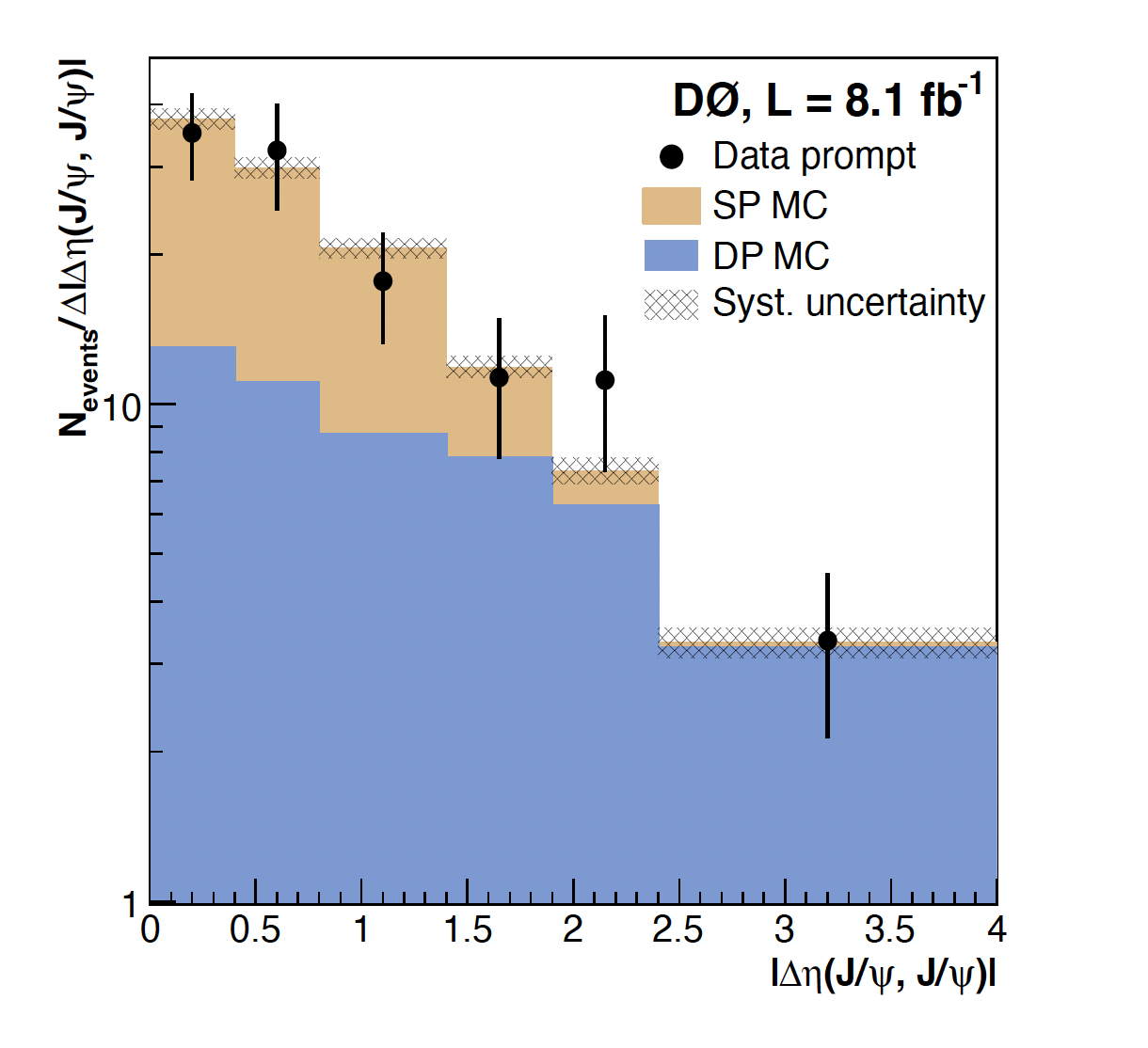}
\caption{The $|\Delta\eta(J/\psi,J/\psi)|$ distribution of background-subtracted double
$J/\psi$ events, with the fitted single-parton (SP) and double-parton (DP) Monte Carlo
templates. The DP contribution dominates the broad, large-$|\Delta\eta|$ tail. Figure
reproduced from \cite{Abazov:2014qba}, Fig.~4.}
\label{fig:tevatron_d0_2014_deta}
\end{figure}

Requiring both $J/\psi$ candidates to satisfy $p_T^{J/\psi}>4$~GeV/$c$ and
$|\eta^{J/\psi}|<2$, a maximum-likelihood fit to the single-$J/\psi$ decay-length ($c\tau$)
distribution (Fig.~\ref{fig:tevatron_d0_2014_ctau}) gave a prompt fraction of
$0.814\pm0.009$ and a fiducial prompt single-$J/\psi$ cross section
\begin{equation}
\sigma(J/\psi) = 23.9\pm4.6\,(\mathrm{stat.})\pm3.7\,(\mathrm{syst.})\;\mathrm{nb},
\label{eq:d0_2014_single_jpsi}
\end{equation}
in agreement with the $k_T$-factorization calculation discussed in the
original analysis~\cite{Abazov:2014qba}.
This single-$J/\psi$ cross section is not the final physics observable of the
section; it is the measured input required to convert the double-parton yield
into $\sigma_{\rm eff}$. After double-$J/\psi$
event selection and background modeling using a two-dimensional fit to the dimuon-pair mass
plane (Fig.~\ref{fig:tevatron_d0_2014_massplane}), the prompt double-$J/\psi$ fraction was found to be
$f_{\rm prompt}=0.592\pm0.101$ via a fit to the two-dimensional $c\tau$ distribution. The SP
and DP contributions were then separated using the pseudorapidity difference between the
two $J/\psi$ candidates, $|\Delta\eta(J/\psi,J/\psi)|$, which is expected to be broader for
DP events, in which the two $J/\psi$ mesons originate from kinematically uncorrelated
scatterings (Fig.~\ref{fig:tevatron_d0_2014_deta}). This fit gives
$f^{\rm DP}=0.42\pm0.12$ and $f^{\rm SP}=0.58\pm0.12$, and, combined with the measured double-$J/\psi$
cross section, yields the individual SP and DP cross sections
\begin{equation}
\begin{split}
\sigma_{\rm DP}(J/\psi J/\psi) = 59\pm6\,(\mathrm{stat.})\pm22\,(\mathrm{syst.})\;\mathrm{fb},
\\
\sigma_{\rm SP}(J/\psi J/\psi) = 70\pm6\,(\mathrm{stat.})\pm22\,(\mathrm{syst.})\;\mathrm{fb}.
\end{split}
\label{eq:d0_2014_sp_dp}
\end{equation}
The measured $\sigma_{\rm SP}$ is consistent with both leading-order and next-to-leading-order
NRQCD predictions and with the $k_T$-factorization calculation, providing an independent
cross-check of the color-octet/color-singlet mixture in the ${}^3S_1^{[8]}$-dominated $gg\to
J/\psi J/\psi$ process at moderate $p_T$.

Combining the DP double-$J/\psi$ cross section with the single-$J/\psi$
cross section of Eq.~\eqref{eq:d0_2014_single_jpsi} defines the effective
cross section
$\sigma_{\rm eff}=\tfrac12\,\sigma(J/\psi)^2/
\sigma_{\rm DP}(J/\psi J/\psi)$. It characterizes the transverse overlap
of the two independent parton scatterings and was extracted for the first
time in a purely $gg$-initiated double-parton channel:
\begin{equation}
\sigma_{\rm eff} = 4.8\pm0.5\,(\mathrm{stat.})\pm2.5\,(\mathrm{syst.})\;\mathrm{mb}.
\label{eq:d0_2014_sigmaeff}
\end{equation}
This value agrees with the AFS 4-jet result ($\approx5$~mb) but is markedly lower than the
$\sigma_{\rm eff}\sim12$--$21$~mb extracted from $q\bar q^{(\prime)}$- and $qg$-initiated
double-parton channels such as $\gamma+3$-jet, $W+2$-jet, and 4-jet production at CDF, D0,
ATLAS, and CMS. Because the double-$J/\psi$ and 4-jet final states are dominated by $gg$
initial states while the $\gamma(W)$+jets channels probe $q\bar q^{\prime}$ and $qg$
scattering, this pattern indicates that gluons occupy a smaller average transverse region of
the proton than quarks -- a conclusion in qualitative agreement with the pion-cloud picture
of the nucleon and a result that would be sharpened considerably by the $J/\psi+\Upsilon$
measurement discussed next.

\subsection{Evidence for Simultaneous $J/\psi+\Upsilon$ Production}
\label{sec:tevatron_d0_jpsiupsilon}

The double-parton-scattering program of Section~\ref{sec:tevatron_d0_doublejpsi} was
extended by D0, using the same $8.1$~fb$^{-1}$ data set, to the first search for
\emph{simultaneous} production of a $J/\psi$ and an $\Upsilon(1S,2S,3S)$ meson in the same
$p\bar p$ collision~\cite{Abazov:2015fbq}. Unlike double $J/\psi$ or double-quarkonium
processes at lower invariant mass, $J/\psi+\Upsilon$ production is expected to receive a
negligible single-parton contribution -- suppressed both by additional powers of
$\alpha_s$ and by the smallness of the color-octet matrix elements required to produce a
$c\bar c$ and a $b\bar b$ pair from the same short-distance vertex. The analysis
therefore assumes that double-parton scattering dominates. This assumption is
supported by the flat azimuthal correlation discussed below, but the quoted
effective cross section should be interpreted within that DP-dominated
model~\cite{Diehl:2011yj,Blok:2013bpa}.

The quoted signal concerns prompt $J/\psi+\Upsilon$ production. Events in which
either meson originates from a long-lived $b$-hadron are treated as
non-prompt background. The analysis does not explicitly tag a possible
additional $b\bar b$ pair, however; such an associated final state would remain
part of the prompt sample and is assumed to be small within the DP-dominated
model.

\begin{figure}[htbp]
\centering
% Source: Abazov et al. 2016 (arXiv:1511.02428), Fig. 2, p. 6 -- KEY FIGURE
\includegraphics[width=0.6\textwidth]{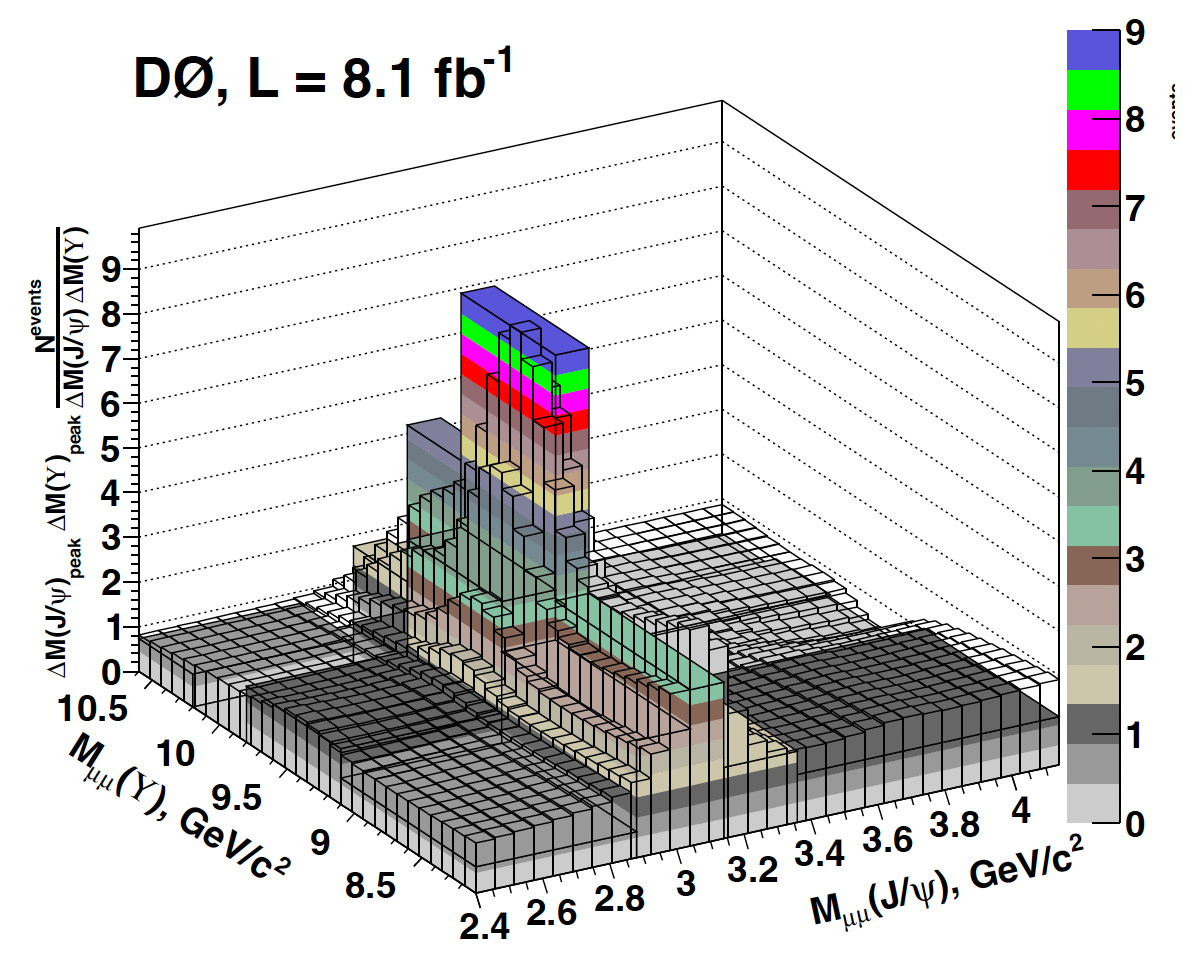}
\caption{Dimuon invariant-mass distribution in data for the two muon pairs,
$M_{\mu\mu}(J/\psi)$ and $M_{\mu\mu}(\Upsilon)$, divided by bin area, together with the
two-dimensional signal-plus-background fit surface, shown as the colored surface,
used to extract the 21 observed
$J/\psi+\Upsilon$ candidates. Figure reproduced from
Ref.~\cite{Abazov:2015fbq}, Fig.~2.}
\label{fig:tevatron_d0_2016_massplane}
\end{figure}

\begin{figure}[htbp]
\centering
% Source: Abazov et al. 2016 (arXiv:1511.02428), Fig. 3, p. 6
\includegraphics[width=0.55\textwidth]{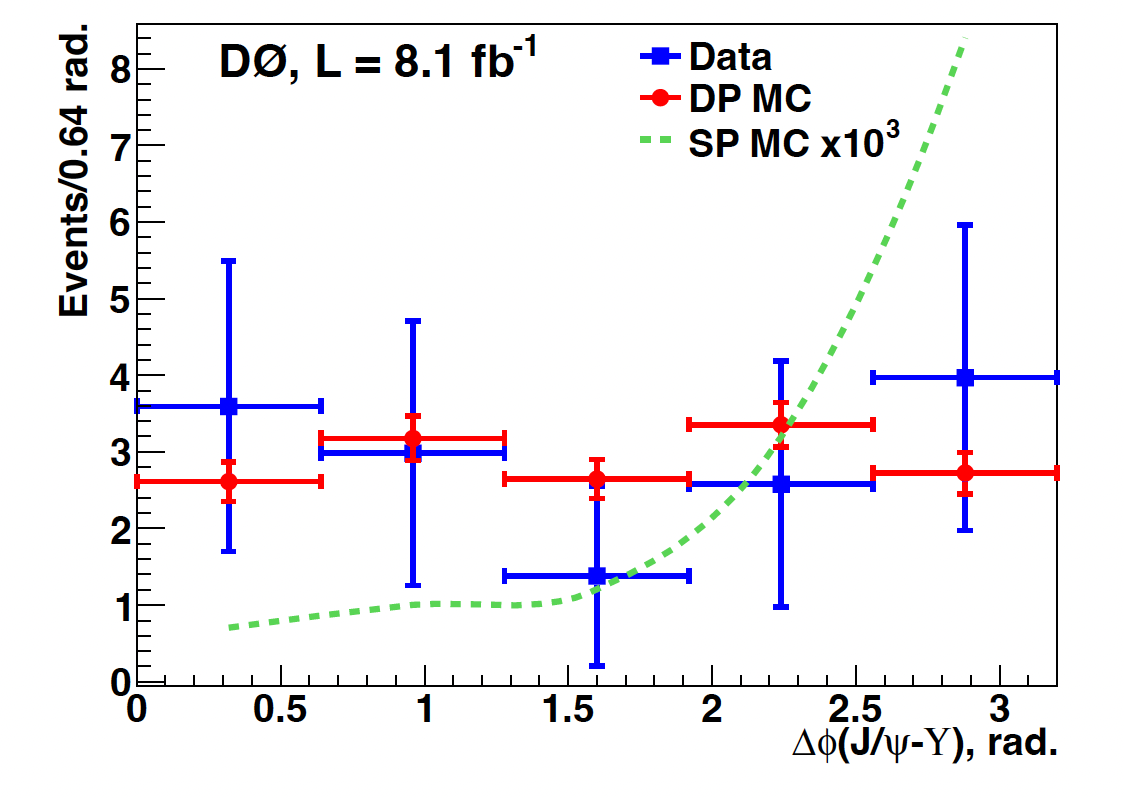}
\caption{Distribution of the azimuthal angle between the $J/\psi$ and $\Upsilon$
candidates, $\Delta\phi(J/\psi,\Upsilon)$, after background subtraction, compared to the
flat double-parton-scattering (DP) Monte Carlo prediction and the single-parton
(SP) Monte Carlo prediction rescaled to the same total yield so that its angular
shape can be compared directly. Figure reproduced from \cite{Abazov:2015fbq},
Fig.~3.}
\label{fig:tevatron_d0_2016_dphi}
\end{figure}

Events with two muon pairs simultaneously satisfying the $J/\psi$ ($2.4<M_{\mu\mu}<4.2$~GeV)
and $\Upsilon$ ($8<M_{\mu\mu}<12$~GeV) mass windows, with $p_T^\mu>2$~GeV/$c$ and
$|\eta^\mu|<2.0$, yielded 21 candidate events (Fig.~\ref{fig:tevatron_d0_2016_massplane}).
A two-dimensional fit to the $M_{\mu\mu}(J/\psi)$--$M_{\mu\mu}(\Upsilon)$ plane, using the
same combinatorial-background parameterizations validated in the single-quarkonium samples,
estimated $14.5\pm4.6\,(\mathrm{stat.})\pm3.4\,(\mathrm{syst.})$ signal events, corresponding
to a prompt $J/\psi+\Upsilon$ signal of $12.0\pm3.8\,(\mathrm{stat.})\pm2.8\,(\mathrm{syst.})$
events. The probability that this signal arose from background fluctuations alone was
$6.3\times10^{-4}$, corresponding to \textbf{3.2 standard deviations of evidence} for
prompt $J/\psi+\Upsilon$ production -- the first report of this final state at a
hadron collider. As a cross-check of the assumed DP origin, the distribution of the azimuthal
angle between the $J/\psi$ and $\Upsilon$ candidates, $\Delta\phi(J/\psi,\Upsilon)$
(Fig.~\ref{fig:tevatron_d0_2016_dphi}), was found to be consistent with the flat, uncorrelated
shape predicted by DP Monte Carlo, substantiating the assumption that DP scattering
dominates the observed signal.

Using the measured prompt single-$J/\psi$ cross section, $\sigma(J/\psi)=28\pm7$~nb, needed to
predict the uncorrelated DP yield, and a
single-$\Upsilon$ cross section, $\sigma(\Upsilon)=2.1\pm0.3$~nb, extrapolated from the
earlier D0 $\Upsilon(1S)$ cross-section measurement and increased for the $\Upsilon(2S,3S)$
contributions, the DP cross section for simultaneous $J/\psi+\Upsilon$ production was
determined to be
\begin{equation}
\sigma_{\rm DP}(J/\psi+\Upsilon) = 27\pm9\,(\mathrm{stat.})\pm7\,(\mathrm{syst.})\;\mathrm{fb}.
\label{eq:d0_2016_sigma_dp}
\end{equation}
Under the assumption that this process is dominated by DP scattering, the corresponding effective cross
section is
\begin{equation}
\sigma_{\rm eff} = 2.2\pm0.7\,(\mathrm{stat.})\pm0.9\,(\mathrm{syst.})\;\mathrm{mb},
\label{eq:d0_2016_sigmaeff}
\end{equation}
lower than the value extracted from double-$J/\psi$ production
[Eq.~\eqref{eq:d0_2014_sigmaeff}] and substantially lower than the $\sigma_{\rm eff}$ values
measured in $q\bar q^{(\prime)}$- and $qg$-dominated double-parton channels. Because both the
double-$J/\psi$ and $J/\psi+\Upsilon$ channels are $gg$-initiated, this progressively lower
$\sigma_{\rm eff}$ points in the same direction as the conclusion of
Section~\ref{sec:tevatron_d0_doublejpsi}: gluons appear to occupy a smaller transverse region
of the proton than quarks, in qualitative agreement with the pion-cloud picture of the
nucleon and providing quantitative input to modern parametrizations of the transverse gluon
density used in multi-parton-interaction modeling at the LHC
(Section~\ref{sec:lhc})~\cite{Abazov:2015fbq,Diehl:2011yj,Blok:2013bpa}.
The two central values differ by only a few times their individual
uncertainties, so the ordering is suggestive rather than statistically
decisive.

\subsection{CDF Studies of the \(X(3872)\): Dipion Mass Spectrum and Quantum Numbers}
\label{sec:tevatron_x3872}

\begin{figure}[htbp]
\centering
% Source: Abulencia et al. 2006 (hep-ex/0512074), Fig. 3
\includegraphics[width=0.65\textwidth]{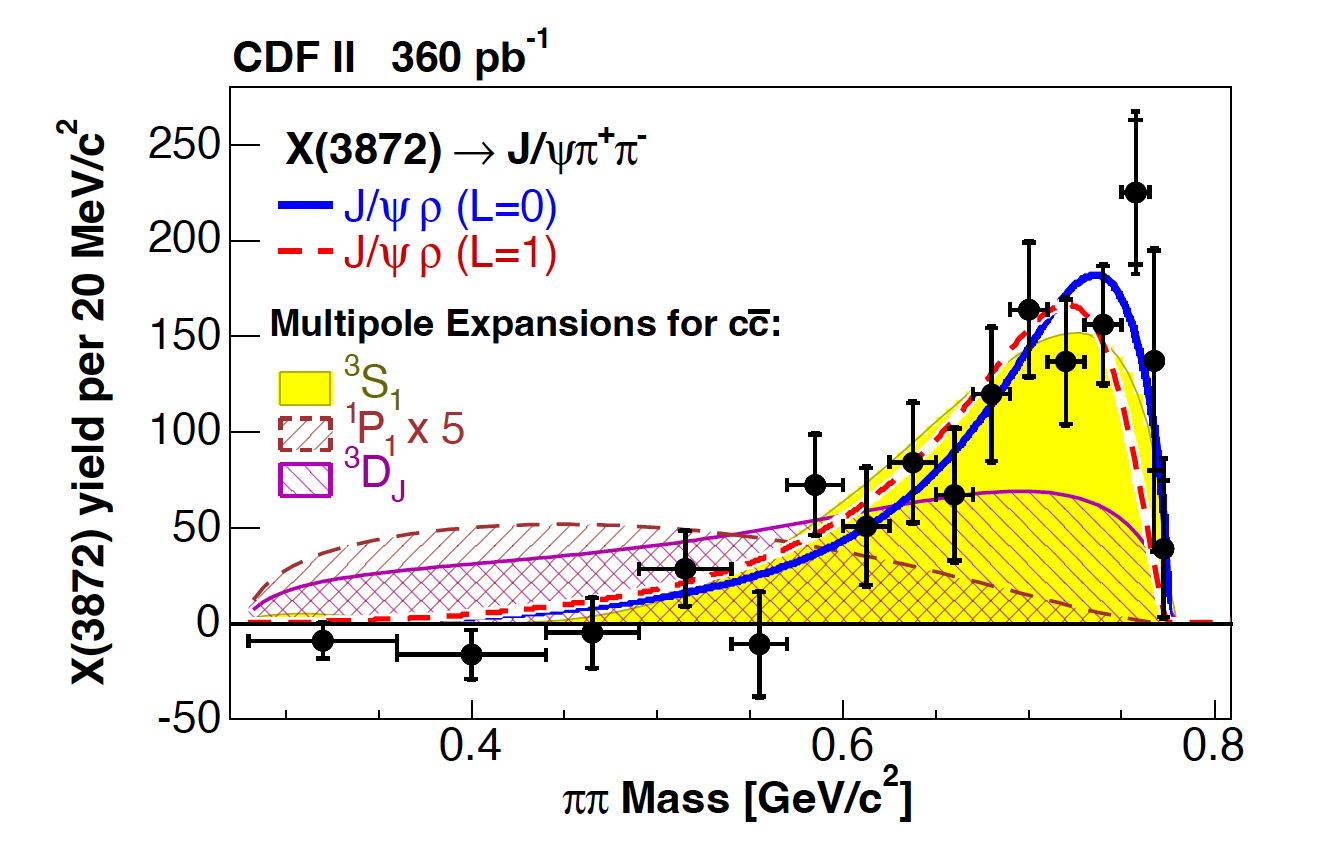}
\caption{Efficiency-corrected dipion mass spectrum of
\(X(3872)\to J/\psi\pi^+\pi^-\) compared with multipole-expansion predictions
for \(C\)-odd charmonia and with \(J/\psi\rho\) models (\(L=0\) and \(L=1\)).
Figure reproduced from Ref.~\cite{Abulencia:2005gc}, Fig.~3.}
\label{fig:tevatron_cdf_x3872_mpipi}
\end{figure}

Complementary to the inclusive production and polarization program, CDF
performed detailed spectroscopic measurements of the exotic candidate
\(X(3872)\) in the exclusive decay \(X(3872)\to J/\psi\pi^+\pi^-\). Using
\(360\,\mathrm{pb}^{-1}\) of Tevatron Run~II data, the collaboration extracted
the efficiency-corrected dipion mass spectrum by fitting the
\(J/\psi\pi^+\pi^-\) invariant-mass distribution in slices of
\(m_{\pi\pi}\). The resulting spectrum rose strongly toward the kinematic
endpoint and was incompatible with the multipole-expansion predictions for the
\(C\)-odd charmonium assignments \(^3D_J\) and \(^1P_1\). The only conventional
\(C\)-odd hypothesis that described the data was \(^3S_1\), but no unassigned
\(^3S_1\) state existed near \(3872\,\mathrm{MeV}/c^2\). In contrast, the
spectrum was well reproduced by the hypothesis that the pions originated from
a \(\rho^0\), with optional \(\rho\)--\(\omega\) interference, implying
\(C=+1\) for the \(X(3872)\). Models with relative orbital angular momentum
\(L=0\) or \(L=1\) between the final-state \(J/\psi\) and \(\rho\) both provided
acceptable descriptions once interference was allowed
(Fig.~\ref{fig:tevatron_cdf_x3872_mpipi}).

A subsequent angular analysis with \(780\,\mathrm{pb}^{-1}\) further constrained the full set of quantum numbers~\cite{Abulencia:2007zz}. The decay chain \(X(3872)\to J/\psi(\pi^+\pi^-)\), \(J/\psi\to\mu^+\mu^-\) was modeled with sequential two-body amplitudes, retaining only the lowest allowed partial waves and averaging over production polarization. A three-dimensional binned fit to the angular variables \(\lvert\cos\theta_{J/\psi}\rvert\), \(\lvert\cos\theta_{\pi\pi}\rvert\) and \(\lvert\Delta\Phi-\pi\rvert-\pi/2\) was performed in twelve bins, with the \(X(3872)\) yield extracted from a Gaussian-plus-polynomial fit to the \(J/\psi\pi^+\pi^-\) mass spectrum in each bin. Of all \(J^{PC}\) hypotheses up to spin~2 (and \(C\)-odd spin~3), only \(J^{PC}=1^{++}\) and \(2^{-+}\) were consistent with the CDF data. The same angular framework applied to the high-statistics \(\psi(2S)\) control sample recovered the known \(1^{--}\) assignment, confirming the analysis method.

Together, these CDF results established \(C=+1\) and restricted the possible
\(J^{PC}\) values of the \(X(3872)\) to \(1^{++}\) or \(2^{-+}\). A later LHCb
amplitude analysis resolved the ambiguity and established
\(J^{PC}=1^{++}\)~\cite{LHCb:2013x3872JPC}. This quantum-number determination is
the concrete payoff of the Tevatron spectroscopy program: it selects the
allowed molecular, tetraquark, and coupled-channel descriptions and supplies a
benchmark for the LHCb and Belle studies discussed later.

CDF also reported evidence for a narrow $J/\psi\phi$ threshold structure in
exclusive $B^+\to J/\psi\phi K^+$ decays, with a mass near
$4143~\mathrm{MeV}/c^2$ and a significance exceeding $3.8\sigma$
\cite{Aaltonen:2009tz}. The state is not needed for the inclusive-production
discussion above, but it is important here because its mass lies close to the
$J/\psi\phi$ threshold and its interpretation therefore probes the same
threshold-dynamics issues as the $X(3872)$. It also illustrates why
exclusive $B$ decays, rather than untagged inclusive samples, are essential for
resolving nearby exotic candidates.

\subsection{Synthesis: The Tevatron Legacy}
\label{sec:tevatron_summary}

Taken together, the measurements reviewed in this section define a coherent, and in several respects still unresolved, picture of Tevatron-era quarkonium physics. On the cross-section side, the CDF and D0 measurements of Sections~\ref{sec:tevatron_jpsi_xsec}--\ref{sec:tevatron_d0_smallangle}
and~\ref{sec:tevatron_psi2s_xsec} -- spanning inclusive $J/\psi$ and $b$-hadron production
from central to forward rapidity, the feed-down-free $\psi(2S)$ channel, and the full
$\Upsilon(nS)$ family -- establish that NRQCD color-octet fits, once tuned to the data,
provide an adequate quantitative description of the $p_T$-differential production rates
across nearly two orders of magnitude in cross section, the full measured rapidity range
$0\leq|y|\lesssim3.7$, and up to $p_T=30$~GeV/$c$. The complementary D0 decomposition of
$J/\psi$ sources by isolation and radiative $\chi_c$ tagging (Section~\ref{sec:tevatron_d0_jpsi_run1})
further showed, already in Run~I, that a $\chi_c$-dominated direct-production picture could
not account for the observed prompt $J/\psi$ rate, an early hint of the missing
color-octet contribution that NRQCD would later supply.

For polarization, the picture is less settled. The CDF charmonium measurement of
Section~\ref{sec:tevatron_charmonium_pol} found a statistically significant
longitudinal trend for the $J/\psi$, while the $\psi(2S)$ data were
compatible with the same trend but remained statistically limited. This is
opposite to the leading-order NRQCD expectation driven by
$^3S_1^{[8]}$ gluon fragmentation. The bottomonium sector adds an unresolved
experimental layer: whereas the original CDF $\Upsilon(1S)$ measurement
(Section~\ref{sec:tevatron_upsilon}) found polarization consistent with zero, the
higher-statistics D0 measurement (Section~\ref{sec:tevatron_d0_upsilon_pol}) found
significant, $p_T$-dependent \emph{longitudinal} polarization for the same state and a
similar kinematic range, with no identified explanation for the disagreement between the two
experiments. This combination of successful cross-section descriptions alongside a
qualitative and, in the case of $\Upsilon(1S)$, even experiment-to-experiment
polarization discrepancy is precisely the ``quarkonium polarization puzzle'' introduced in
Section~\ref{sec:introduction}.

The D0 double-quarkonium program of
Sections~\ref{sec:tevatron_d0_doublejpsi}--\ref{sec:tevatron_d0_jpsiupsilon} opened an
entirely new observable class at the Tevatron: multi-parton interactions probed through
simultaneous $gg$-initiated quarkonium production. The steadily decreasing effective cross
section extracted from double $J/\psi$
($\sigma_{\rm eff}=4.8\pm0.5\pm2.5$~mb), and $J/\psi+\Upsilon$
($\sigma_{\rm eff}=2.2\pm0.7\pm0.9$~mb) production -- all lower than the values measured in
quark-initiated double-parton channels -- points to a more compact transverse
gluon distribution than that of quarks, within the DP-dominated interpretation
of these channels. That conclusion is relevant for modern
multi-parton-interaction and underlying-event models used throughout the LHC
program (Section~\ref{sec:lhc}).

The Tevatron spectroscopy program closes the section with a different kind of
legacy. The CDF $X(3872)$ measurements established its positive charge parity
and narrowed its quantum numbers, later resolved to \(1^{++}\) by LHCb, while
the $J/\psi\phi$ threshold structure demonstrated that precision exclusive
reconstruction can expose narrow structures hidden in inclusive samples. These
results do not directly determine the LDMEs, but they identify the
threshold and feed-down dynamics that a purely inclusive production model
cannot ignore.

Together,
these CDF and D0 results, with their substantially larger data sets, extended $p_T$ and
rapidity reach, improved treatment of feed-down and higher-order QCD corrections, and the
entirely new multi-quarkonium observable, constitute the empirical foundation that the
subsequent generation of $e^+e^-$-collider (Section~\ref{sec:eecolliders}) and LHC
(Section~\ref{sec:lhc}) measurements was designed to extend and, where possible, resolve.

%% file: D_EE_Colliders.tex
\section{Quarkonium Studies at $e^+e^-$ Colliders:\\ Belle, BaBar, Belle~II, BESIII, and LEP}
\label{sec:eecolliders}

\subsection{The $e^+e^-$ Program: A Clean Complement to Hadroproduction}
\label{sec:ee_intro}
The $e^+e^-$ program reviewed here addresses questions that are related to,
but not identical with, those of hadroproduction. At the $\Upsilon(4S)$
factories, the dominant quarkonium samples come from $B$ decays and from
continuum annihilation into a virtual photon; at BESIII, direct annihilation
and hadronic transitions provide access to charmonium and exotic states; and at
LEP, prompt quarkonium produced in hadronic $Z$ decays tests gluon
fragmentation with a different initial state and virtuality. These environments
therefore provide absolute normalizations, clean quantum-number assignments,
and tests of final-state hadronization that cannot be replaced by a
hadron-collider measurement alone. This section includes the OPAL LEP result
for that reason; a complete LEP program is beyond its scope.

Three complementary facilities have driven the lower-energy program.
The asymmetric-energy KEKB collider, operating at and near the $\Upsilon(4S)$ resonance, hosted the Belle experiment from 1999 to 2010, accumulating an integrated luminosity of approximately $1$~ab$^{-1}$; its successor, SuperKEKB, has hosted Belle~II since 2019 with a target integrated luminosity of $50$~ab$^{-1}$~\cite{Kou:2018nap}. 
Its PEP-II counterpart hosted BaBar from 1999 to 2008 and collected
approximately $0.53$~ab$^{-1}$ at and near the $\Upsilon(4S)$.
The Beijing Electron--Positron Collider (BEPC and its upgrade BEPCII), operating from $\sqrt{s}=2$ to nearly $5$~GeV, has hosted the BES, BESII, and BESIII spectrometers, giving direct access to charmonium resonances and the open-charm threshold region through $e^+e^-$ annihilation into
hadrons.
Three physics capabilities distinguish this program from the hadroproduction studies of Section~\ref{sec:tevatron}. 
First, $B$-meson decays at the $\Upsilon(4S)$ and direct $e^+e^-$ annihilation in the charm and bottomonium regions allow exclusive reconstruction of quarkonium states with photon or pion tags, enabling precision mass, width, and branching-fraction measurements essentially free of the combinatorial background inherent to hadron colliders.
Second, exclusive double-quarkonium production in $e^+e^-$ annihilation into a single virtual photon provides a theoretically clean test of production mechanisms, complementary to the inclusive production studied at the Tevatron and LHC. 
Third, the scan capability of both $e^+e^-$ machines -- running at a dense
grid of center-of-mass energies -- makes possible line-shape measurements of
vector states directly produced in annihilation, a capability with no
hadron-collider analogue.

\subsection{Prompt Charmonium Production Cross Sections}
\label{sec:prompt_charmonium_production}

Inclusive prompt production of $J/\psi$ and $\psi(2S)$ in $e^+e^-$
annihilation offers a clean laboratory in which the initial state is a single
virtual photon of known energy and quantum numbers. Here "inclusive" means
that only the charmonium candidate is reconstructed while the rest of the event
is unconstrained. The first measurement below the $Z$ resonance came from
BaBar, using $23.3~\mathrm{fb}^{-1}$ at and below the $\Upsilon(4S)$; prompt
production was isolated from $B$ decays by requiring the charmonium momentum in
the $e^+e^-$ center-of-mass frame, $p^*$, to exceed
$2~\mathrm{GeV}/c$~\cite{BaBar:2001promptJpsi}.
The inclusive cross section was measured to be
$\sigma(e^+e^-\to J/\psi X)=2.52\pm0.21\pm0.21~\mathrm{pb}$. BaBar also
reported the production-angle coefficient $A=1.5\pm0.6$ in the convention
$d\sigma/d\cos\theta^*\propto1+A\cos^2\theta^*$ for the $J/\psi$ production
angle $\theta^*$ in the center-of-mass frame, as well as a polarization
parameter of $\alpha=-0.80\pm0.09$ for $p^*>3.5~\mathrm{GeV}/c$~\cite{BaBar:2001promptJpsi}.
Belle then analyzed $32.4~\mathrm{fb}^{-1}$ at and below the $\Upsilon(4S)$~\cite{Belle:2002promptcharmonia}, finding $\sigma(e^+e^-\to J/\psi X)=1.47\pm0.10\pm0.13~\mathrm{pb}$, about $40\%$ below the BaBar value, and, after $\psi(2S)$ feed-down subtraction, a first direct-production cross section of $0.72\pm0.08^{+0.13}_{-0.17}~\mathrm{pb}$ for $p^*>2~\mathrm{GeV}/c$.
The soft momentum spectrum showed no sign of an endpoint rise. Belle measured
$\alpha=-0.4\pm0.1$, favoring longitudinal spin alignment. This comparison is
made to the color-singlet two-gluon contribution after the initial virtual
photon has created the heavy-quark pair; it is not a direct measurement of the
gluon-fusion mechanism that dominates inclusive hadroproduction.
Together, the BaBar and Belle measurements established the absolute scale of
prompt charmonium production in the continuum and constrained the relative
size of the allowed short-distance
contributions~\cite{BaBar:2001promptJpsi,Belle:2002promptcharmonia,Butenschoen:2011yh}.
BESIII has now mapped the near-threshold regime directly, measuring inclusive prompt $J/\psi$ and $\psi(2S)$ production from $3.808$ to $4.951~\mathrm{GeV}$ with $22~\mathrm{fb}^{-1}$. The measured averages are $14.0\pm1.7\pm3.1~\mathrm{pb}$ between $4.527$ and $4.951~\mathrm{GeV}$ and $15.3\pm3.0~\mathrm{pb}$ between $4.843$ and $4.951~\mathrm{GeV}$ in resonance-free windows~\cite{BESIII:2025inclusiveJpsi}. These rates provide the local normalization with which the resonance-enhanced and threshold regions can be compared.

\subsection{Prompt Quarkonium Production in Hadronic $Z$ Decays at LEP}
\label{sec:ee_lep}
OPAL measured prompt $J/\psi$ and $\psi(2S)$ production in hadronic $Z^0$
decays, where the charmonium is produced from gluons and heavy quarks rather
than from a single virtual photon~\cite{OPAL:1996jpsi}. The large $Z$-pole
sample and the clean dilepton final state made it possible to separate prompt
quarkonium from $b$-hadron feed-down and to study the fragmentation momentum
spectrum. These measurements test the color-octet and gluon-fragmentation
components in an environment with much larger virtuality than the $B$-factory
continuum, while remaining distinct from the hadronic initial state of
Tevatron and LHC production.

\subsection{Exclusive Double-Charmonium Production}
\label{sec:ee_double_charmonium}
Exclusive double-charmonium production has long played a special role as one of the cleanest tests of production mechanisms. 
The initial state of $e^+e^-$ annihilation is elementary: a single virtual photon with precisely known energy and quantum numbers, so that the short-distance creation of the heavy-quark pair factorizes cleanly from the long-distance quarkonium wave functions.
The first observation was reported by the Belle Collaboration in 2002 at $\sqrt{s}=10.58$~GeV~\cite{Belle:2002doublecc}. 
By reconstructing the $J/\psi$ and measuring the recoil mass, Belle obtained $\sigma(e^+e^-\to J/\psi+\eta_c)\times\mathcal{B}(\eta_c\to\geq4~\mathrm{charged}) = 33.0^{+7.0}_{-6.0}\pm9.0~\mathrm{fb}$, substantially larger than the leading-order theoretical estimates of $2.3$--$5.5~\mathrm{fb}$.
A subsequent analysis with $155~\mathrm{fb}^{-1}$ confirmed the recoil-mass signals with higher statistics and yielded $\sigma(e^+e^-\to J/\psi+\eta_c)\times\mathcal{B}(\eta_c\to>2~\mathrm{charged}) = 25.6\pm2.8\pm3.4~\mathrm{fb}$~\cite{Belle:2004doublecc}.
The BaBar Collaboration reached a compatible conclusion with an independent data set, obtaining $\sigma(e^+e^-\to J/\psi+\eta_c)\times\mathcal{B}(\eta_c\to>2~\mathrm{charged}) = 17.6\pm2.8^{+1.5}_{-2.1}~\mathrm{fb}$~\cite{BaBar:2005doublecc}.
Higher-order QCD and relativistic corrections later brought theory into
agreement with these measurements, establishing exclusive double-charmonium
production as a benchmark process
\cite{Braaten:2003doublecc,Dong:2013prog,Feng:2024nnlo}.
Belle has also searched for analogous structures in the $J/\psi\eta_c$ channel using the full $955~\mathrm{fb}^{-1}$ data set via initial-state radiation; no significant double-charmonium resonance was observed~\cite{Belle:2023etacJpsi}.

\subsection{Quarkonium Results from Belle II}
\label{sec:belleii_results}

Belle II has made important contributions to bottomonium spectroscopy and production above the $\Upsilon(4S)$ resonance. Using data collected at $\sqrt{s}=10.701$, 10.745, and 10.805 GeV, corresponding to integrated luminosities of 1.6, 9.8, and 4.7~fb$^{-1}$, respectively, the Belle II Collaboration studied $e^+e^-\to\omega\chi_{bJ}(1P)$ ($J=0,1,2$), with $\omega\to\pi^+\pi^-\pi^0$, $\chi_{bJ}\to\gamma\Upsilon(1S)$, and $\Upsilon(1S)\to\ell^+\ell^-$. The $\omega\chi_{b1}$ and $\omega\chi_{b2}$ signals were observed for the first time at $\sqrt{s}=10.745$ GeV, with Born cross sections of $3.6^{+0.7}_{-0.7}\,(\mathrm{stat.})\pm0.5\,(\mathrm{syst.})$ pb and $2.8^{+1.2}_{-1.0}\,(\mathrm{stat.})\pm0.4\,(\mathrm{syst.})$ pb, respectively. The energy dependence of the cross sections, combined with the Belle measurements at higher energies, is consistent with the $\Upsilon(10753)$ resonance, suggesting that its internal structure may differ from that of the $\Upsilon(10860)$ \cite{BelleII:2023OmegaChiB}. Belle II also searched for $e^+e^-\to\gamma\chi_{bJ}$ near $\sqrt{s}=10.746$ GeV, but no significant signal was observed, and upper limits were set on the corresponding Born cross sections \cite{Belle-II:2025nxh}.

A subsequent joint analysis of Belle and Belle II data further investigated the decay patterns of the $\Upsilon(10753)$ and $\Upsilon(10860)$ states in $\chi_{bJ}(1P)\omega$ and $\chi_{bJ}(1P)(\pi^+\pi^-\pi^0)_{\mathrm{non-\omega}}$ final states. The analysis used $142.5~\mathrm{fb}^{-1}$ of Belle data and $19.8~\mathrm{fb}^{-1}$ of Belle II data. The results indicate that the $\Upsilon(10753)$ decays into $\chi_{bJ}\omega$, whereas the $\Upsilon(10860)$ preferentially decays into the non-$\omega$ three-pion final state. The mass and width of the $\Upsilon(10753)$ were measured to be $10756.1\pm3.4\pm2.7~\mathrm{MeV}/c^2$ and $32.2\pm11.3\pm14.9~\mathrm{MeV}$, respectively. These different decay patterns provide further experimental information on the internal structures and transition dynamics of the bottomonium-like states above the open-bottom threshold \cite{BelleBelleII:2026ChiBOmega}.

Belle II has also extended charmonium studies through initial-state-radiation (ISR) measurements of $e^+e^-\to h^+h^-J/\psi$ ($h=\pi,K,p$), using a data sample corresponding to an integrated luminosity of $427.9~\mathrm{fb}^{-1}$ \cite{Belle-II:2026ISRhhJpsi}. The cross sections for $e^+e^-\to\pi^+\pi^-J/\psi$ and $K^+K^-J/\psi$ were measured over broad center-of-mass energy ranges and found to be consistent with previous results. The process $e^+e^-\to p\bar{p}J/\psi$ was investigated for the first time, but no significant signal was observed. In the $\pi^+\pi^-J/\psi$ channel, the cross-section distribution exhibits the enhancement associated with the $Y(4230/4320)$ region, while a significant signal of the $Z_c(3900)^\pm$ was observed in the $\pi^\pm J/\psi$ invariant-mass distribution. No significant structure was found in the $K^\pm J/\psi$ system. These measurements provide complementary information on charmonium-like states and their production in ISR processes.

\begin{figure}[htbp]
\centering
\includegraphics[width=0.75\textwidth]{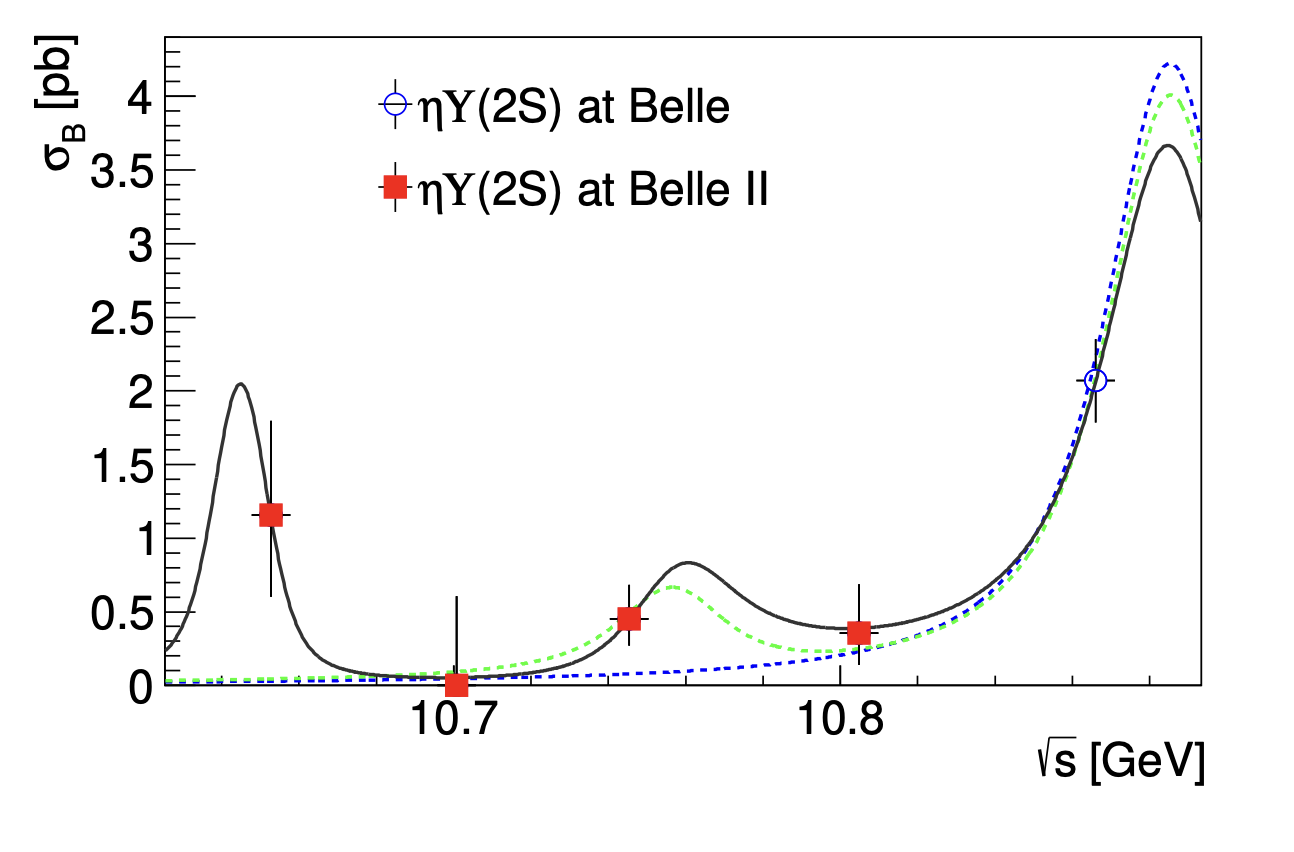}
\caption{Born cross-sections for $e^+e^-\to\eta\Upsilon(2S)$ with fit results overlaid. Points with error bars show measured cross sections, the solid curve is the nominal fit result with $\Upsilon(5S)$, $\Upsilon(10753)$ and a state near $B^*\bar{B}^*$ threshold, the green dashed curve is the fit result with $\Upsilon(5S)$ and $\Upsilon(10753)$, and the blue dashed curve is the fit result with only $\Upsilon(5S)$. The new state near the $B^*\bar{B}^*$ threshold is decoupled from the other two states, with only its amplitude contributing to the fit. Figure reproduced from Ref.~\cite{BelleII:2026EtaUpsilon2S}.}
\label{fig:ee_bes_BornCrossSection}
\end{figure}

Belle II has further investigated bottomonium transitions near the $\Upsilon(10753)$ region through the processes $e^+e^-\to\eta\Upsilon(1S)$, $\eta\Upsilon(2S)$, and $\gamma X_b$, using $19.6~\mathrm{fb}^{-1}$ of data collected at four center-of-mass energy points. A signal of $e^+e^-\to\eta\Upsilon(2S)$ was observed with a combined significance exceeding $6\sigma$, while the significance at $\sqrt{s}=10.653$ GeV reached $4.2\sigma$. The measured Born cross section at 10.653 GeV is higher than that at 10.745 GeV, and the energy-dependent analysis provides evidence for a possible additional structure near the $B^*\bar{B}^*$ threshold with a significance of $3.2\sigma$. Figure~\ref{fig:ee_bes_BornCrossSection} shows the measured Born cross sections for $e^+e^-\to\eta\Upsilon(2S)$ together with the fits under different resonant-structure hypotheses. No significant signals were observed for $e^+e^-\to\eta\Upsilon(1S)$ or $\gamma X_b$, and upper limits were determined. These measurements provide additional information on the energy dependence of bottomonium production near the $\Upsilon(10753)$ region and motivate further studies of the nearby open-bottom thresholds \cite{BelleII:2026EtaUpsilon2S}.

\subsection{Precision Charmonium Transitions and Branching Fractions at BESIII}
\label{sec:ee_bes_transitions}
\begin{figure}[htbp]
\centering
\includegraphics[width=0.65\textwidth]{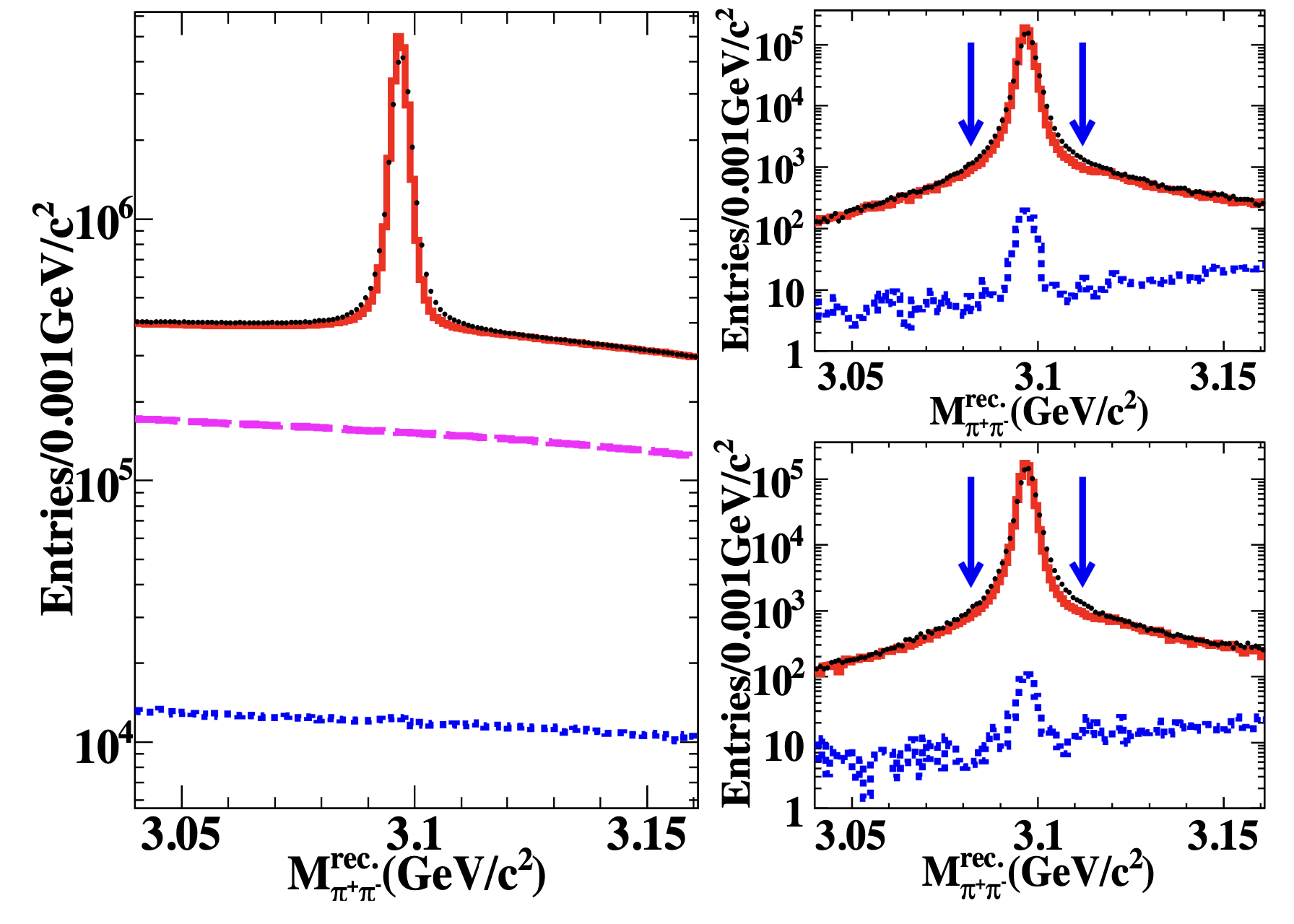}
\caption{Dipion recoil-mass spectrum used by BESIII to isolate $\psi(2S)\to\pi^+\pi^-J/\psi$
decays in a sample of $(106.41\pm0.86)\times10^6$ $\psi(2S)$ events. Figure
reproduced from M.~Ablikim \textit{et al.} (BESIII Collaboration)
\cite{Ablikim:2013wzq}, Fig.~1.}
\label{fig:ee_bes_psi2s_pipi}
\end{figure}

BESIII, operating with samples of $\psi(2S)$ decays two orders of magnitude
larger than those available to earlier BES and BESII analyses, has transformed
the precision of charmonium transition spectroscopy. Using
$(106.41\pm0.86)\times10^6$ $\psi(2S)$ events, BESIII performed a simultaneous
measurement of the dominant hadronic transition and the $J/\psi$ leptonic
branching fractions, obtaining
$\mathcal{B}[\psi(2S)\to\pi^+\pi^-J/\psi]=(34.98\pm0.02\pm0.45)\%$,
$\mathcal{B}[J/\psi\to e^+e^-]=(5.983\pm0.007\pm0.037)\%$, and
$\mathcal{B}[J/\psi\to\mu^+\mu^-]=(5.973\pm0.007\pm0.038)\%$
\cite{Ablikim:2013wzq} (Fig.~\ref{fig:ee_bes_psi2s_pipi}). Because the
$\pi^+\pi^-J/\psi$ transition is the single most common $\psi(2S)$ decay mode
and is routinely used to tag $J/\psi$ samples, this measurement anchors the
normalization of a large fraction of subsequent charmonium branching-fraction
determinations, including the Tevatron analyses of
Section~\ref{sec:tevatron}.

A second cornerstone measurement addresses the radiative $E1$ transitions
$\psi(2S)\to\gamma\chi_{cJ}$ ($J=0,1,2$). Using the same sample, BESIII measured
$\mathcal{B}[\psi(2S)\to\gamma\chi_{c0}]=(9.389\pm0.014\pm0.332)\%$,
$\mathcal{B}[\psi(2S)\to\gamma\chi_{c1}]=(9.905\pm0.011\pm0.353)\%$, and
$\mathcal{B}[\psi(2S)\to\gamma\chi_{c2}]=(9.621\pm0.013\pm0.272)\%$, together
with the product branching fractions for the sequential decay chain
$\psi(2S)\to\gamma\chi_{cJ}\to\gamma\gamma J/\psi$
\cite{Ablikim:2017pkk}. The precision is not an
end in itself: the absolute rates test the charmonium wave functions and
electromagnetic form factors, while the $\chi_{cJ}$ feed-down probabilities are
required to convert an inclusive $J/\psi$ yield into a prompt or direct sample.
They are therefore direct inputs to the production comparisons in
Sections~\ref{sec:tevatron} and~\ref{sec:lhc}.

\subsection{Polarization and Angular Distributions in Charmonium Decays}
\label{sec:ee_polarization}
BESIII has also used its $J/\psi$ sample to study spin correlations in the
decay $J/\psi\to\Lambda\bar{\Lambda}$. With $1.31\times10^9$ $J/\psi$ events,
it reported the first observation of $\Lambda$ and $\bar{\Lambda}$
polarization from the coherent production of the pair, together with the
relative phase between the hadronic form factors,
$\Delta\Phi=(42.4\pm0.6\pm0.5)^\circ$, and improved weak-decay
parameters~\cite{BESIII:2019baryonpol}. This observable is valuable for
testing the spin structure of the charmonium decay and for measuring
hyperon-decay parameters, but it does not by itself determine the polarization
of prompt charmonium produced in hadron collisions. It is therefore included as
a precision QCD and spin-physics result, not as a substitute for the
hadroproduction polarization measurements.

\subsection{Charged Exotic States and Threshold Dynamics}
\label{sec:ee_exotics}
\begin{figure}[htbp]
\centering
\includegraphics[width=0.52\textwidth]{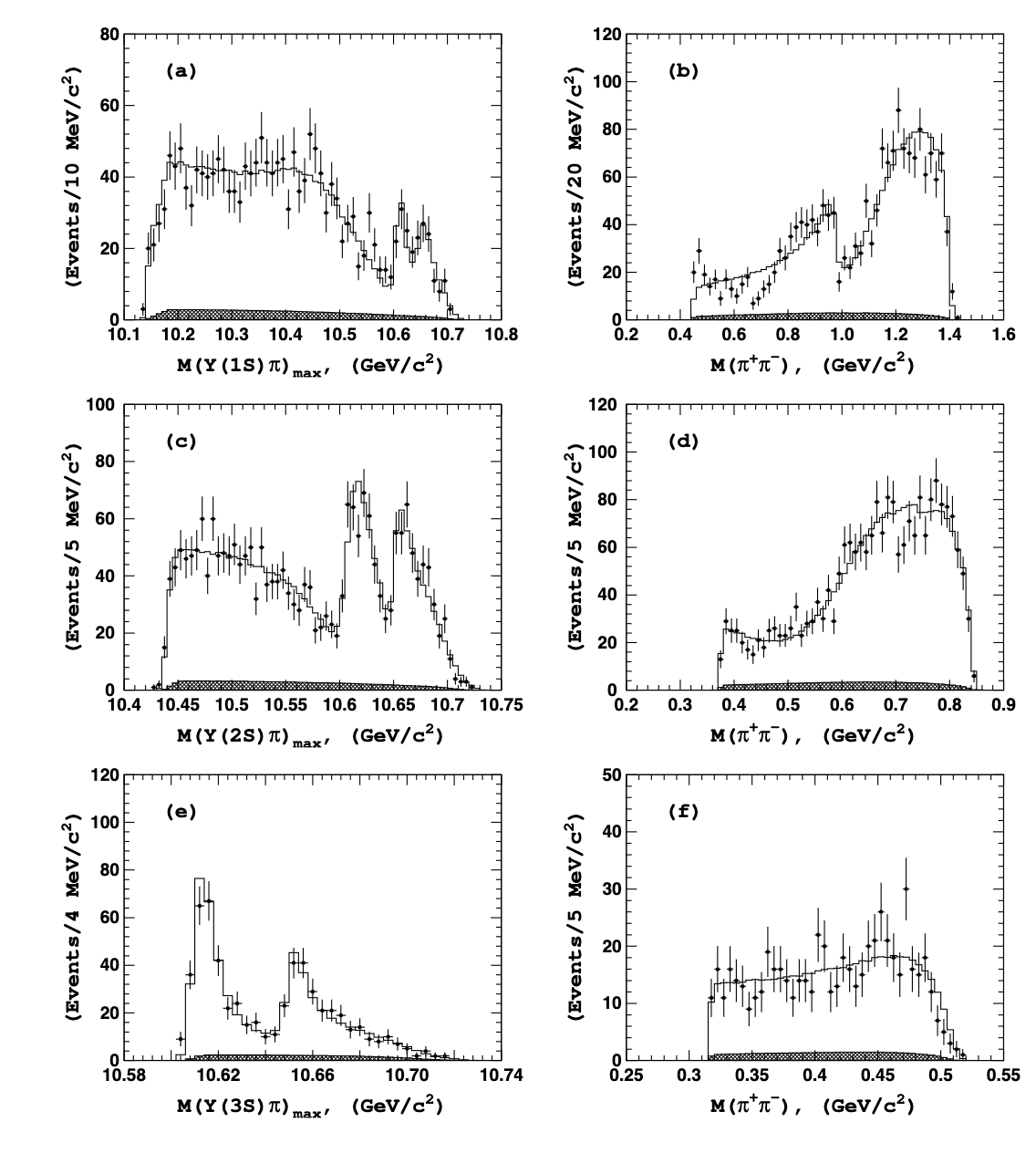}
\caption{Invariant-mass distributions used by Belle to identify the charged
bottomonium-like states $Z_b(10610)^+$ and $Z_b(10650)^+$ in
$\Upsilon(5S)\to\pi^+\pi^-\Upsilon(nS)$ decays. Panels (a), (c), and (e) show
the $M(\Upsilon(nS)\pi^\pm)_{\max}$ distributions for $n=1,2,3$; panels (b),
(d), and (f) show the corresponding $\pi^+\pi^-$ distributions. Points are
data and the histogram is the fit. Figure reproduced from
\cite{Bondar:2011xe}.}
\label{fig:ee_belle_zb_massspectra}
\end{figure}

The clean initial state and low background of $e^+e^-$ annihilation are
particularly useful for identifying states that cannot be assigned to a simple
$Q\bar{Q}$ multiplet. BESIII observed the charged $Z_c(3900)^+$ in
$e^+e^-\to\pi^+\pi^-J/\psi$ near $4.26$~GeV and the
$Z_c(4020)^+$ in the $\pi^\pm h_c$ channel
\cite{Ablikim:2013mio,Ablikim:2013xfr}. The non-zero charge and the decay into
quarkonium plus a light pion exclude a pure $c\bar{c}$ interpretation. Belle
reported the corresponding bottomonium-like pair
$Z_b(10610)^+$ and $Z_b(10650)^+$ in
$\Upsilon(5S)\to\pi^+\pi^-\Upsilon(nS)$ decays~\cite{Bondar:2011xe}; the
measured mass spectra are shown in
Fig.~\ref{fig:ee_belle_zb_massspectra}. Their masses lie close to the
$B\bar B^*$ and $B^*\bar B^*$ thresholds, respectively, which makes
threshold dynamics a natural part of any interpretation.

The BESIII production scan of $e^+e^-\to\pi^+\pi^-J/\psi$ shows that the broad
$Y(4260)$-like enhancement contains a sharply structured peak rather than a
single featureless resonance~\cite{Ablikim:2016qzc}. This is important for
production studies because a resonance fitted in a scan reflects both the
short-distance vector-current production and the rescattering of its decay
products. The same caution applies to $B$-factory measurements of the
$X(3872)$, discovered by Belle in
$B^\pm\to K^\pm\pi^+\pi^-J/\psi$~\cite{Choi:2003ue}; its quantum numbers are
now established as \(J^{PC}=1^{++}\) by LHCb~\cite{LHCb:2013x3872JPC}, and its
threshold proximity makes molecular and compact-tetraquark descriptions
quantitatively different rather than merely semantic.

For the present review, exotic states provide two concrete inputs. First,
their charged members demonstrate that the conventional $Q\bar{Q}$ basis is
incomplete. Second, their decays and line shapes constrain the final-state
interactions and feed-down that also affect inclusive quarkonium yields. They
do not replace the direct LDME tests based on conventional $J/\psi$,
$\psi(2S)$, and $\Upsilon(nS)$ production, and the review uses them at that
level of relevance.

\subsection{Synthesis: The $e^+e^-$-Collider Contribution to Quarkonium Physics}
\label{sec:ee_summary}
The $e^+e^-$-collider measurements reviewed in this section supply absolute
branching fractions and exclusive production cross sections that are required
to normalize, but do not directly reproduce, the inclusive hadroproduction
results of Section~\ref{sec:tevatron}. Precision determinations of
$\mathcal{B}[\psi(2S)\to\pi^+\pi^-J/\psi]$, the radiative transitions
$\psi(2S)\to\gamma\chi_{cJ}$, and the continuum prompt $J/\psi$ and $\psi(2S)$
cross sections provide the long-distance inputs used in absolute
hadron-collider analyses.
Exclusive double-charmonium production further offers a theoretically clean benchmark process whose higher-order corrections have been shown to reconcile early discrepancies between data and leading-order calculations. 
The charged $Z_c$ and $Z_b$ states and the threshold scans additionally show
that the hadron spectrum contains dynamics outside a pure $Q\bar{Q}$ basis.
Together these results define a high-precision, low-background reference
against which the more complex hadroproduction environment of the Tevatron and
LHC can be calibrated and interpreted.

%% file: E_LHC.tex
\section{Quarkonium Physics at the LHC}
\label{sec:lhc}

The LHC tests quarkonium production over the broadest kinematic and nuclear
environment available at a single facility. Proton-proton ($pp$) data provide
the high-luminosity reference samples; proton-lead ($p$Pb) data isolate
initial- and final-state cold-nuclear-matter effects; and lead-lead (PbPb)
data probe the hot quark--gluon plasma (QGP). The four major experiments are
complementary. ALICE measures low-$p_T$ charmonium and has dedicated
heavy-ion capabilities; ATLAS and CMS reconstruct electrons and muons over
large central acceptances and operate high-rate multilevel triggers; and LHCb
covers the forward region with a flexible software trigger, precise vertexing,
and powerful particle identification. This section follows the same questions
as the earlier chapters: whether a common production description survives the
LHC kinematic reach, how the spin observables behave, and how the medium and
final-state interactions alter the observed yields.

\subsection{Production Cross Sections and Polarization}
\label{sec:lhc_production}

\begin{figure}[htbp]
\centering
\includegraphics[width=0.76\textwidth]{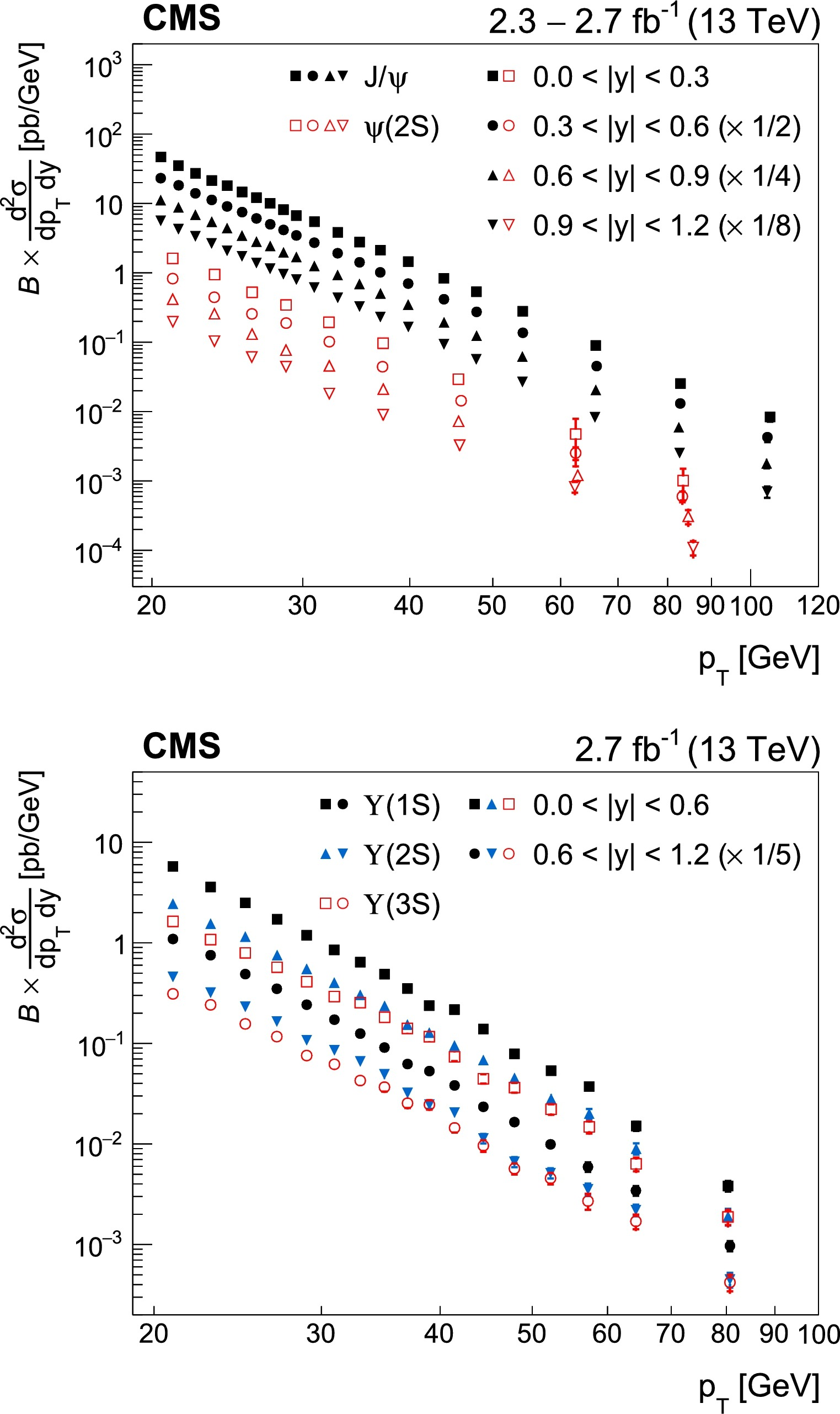}
\caption{CMS measurements of
$\mathcal{B}\,d^2\sigma/(dp_T\,dy)$ in $pp$ collisions at
$\sqrt{s}=13$~TeV. The upper panel shows prompt $J/\psi$ (black) and
$\psi(2S)$ (red) results for four $|y|$ intervals; the lower panel shows
$\Upsilon(1S)$, $\Upsilon(2S)$, and $\Upsilon(3S)$ results in two $|y|$
intervals. Symbols with different scales correspond to the rapidity bins
identified in the legends. Figure reproduced from
\cite{CMS:2018quarkonium}.}
\label{fig:lhc_jpsi_xsec}
\end{figure}

CMS measured prompt $J/\psi$, $\psi(2S)$, $\Upsilon(1S)$, $\Upsilon(2S)$, and
$\Upsilon(3S)$ production in $pp$ collisions at $\sqrt{s}=13$~TeV with
$2.3$--$2.7~\mathrm{fb}^{-1}$ of data~\cite{CMS:2018quarkonium}. The
differential cross sections, multiplied by the relevant dilepton branching
fraction, are shown in Fig.~\ref{fig:lhc_jpsi_xsec} for several rapidity
intervals, using the acceptance- and efficiency-corrected yield definition of
Eq.~\eqref{eq:differential_cross_section}. They extend the Tevatron measurements by roughly an order of
magnitude in $p_T$ and provide a common normalization for the higher-energy
measurements discussed below.

\begin{figure}[htbp]
\centering
\includegraphics[width=0.76\textwidth]{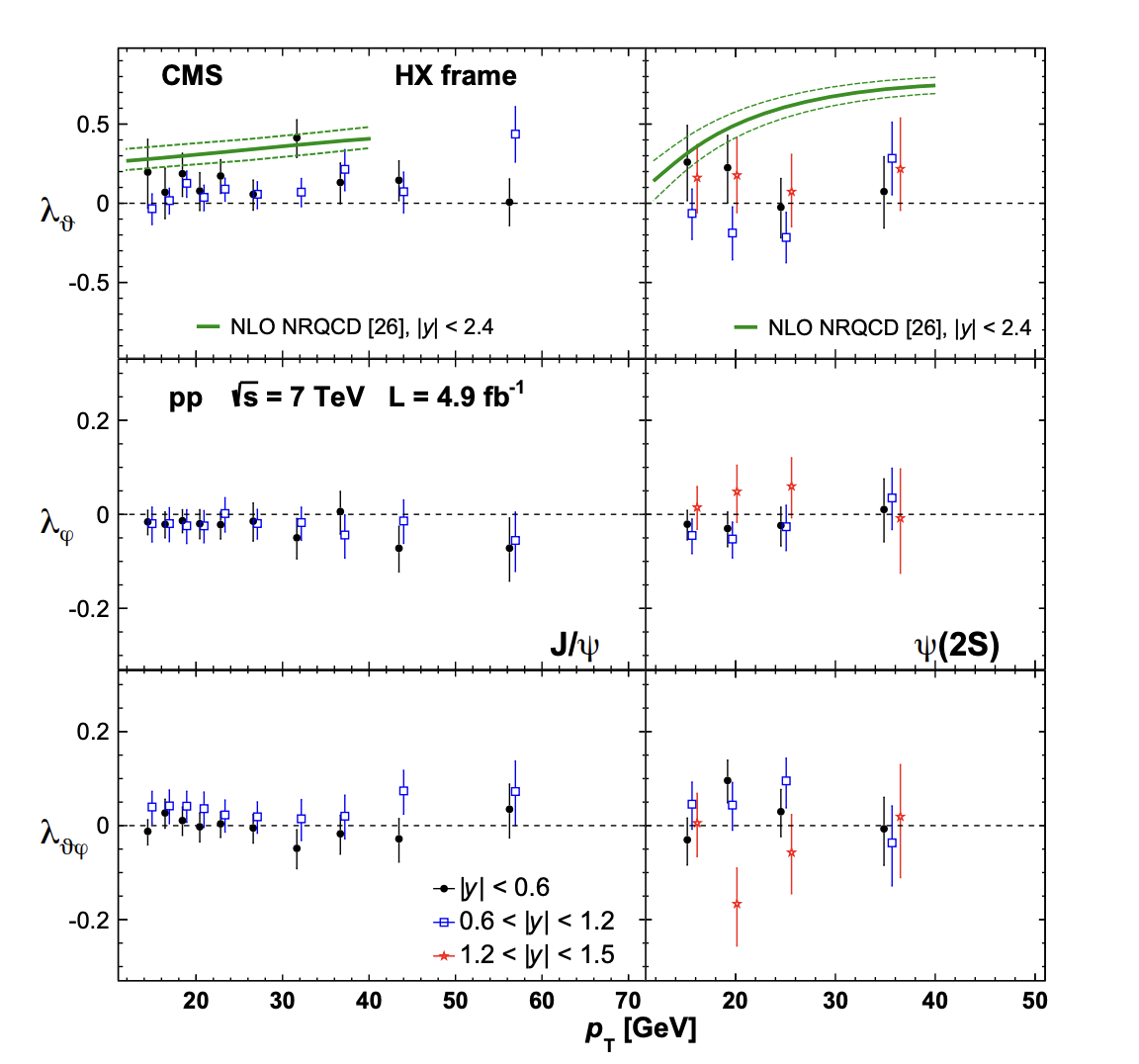}
\caption{Polarization parameters $\lambda_\theta$, $\lambda_\phi$, and $\lambda_{\theta\phi}$ measured in the helicity frame for prompt $J/\psi$ (left) and $\psi(2S)$ (right) mesons produced in $pp$ collisions at $\sqrt{s}=7$ TeV, as functions of $p_T$ in different rapidity intervals. The measurements are compared with NLO NRQCD predictions for $\lambda_\theta$. Figure reproduced from Ref.~\cite{CMS:2013JpsiPsi2SPolarization}.}
\label{fig:cms_jpsi_psi2s_pol1}
\end{figure}

CMS also measured the polarizations of prompt $J/\psi$ and $\psi(2S)$ mesons in $pp$ collisions at $\sqrt{s}=7$ TeV using a data sample corresponding to an integrated luminosity of $4.9~\mathrm{fb}^{-1}$. The polarization parameters $\lambda_\theta$, $\lambda_\phi$, and $\lambda_{\theta\phi}$, together with the frame-invariant parameter $\widetilde{\lambda}$, were extracted from the dimuon decay angular distributions in three polarization frames. The measurements cover $14<p_T<70~\mathrm{GeV}$ for $J/\psi$ and $14<p_T<50~\mathrm{GeV}$ for $\psi(2S)$, with several rapidity intervals. No evidence of large transverse or longitudinal polarizations was observed in the explored kinematic regions, and the measured polarization parameters were generally close to zero. Figure~\ref{fig:cms_jpsi_psi2s_pol1} shows the measured polarization parameters in the helicity (HX) frame as a function of $p_T$ across different rapidity bins, compared with next-to-leading-order (NLO) NRQCD predictions. These results provided important constraints on NRQCD polarization predictions, particularly for $\psi(2S)$, which is not affected by feed-down from higher charmonium states \cite{CMS:2013JpsiPsi2SPolarization}.

\begin{figure}[htbp]
\centering
\includegraphics[width=0.95\textwidth]{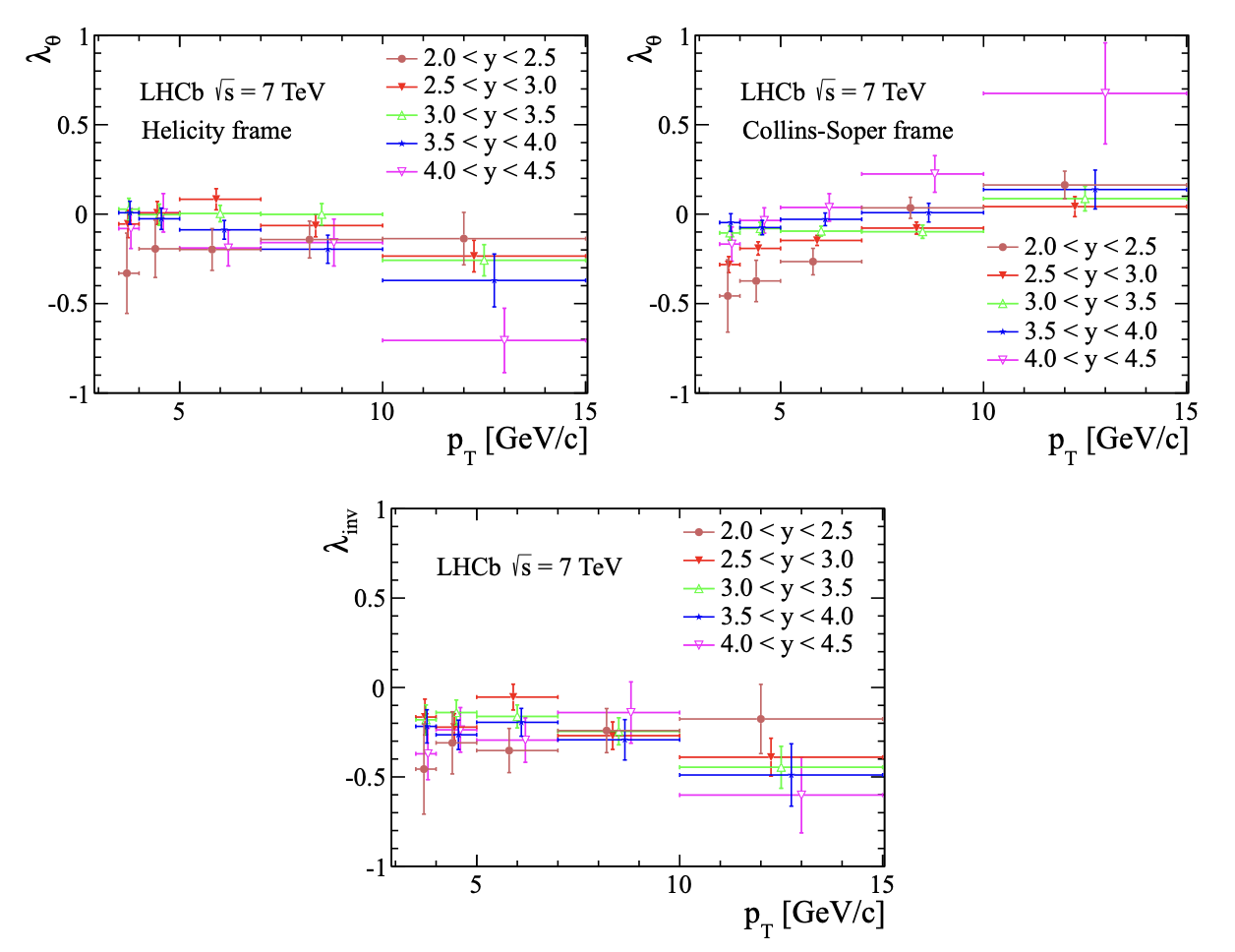}
\caption{Polarisation parameters for prompt $\psi(2S)$ mesons as a function of $p_{\mathrm{T}}$, in five rapidity intervals, (\textit{top left}) $\lambda_\theta$ and (\textit{bottom}) $\lambda_{\mathrm{inv}}$, measured in the helicity frame, and (\textit{top right}) $\lambda_\theta$ in the Collins-Soper frame. The uncertainties on data points are the sum in quadrature of statistical and systematic uncertainties. The horizontal bars represent the width of the $p_{\mathrm{T}}$ bins for the $\psi(2S)$ meson. The data points for each rapidity interval are displaced horizontally to improve visibility. Figure reproduced from Ref.~\cite{LHCb:2014Psi2SPolarization}.}
\label{fig:lhcb_psi2s_pol}
\end{figure}

Complementary forward-rapidity measurements were performed by LHCb, which measured the polarization of prompt $\psi(2S)$ mesons in $pp$ collisions at $\sqrt{s}=7$ TeV using a data sample corresponding to an integrated luminosity of $1.0~\mathrm{fb}^{-1}$. The measurement covers $3.5<p_T<15~\mathrm{GeV}/c$ and $2.0<y<4.5$, providing complementary kinematic coverage to the CMS measurements at mid-rapidity. The polarization parameters were determined from the dimuon angular distributions in the helicity and Collins--Soper frames. No significant polarization was observed in the helicity frame, where $\lambda_\theta$ is consistent with zero, while the frame-invariant polarization parameter is consistent with a negative polarization. Figure~\ref{fig:lhcb_psi2s_pol} shows the measured $\lambda_\theta$ in both frames together with $\lambda_{\mathrm{inv}}$ in the helicity frame as a function of $p_T$ across five rapidity intervals. The measured results disfavour the NLO color-singlet-model predictions, while the NRQCD calculations provide a reasonable description of the data at low $p_T$ but do not reproduce the predicted increase toward transverse polarization at higher $p_T$ \cite{LHCb:2014Psi2SPolarization}.

\begin{figure}[htbp]
\centering
\includegraphics[width=0.85\textwidth]{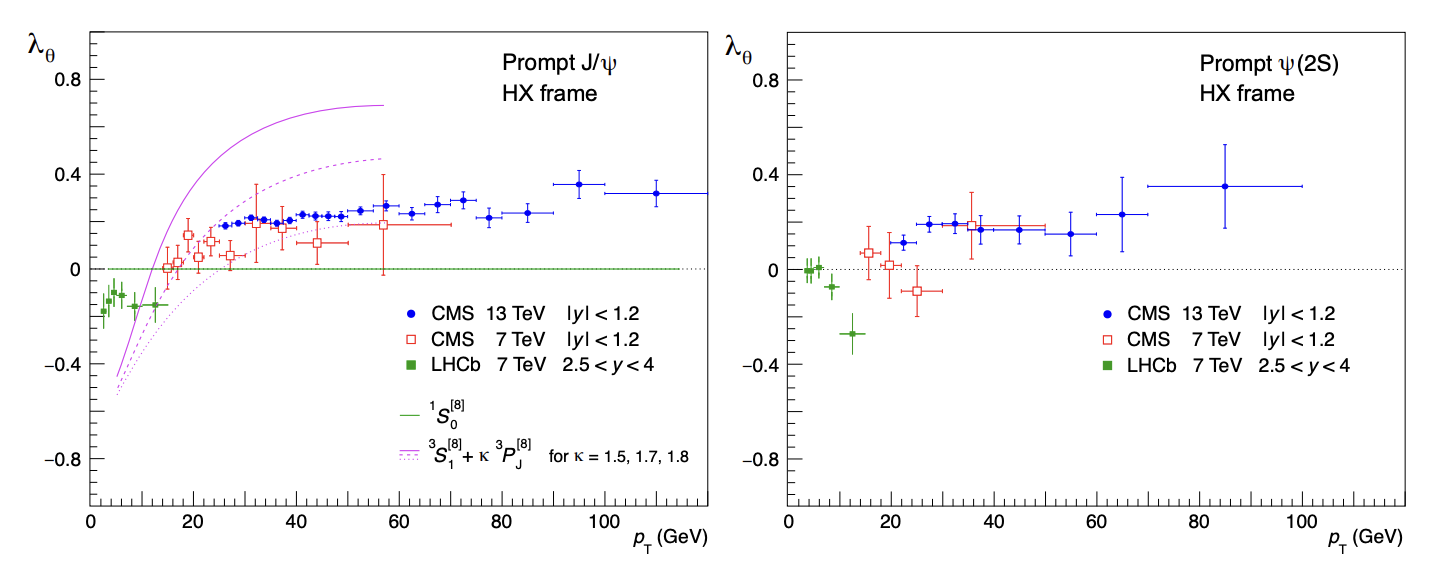}
\caption{The polar anisotropy parameter $\lambda_\theta$ measured in the helicity frame as a function of $p_T$ for prompt $J/\psi$ (left) and $\psi(2S)$ (right) mesons produced in $pp$ collisions at $\sqrt{s}=13$ TeV. The CMS measurements are compared with previous measurements at $\sqrt{s}=7$ TeV by CMS and LHCb. Figure reproduced from Ref.~\cite{CMS:2024PromptNonpromptPol}.}
\label{fig:cms_jpsi_psi2s_pol}
\end{figure}

CMS subsequently measured the polarizations of prompt and non-prompt $J/\psi$ and $\psi(2S)$ mesons in $pp$ collisions at $\sqrt{s}=13$ TeV using data collected in 2017 and 2018, corresponding to an integrated luminosity of $103.3~\mathrm{fb}^{-1}$. The polar anisotropy parameter $\lambda_\theta$ was measured in the helicity frame for $|y|<1.2$, covering $25<p_T<120~\mathrm{GeV}$ for $J/\psi$ and $20<p_T<100~\mathrm{GeV}$ for $\psi(2S)$. The prompt measurements show no evidence of strong transverse polarization, with $\lambda_\theta$ approaching approximately $0.3$ at high $p_T$, rather than the value near $+1$ expected if the transverse color-octet contributions became dominant. As shown in Fig.~\ref{fig:cms_jpsi_psi2s_pol}, the measured $p_T$ dependence of $\lambda_\theta$ for prompt $J/\psi$ and $\psi(2S)$ at $\sqrt{s}=13$ TeV aligns smoothly with the earlier CMS and LHCb measurements at $\sqrt{s}=7$ TeV. The non-prompt $J/\psi$ and $\psi(2S)$ polarizations exhibit compatible trends, approaching $\lambda_\theta\approx-0.2$ for $p_T\gtrsim30~\mathrm{GeV}$, consistent with predictions based on their predominant production through two-body $B$-meson decays. Together with the earlier CMS and LHCb measurements at $\sqrt{s}=7$ TeV, these results reveal a significant $p_T$ dependence of prompt charmonium polarization and provide important constraints on the relative contributions of the color-octet production channels \cite{CMS:2024PromptNonpromptPol}.

These measurements continue a concrete empirical chain. The Tevatron exposed
the excess over the leading-order color-singlet rate and the first
polarization tension; $e^+e^-$ experiments fixed the branching fractions and
feed-down inputs; and the LHC now extends the same tests to higher $p_T$ and
new collision systems. NRQCD factorization, introduced in
Section~\ref{sec:introduction}, relates
these differential spectra to short-distance coefficients and universal
long-distance matrix elements (LDMEs). The leading color-singlet and
color-octet terms have been tested against the Tevatron and LHC data using
global fits~\cite{Bodwin:1994jh,Butenschoen:2010rq}. The resulting
cross-section descriptions are generally successful once feed-down and
higher-order corrections are included, but they do not uniquely select the
corresponding spin observables.

Polarization supplies complementary sensitivity to the underlying production mechanisms. ALICE measured the polarization of inclusive \(J/\psi\) mesons through their dimuon decays in \(pp\) collisions at \(\sqrt{s}=7\) TeV, covering \(2.5<y<4\) and \(2<p_T<8\) GeV/\(c\)~\cite{ALICE:2012jpsipol}. Unlike the CMS and LHCb measurements discussed above, this measurement did not separate prompt \(J/\psi\) production from non-prompt contributions originating from \(B\)-hadron decays. Consequently, the measured polarization parameters characterize the inclusive \(J/\psi\) sample and should not be interpreted as a direct measurement of prompt \(J/\psi\) polarization. In the helicity frame, \(\lambda_\theta\) indicated longitudinal polarization at low \(p_T\) and became consistent with zero as \(p_T\) increased, while the other parameters were consistent with zero within the quoted uncertainties. The inclusive nature of this measurement should be taken into account when comparing its results with prompt polarization measurements from CMS, LHCb, and the Tevatron. Such comparisons require careful consideration of the different production components, kinematic ranges, and polarization frames.

\subsection{High-$p_T$ Bottomonium at $\sqrt{s}=13.6$~TeV}
\label{sec:lhc_upsilon136}

CMS has measured the differential $\Upsilon(1S)$, $\Upsilon(2S)$, and
$\Upsilon(3S)$ cross sections in $pp$ collisions at
$\sqrt{s}=13.6$~TeV using $37.4\pm0.5~\mathrm{fb}^{-1}$ collected in 2022
\cite{CMS-BPH-24-004}. The analysis uses the dimuon channel for
$20<p_T<200~\mathrm{GeV}$ in $|y|<0.6$ and
$0.6<|y|<1.2$. The high-$p_T$ reach and the two rapidity bins make this sample
especially useful for discriminating among LDME sets that have similar
integrated rates but different $p_T$ and rapidity shapes.

\begin{figure}[htbp]
\centering
\includegraphics[width=0.45\textwidth]{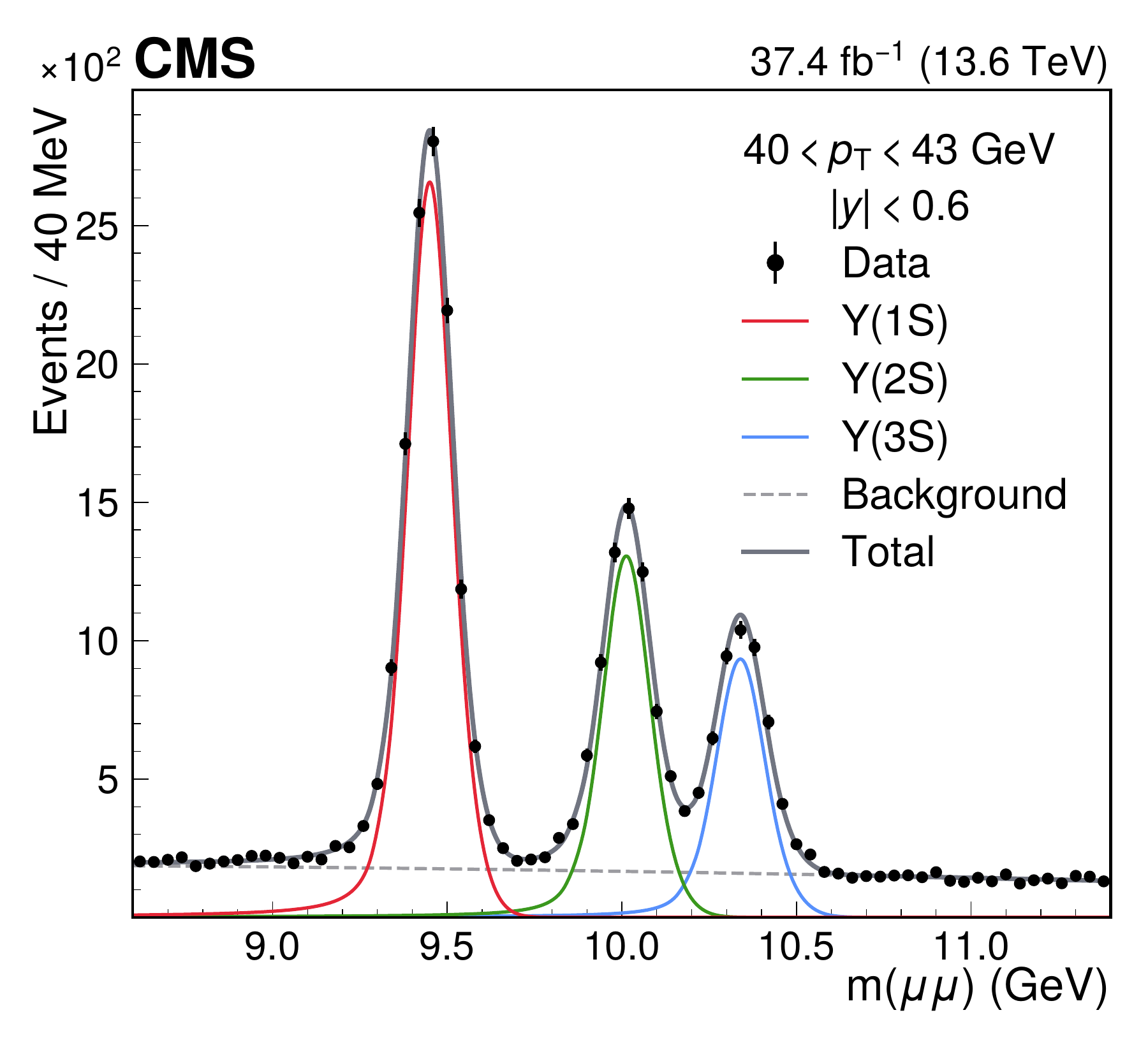}
\includegraphics[width=0.45\textwidth]{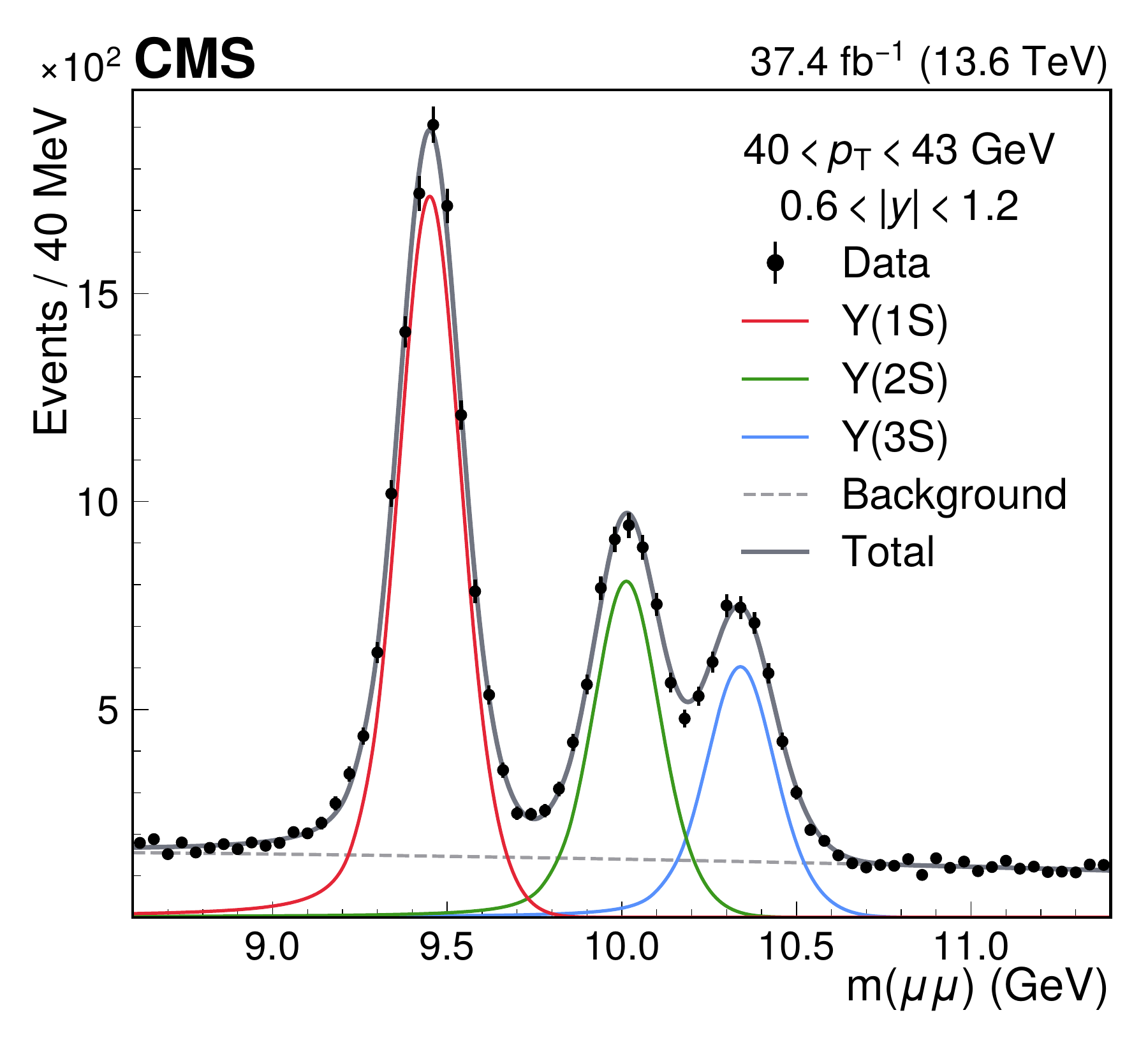}
\caption{Examples of dimuon invariant-mass fits in one high-$p_T$ bin for
$|y|<0.6$ (left) and $0.6<|y|<1.2$ (right) at
$\sqrt{s}=13.6$~TeV. Shown are the measured spectrum, the fitted
$\Upsilon(1S)$, $\Upsilon(2S)$, and $\Upsilon(3S)$ contributions, and the
background model. The acceptance baseline assumes unpolarized production.
Figure reproduced from~\cite{CMS-BPH-24-004}.}
\label{fig:cms_mass_fit}
\end{figure}

The signal is extracted from fits to the dimuon mass spectrum; an illustrative
fit in one $p_T$ bin is shown in Fig.~\ref{fig:cms_mass_fit}. The acceptance is
evaluated with a simulated unpolarized sample and the result is quoted after
the corresponding branching fraction and efficiency corrections. The absolute
luminosity uncertainty is $1.4\%$ and is fully correlated across bins; the
yield-extraction and acceptance uncertainties grow toward low $p_T$, while the
dimuon-trigger uncertainty is largest where the two muons overlap in the
detector. These effects are included in the published spectra and are not
repeated for each data point here.

\begin{figure}[htbp]
\centering
\includegraphics[width=0.48\textwidth]{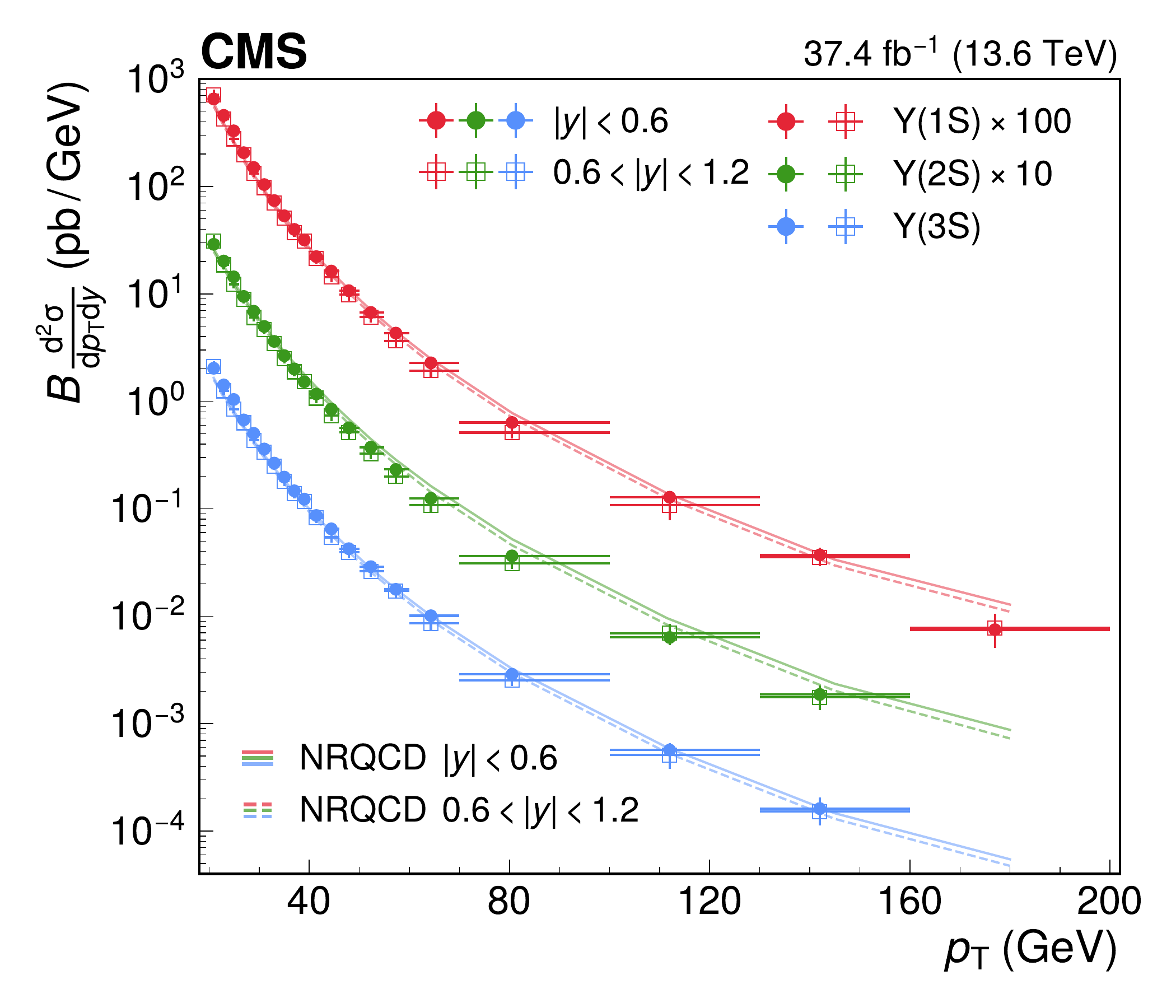}
\includegraphics[width=0.48\textwidth]{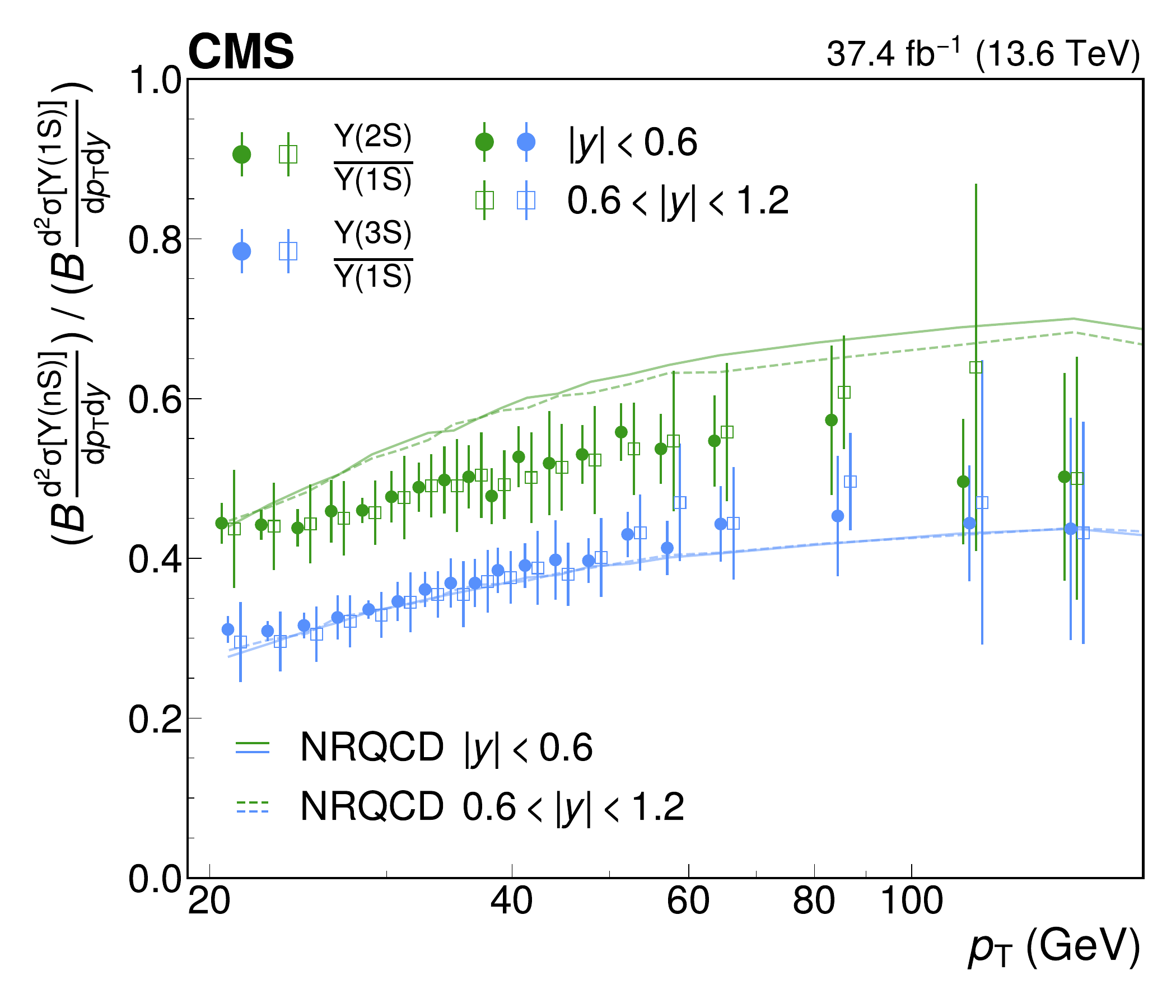}\\[0.5em]
\includegraphics[width=0.64\textwidth]{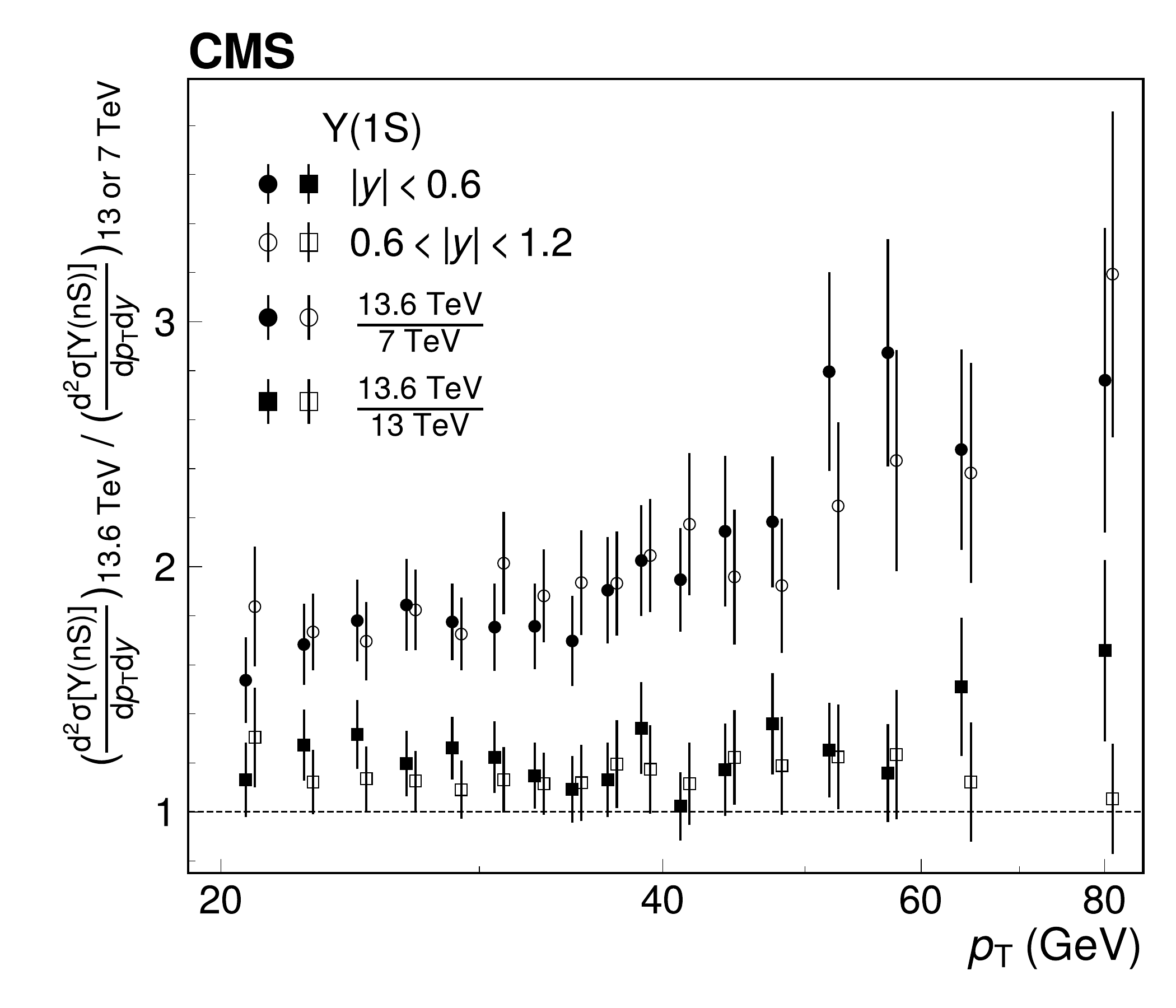}
\caption{Top left: $\Upsilon(nS)$ differential cross sections versus $p_T$ in
$pp$ collisions at $\sqrt{s}=13.6$~TeV. Top right: the excited-to-ground-state
cross-section ratios, which increase at lower-to-intermediate $p_T$ and
gradually approach a plateau at higher $p_T$. Bottom: the
$\Upsilon(1S)$ cross-section ratio between 13.6~TeV and earlier 7~TeV and
13~TeV measurements. Figure reproduced from~\cite{CMS-BPH-24-004}.}
\label{fig:cms_results}
\end{figure}

The resulting spectra and cross-section ratios are summarized in
Fig.~\ref{fig:cms_results}. The two midrapidity intervals have similar yields.
The $\Upsilon(2S)/\Upsilon(1S)$ and $\Upsilon(3S)/\Upsilon(1S)$ cross-section
ratios increase with $p_T$ over the lower-to-intermediate $p_T$ range and
then gradually approach a plateau at higher $p_T$. This relative observable
is particularly constraining because many luminosity and acceptance effects
cancel. The comparison with NRQCD calculations including feed-down from
higher bottomonia~\cite{Butenschoen:2010rq} tests whether the same
non-perturbative inputs describe both the ground and excited states. The data
reach beyond the $p_T$ range of previous 13~TeV measurements and provide a
useful benchmark for future global fits.

Recent CMS and ATLAS analyses also report an excess near the
$t\bar{t}$ production threshold whose shape is compatible with toponium
formation~\cite{CMS:2026toponium,ATLAS:2026toponium}. The observation is
notable because the top-quark lifetime is much shorter than a conventional
quarkonium oscillation period. It extends the bound-state question into a
regime where the ``state'' is a short-lived threshold enhancement rather than a
well-separated resonance, and it provides a new constraint on threshold QCD.

\subsection{LHC Spectroscopy and Exotic States}
\label{sec:lhc_spectroscopy}

The LHC has also become a discovery facility for hadrons that test the
conventional quarkonium picture. LHCb observed $J/\psi p$ resonances consistent
with pentaquark states in
$\Lambda_b^0\to J/\psi K^-p$ decays~\cite{LHCb:2015pentaquark}, established the
narrow doubly charmed tetraquark $T_{cc}^+$ in
$D^0D^0\pi^+$~\cite{LHCb:2021tcc}, and discovered a structure near
$6.9~\mathrm{GeV}$ in the di-$J/\psi$ mass spectrum~\cite{LHCb:2020bwg}. The $J/\psi\phi$ invariant-mass spectrum provides another important
experimental channel for investigating exotic hadron structures.
The experimental measurements, current status, and possible
implications of structures observed in this channel have been
reviewed in Ref.~\cite{Zhou:2025nyj}. CMS
subsequently resolved three interfering structures near
$6.6$, $6.9$, and $7.1~\mathrm{GeV}$, while ATLAS confirmed the
$X(6900)$-like excess~\cite{CMS:2024JJ,ATLAS:2023bft}. These results are
complementary to the conventional production program: the final states contain
two heavy-quark pairs or an extra light-quark pair, so their production and
decay probe multi-parton and threshold dynamics beyond the simplest $Q\bar{Q}$
classification.

For the $X(6900)$, LHCb currently provides a production measurement relative
to inclusive di-$J/\psi$ production rather than an absolute cross section.
The ratio is $[1.1\pm0.4\,(\mathrm{stat.})\pm0.3\,(\mathrm{syst.})]\%$ for
both $J/\psi$ mesons in $p_T<10~\mathrm{GeV}$ and $2<y<4.5$, and rises to
$[2.6\pm0.6\,(\mathrm{stat.})\pm0.8\,(\mathrm{syst.})]\%$ after requiring
$p_T(J/\psi J/\psi)>5.2~\mathrm{GeV}$~\cite{LHCb:2020bwg}. The restriction to
this forward acceptance is important when comparing with central CMS and ATLAS
measurements.

This review does not attempt a complete catalogue of exotic states; the
Particle Data Group provides the broader experimental summary
\cite{ParticleDataGroup:2022pth}. The relevant point here is that spectroscopy
and production are not competing subjects. Spectroscopy fixes the quantum
numbers and line shapes needed to interpret the yields, while production
measurements test whether the same short-distance and hadronization
descriptions can be extended to threshold and multiquark systems.

\subsection{Quarkonium Suppression in Heavy-Ion Collisions}
\label{sec:lhc_heavy_ion}

The underlying suppression observable is the nuclear modification factor
\begin{equation}
R_{AA}(p_T,y)=\frac{d^2N_{AA}/dp_T\,dy}
{\langle N_{\rm coll}\rangle\,d^2N_{pp}/dp_T\,dy},
\label{eq:raa}
\end{equation}
normalized so that $R_{AA}=1$ corresponds to the expectation from the
nucleon--nucleon collision rate in the absence of nuclear effects.
The proton--nucleus analogue $R_{pA}$ replaces the PbPb yield by the $p$Pb
yield while retaining the same $N_{\rm coll}$ normalization; it is therefore
more sensitive to cold-nuclear-matter effects than to a final-state QGP.
Comparisons with $p$Pb, where the initial-state nuclear modification is present
but a large QGP is not expected, separate cold-nuclear-matter effects from
final-state suppression.

\begin{figure}[htbp]
\centering
\includegraphics[width=0.86\textwidth]{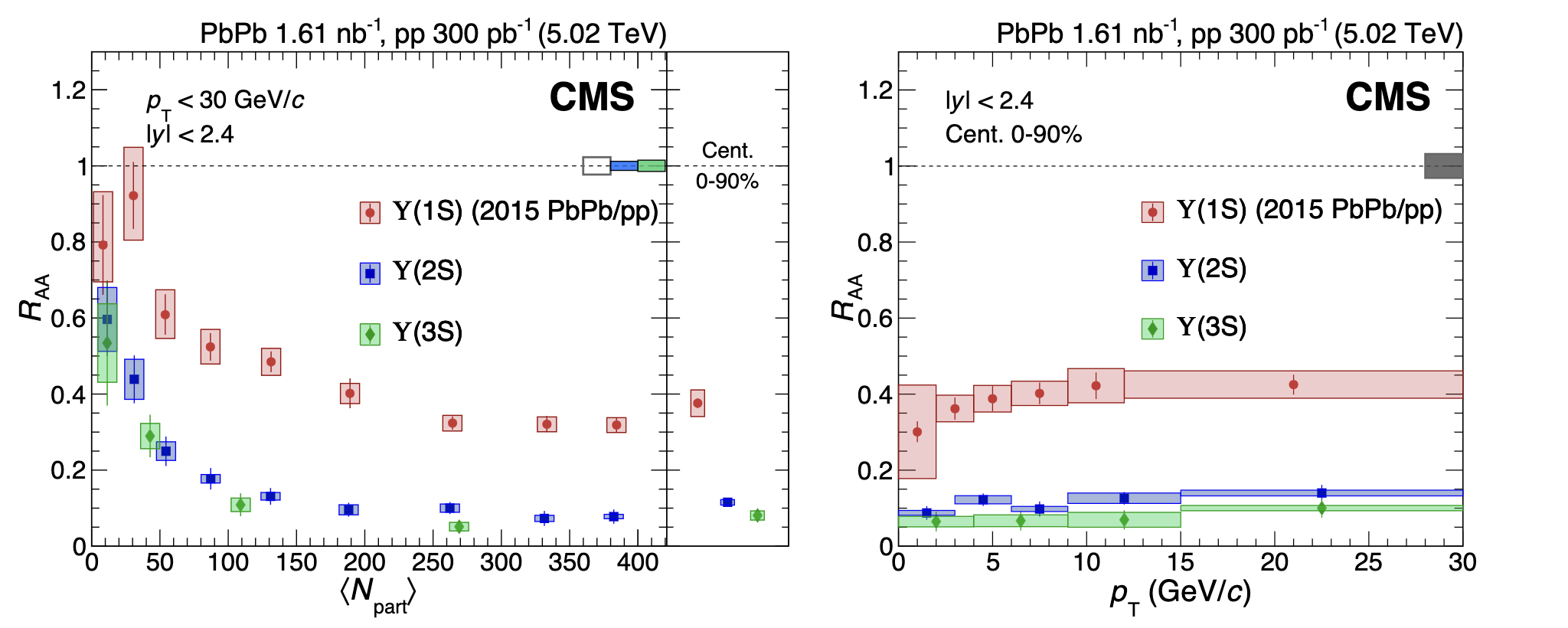}
\caption{Measured $R_{\mathrm{AA}}$ for the $\Upsilon$ states as functions of $\langle N_{\mathrm{part}} \rangle$ (left), showing also the 0--90\% centrality interval, and of $p_{\mathrm{T}}$ (right). The vertical lines and boxes correspond to statistical and systematic uncertainties, respectively. In the left plot, the leftmost box at unity represents the $\mathrm{pp}$ luminosity and $\mathrm{PbPb}$ $N_{\mathrm{MB}}$ combined uncertainties, whereas the second (third) box corresponds to the uncertainty on the $\Upsilon(\mathrm{2S})$ ($\Upsilon(\mathrm{3S})$) $\mathrm{pp}$ yields. The box at unity in the right plot combines the uncertainties of $T_{\mathrm{AA}}$, $\mathrm{pp}$ luminosity, and $\mathrm{PbPb}$ $N_{\mathrm{MB}}$. Figure reproduced from
\cite{CMS:2024upsilon}.}
\label{fig:upsilon_raa}
\end{figure}

CMS measured $\Upsilon(1S)$, $\Upsilon(2S)$, and $\Upsilon(3S)$ production in PbPb collisions at $\sqrt{s_{NN}}=5.02$~TeV using $1.61~\mathrm{nb}^{-1}$ of PbPb and $302~\mathrm{pb}^{-1}$ of $pp$ data~\cite{CMS:2024upsilon}, achieving the observation of the $\Upsilon(3S)$ meson in heavy-ion collisions with a significance exceeding $5\sigma$. The resulting $R_{AA}$ values are shown in Fig.~\ref{fig:upsilon_raa}. The $\Upsilon(1S)$ yield is suppressed by roughly a factor of two to three ($R_{AA} \approx 0.38$), whereas the $\Upsilon(2S)$ is much more strongly suppressed ($R_{AA} \approx 0.10$) and the $\Upsilon(3S)$ exhibits the strongest suppression of all ($R_{AA} \approx 0.080$). The ordering follows the expected decrease of binding energy with radial excitation and is the characteristic sequential-suppression signature. The $p_T$ and rapidity dependences show that the suppression is not a single global normalization shift.

The interpretation is not a simple temperature reading. Feed-down from
$\chi_b$ and excited states modifies the observed ground-state yield;
cold-nuclear-matter effects are present in both PbPb and $p$Pb; and final-state
interactions can break up or regenerate quarkonia. Charmonium observables do not admit a simple single-mechanism interpretation because prompt and non-prompt components have different production and suppression paths, and coalescence of uncorrelated charm quarks can regenerate $J/\psi$ at low $p_T$ \cite{CMS2018psi,BraunMunzinger:2000px}.The LHC data therefore constrain model
combinations rather than proving Debye screening in isolation. This is a more
precise but less dramatic statement than the original sequential-melting
picture.

\subsection{Multi-Quarkonium Production}
\label{sec:lhc_multiquarkonium}

\begin{figure}[htbp]
\centering
\includegraphics[width=0.66\textwidth]{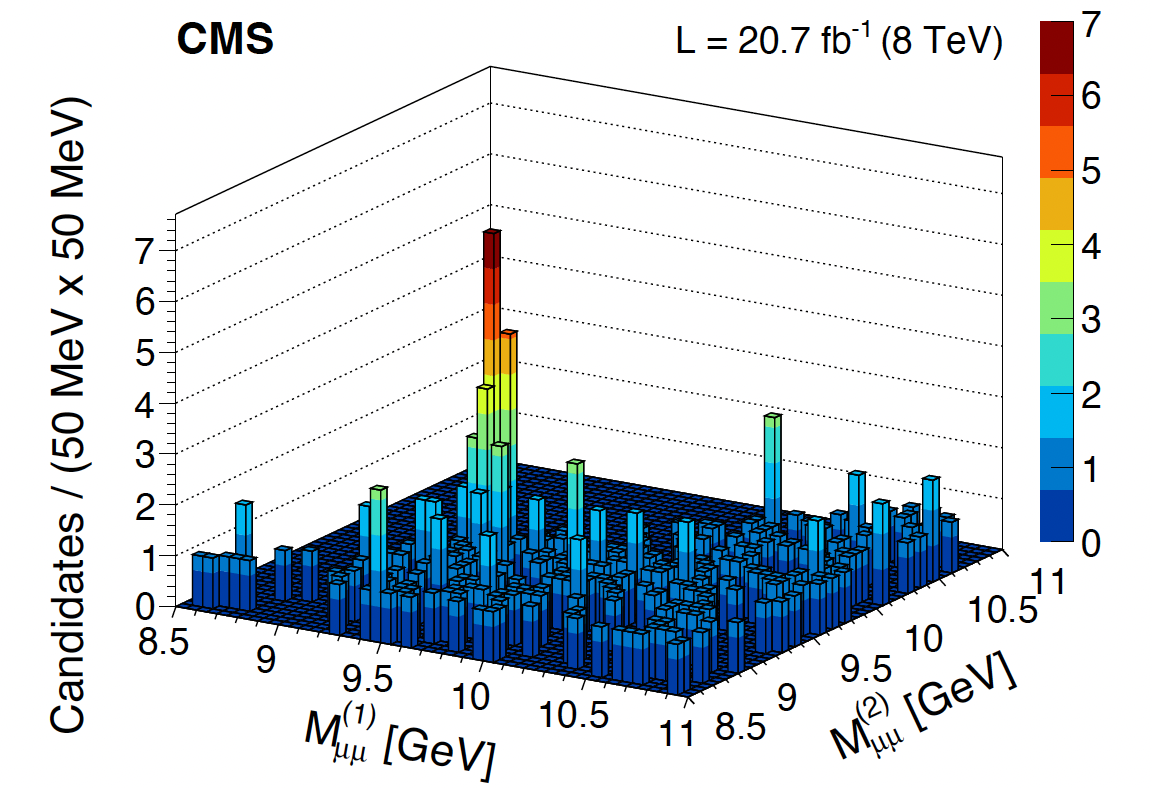}
\caption{Two-dimensional dimuon invariant-mass distribution used to extract
the $\Upsilon(1S)\Upsilon(1S)$ yield in $pp$ collisions at
$\sqrt{s}=8$~TeV. The signal region appears where both dimuon masses are near
the $\Upsilon(1S)$ mass. Figure reproduced from~\cite{CMS:2017vkm}.}
\label{fig:cms_upsilon_pair}
\end{figure}

Double- and triple-quarkonium production provides a probe of multiple parton interactions (MPI) in the same proton--proton collision, including double-parton scattering (DPS) and triple-parton scattering (TPS) contributions. CMS reported the first
observation of $\Upsilon(1S)\Upsilon(1S)$ production at
$\sqrt{s}=8$~TeV using $20.7~\mathrm{fb}^{-1}$; both mesons were reconstructed
in the $\mu^+\mu^-$ channel for $|y|<2.0$, with a $5\sigma$ signal and a
fiducial cross section
\begin{equation}
\sigma[\Upsilon(1S)\Upsilon(1S)]
=68.8\pm12.7\,(\mathrm{stat.})
\pm7.4\,(\mathrm{syst.})
\pm2.8\,(\mathcal{B})~\mathrm{pb},
\label{eq:cms_upsilon_pair}
\end{equation}
where the last uncertainty comes from the dimuon branching fractions
\cite{CMS:2017vkm}. The invariant-mass plane used to isolate the signal is shown
in Fig.~\ref{fig:cms_upsilon_pair}. The dependence on the assumed production
polarization changes the extracted cross section by about
$-38\%$ to $+36\%$, which is a useful reminder that unpolarized acceptance
assumptions propagate into pair-production measurements.

The effective cross section
\begin{equation}
\sigma_{\rm eff}
=\frac{\sigma(A)\,\sigma(B)}{2\,\sigma(AB)}
\label{eq:sigma_eff}
\end{equation}
for identical particles, or without the factor of two for distinguishable
particles, characterizes the transverse overlap of the two parton
distributions. CMS found consistency with other quarkonium-pair channels and
with a smaller effective overlap than typical multi-jet samples. A related CMS
search for simultaneous $\Upsilon(1S)+J/\psi$ production was reported in a
thesis~\cite{Dilsiz:2016thesis}; because it is not a peer-reviewed
publication, it is included only as a complementary cross-check while
unpublished.

\begin{figure}[htbp]
    \centering
    \includegraphics[width=0.95\textwidth]{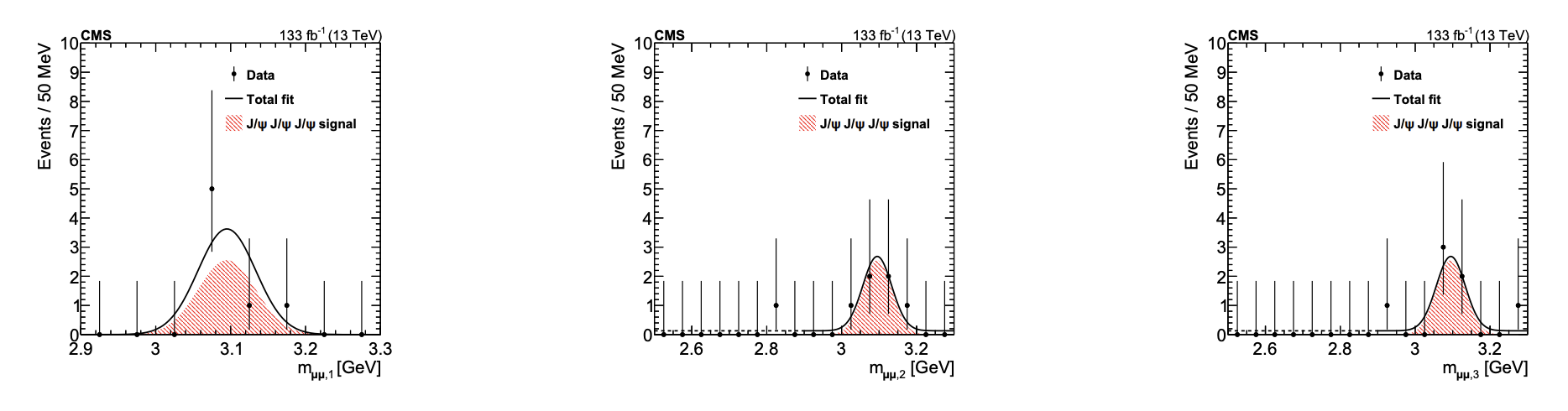}\\[4pt]
    \includegraphics[width=0.35\textwidth]{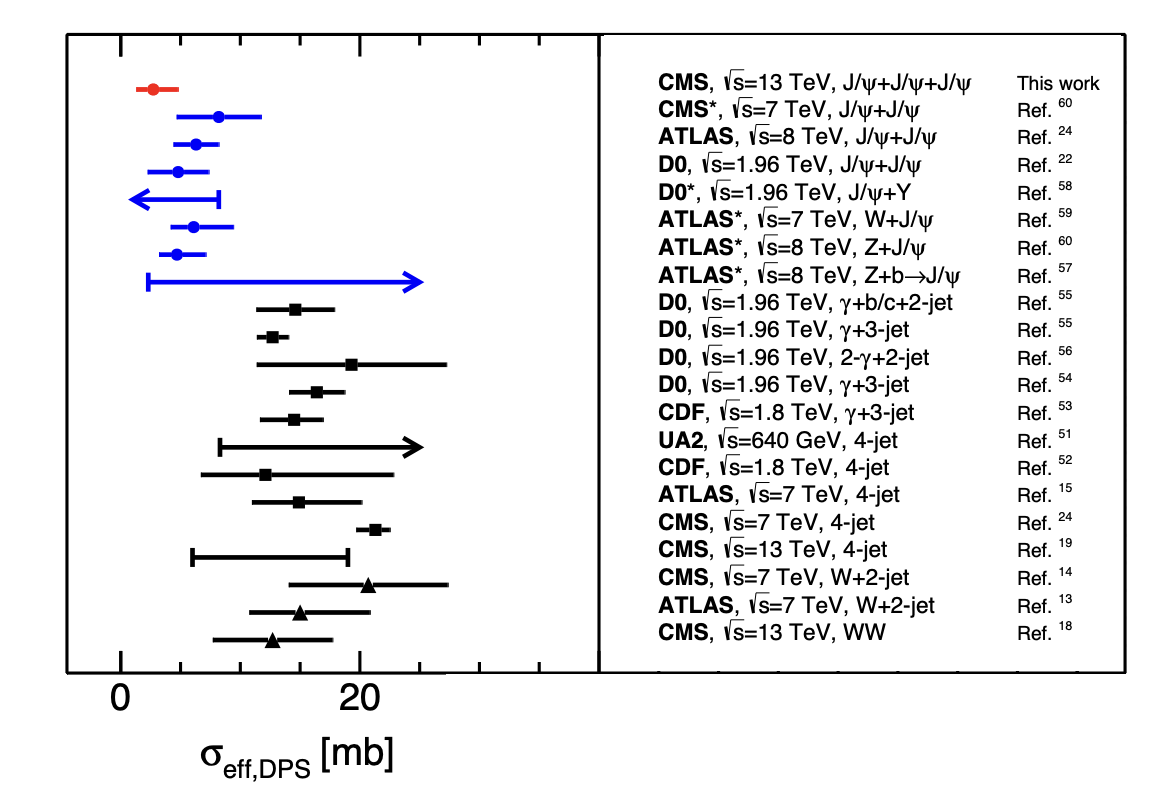}

    \caption{Top: Invariant mass distributions of the three dimuon pairs
    in the selected triple-$J/\psi$ events, together with the fitted
    signal and background contributions. Bottom: Comparison of the effective
    double-parton scattering cross section, $\sigma_{\mathrm{eff,DPS}}$,
    extracted from triple-$J/\psi$ production with measurements from
    other processes. Figure reproduced from
    Ref.~\cite{CMS:2023TripleJpsi}.}

    \label{fig:cms_triple_jpsi}
\end{figure}

CMS further extended the study of multi-quarkonium production by observing the simultaneous production of three $J/\psi$ mesons in $pp$ collisions at $\sqrt{s}=13$ TeV, using a data sample corresponding to an integrated luminosity of $133~\mathrm{fb}^{-1}$. The process was reconstructed through the six-muon final state and observed with a significance above five standard deviations. The inclusive fiducial cross section was measured to be $\sigma(pp\to J/\psi J/\psi J/\psi X)=272^{+141}_{-104}~(\mathrm{stat.})\pm17~(\mathrm{syst.})~\mathrm{fb}$. Under the assumption of factorization of multiple hard-scattering probabilities, the measurement was interpreted as being dominated by double- and triple-parton scattering contributions, with an effective double-parton scattering cross section of $\sigma_{\mathrm{eff,DPS}}=2.7^{+1.4}_{-1.0}~(\mathrm{exp.})^{+1.5}_{-1.0}~(\mathrm{theo.})~\mathrm{mb}$. Figure~\ref{fig:cms_triple_jpsi} shows the invariant mass distributions of the three dimuon pairs from the reconstructed triple-$J/\psi$ candidates, along with a comparison of the extracted $\sigma_{\mathrm{eff,DPS}}$ against measurements from other processes. This observation provides a new experimental handle on multiple parton interactions and the transverse structure of the proton, and demonstrates the potential of triple-quarkonium production for studying triple-parton scattering \cite{CMS:2023TripleJpsi}.

\subsection{The Hidden-Charm State $X(3872)$ in Hadronic Collisions}
\label{sec:lhc_x3872_medium}

The $\chi_{c1}(3872)$, usually denoted $X(3872)$, is a particularly informative
probe of hadronization because competing descriptions predict very different
spatial sizes. Compact diquark--antidiquark models place its radius in the range
$0.3$--$1.3~\mathrm{fm}$, while a loosely bound
$D^0\bar{D}^{*0}$ molecule can extend to about
$10~\mathrm{fm}$~\cite{Maiani2005,Tornqvist2004,Braaten2005}. A larger object
has a larger geometric cross section for breakup by nearby medium
partons or hadrons, but it is also easier to recreate by the coalescence of
those constituents~\cite{Ferreiro2015,Esposito2021,Cho2011a,Cho2020,Andronic2019}.
``Co-moving'' here means constituents with similar
longitudinal velocity in the collision medium, not merely unrelated particles
that happen to pass nearby. Breakup and regeneration therefore have different
multiplicity and density dependences, and neither can be inferred from a
single ratio.

\begin{figure}[htbp]
\centering
\includegraphics[width=0.62\textwidth]{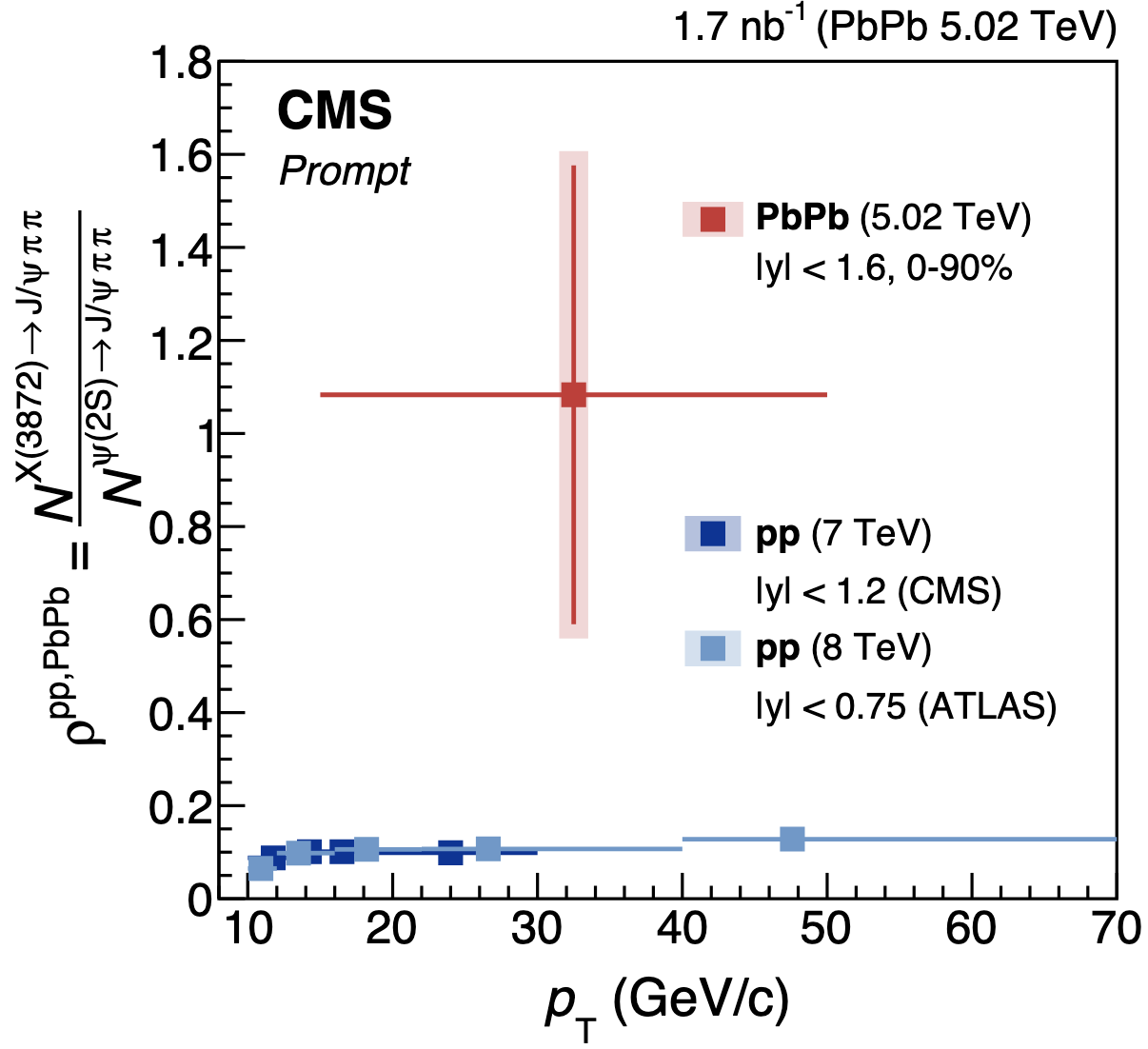}
\caption{Prompt $X(3872)$-to-$\psi(2S)$ yield ratio in PbPb collisions at
$\sqrt{s_{NN}}=5.02$~TeV, shown together with the corresponding lower values
from $pp$ data. Vertical bars and boxes represent statistical and systematic
uncertainties, respectively. Figure reproduced from~\cite{CMS2022x3872}.}
\label{fig:cms_pbpb_ratio}
\end{figure}

To separate structure-dependent dynamics from overall suppression, experiments
form the yield ratio
\begin{equation}
\rho_{\rm sys}
=\frac{N_{\chi_{c1}(3872)\to J/\psi\pi^+\pi^-}}
{N_{\psi(2S)\to J/\psi\pi^+\pi^-}}\bigg|_{\rm sys},
\label{eq:rho_x3872}
\end{equation}
with the two decays reconstructed in the same final state and kinematic region.
CMS found evidence for $X(3872)$ production in PbPb collisions with a
significance of $4.2\sigma$ and
$\rho_{\rm PbPb}=1.08\pm0.49\,(\mathrm{stat.})
\pm0.52\,(\mathrm{syst.})$~\cite{CMS2022x3872}. For comparison, the corresponding
$pp$ ratio is of order $0.1$; the PbPb enhancement is shown in
Fig.~\ref{fig:cms_pbpb_ratio}. If one tentatively combines this yield ratio with the measured $R_{\mathrm{AA}}[\psi(2S)]=0.142\pm0.061\pm0.020$~\cite{CMS2018psi}, the resulting central value would exceed unity, but large experimental uncertainties and differences in kinematic/phase-space coverage prevent a quantitative determination of $R_{\mathrm{AA}}[X(3872)]$ and preclude claiming a firm enhancement of $R_{\mathrm{AA}}[X(3872)]$ itself. Rather, the enhancement is directly observed in the yield ratio $\rho_{\mathrm{PbPb}}$ relative to $pp$.

\begin{figure}[htbp]
\centering
\includegraphics[width=0.72\textwidth]{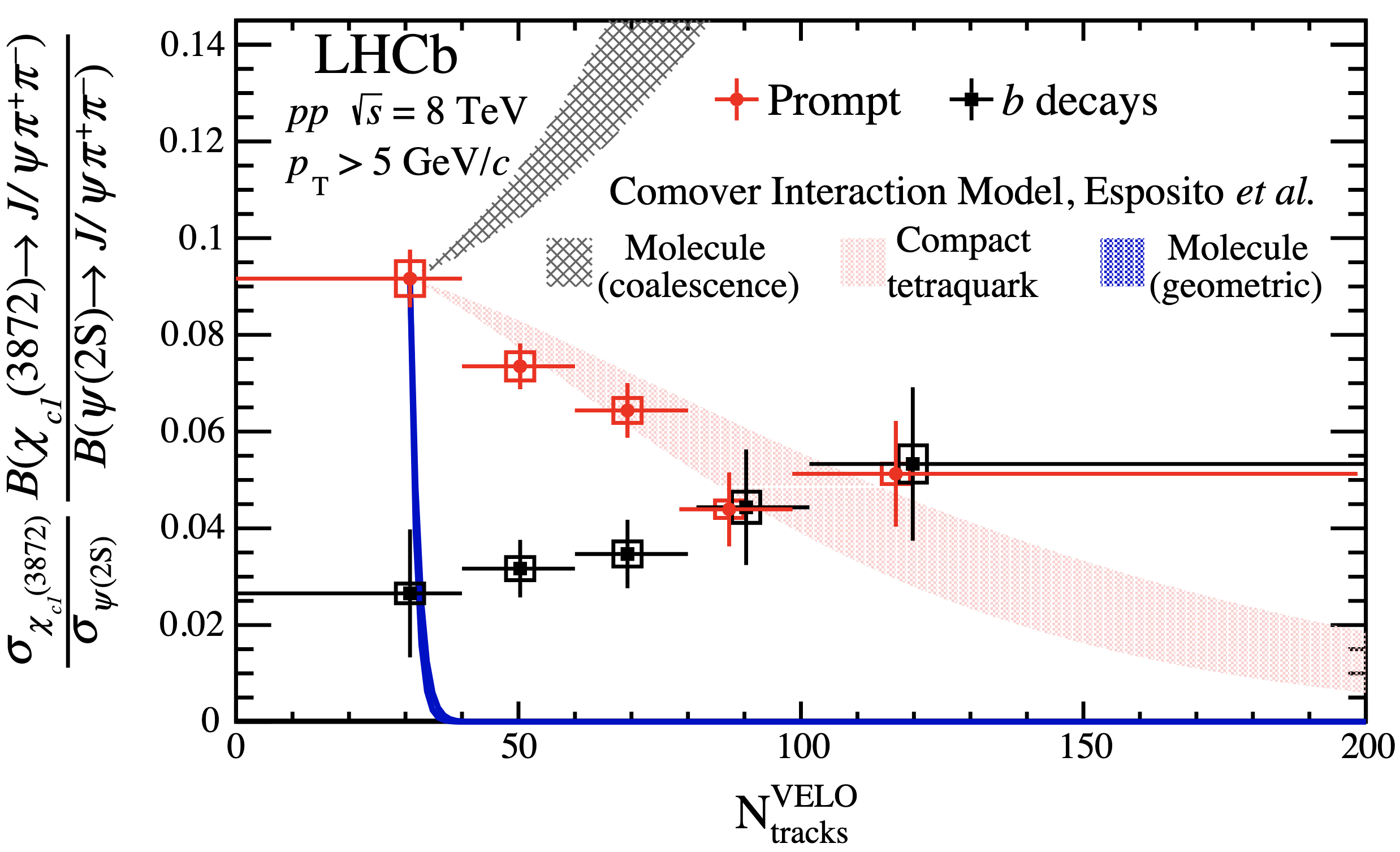}
\caption{Ratio of prompt $\chi_{c1}(3872)$ and $\psi(2S)$ cross sections in
the $J/\psi\pi^+\pi^-$ final state as a function of the number of reconstructed
VELO tracks in $pp$ collisions at $\sqrt{s}=8$~TeV. Vertical bars (boxes)
show bin-by-bin (correlated) uncertainties, and horizontal bars show the bin
widths. Figure reproduced from~\cite{LHCb2021mult}.}
\label{fig:lhcb_mult_ratio}
\end{figure}

LHCb studied the same ratio as a function of charged-particle multiplicity in
$pp$ collisions and found a significant decrease at the $5\sigma$
level~\cite{LHCb2021mult}; the measured trend and the comover-model predictions
are shown in Fig.~\ref{fig:lhcb_mult_ratio}. The trend favors a compact state
with a breakup cross section somewhat larger than that of the $\psi(2S)$, but it
does not by itself exclude a molecular component in a more complete
transport description. In $p$Pb collisions, LHCb measured the ratio to rise
from $0.070\pm0.009\pm0.003$ in $pp$ to
$0.26\pm0.08\pm0.05$ in the forward configuration and
$0.23\pm0.14\pm0.10$ in the backward configuration
\cite{LHCb2024pPb}. The corresponding nuclear modification factors are $R_{p\mathrm{A}}=2.6\pm0.8\pm0.8$ and $2.9\pm1.8\pm1.6$ in the forward and backward configurations, respectively, though with sizable uncertainties; they indicate a weaker suppression (or larger yield relative to binary-collision scaling) than that of the more strongly suppressed $\psi(2S)$ in the same system~\cite{LHCb2016psi2S}.

The resulting pattern---an increasing $X(3872)$-to-$\psi(2S)$ ratio from
$pp$ to $p$Pb to PbPb---is difficult to reproduce with a single universal
breakup cross section. Coalescence, statistical hadronization, and compact
tetraquark descriptions can each explain parts of the
trend~\cite{Cho2011a,Cho2020,Andronic2019,Guo2023,Braaten2021}, but the available
uncertainties and the differences in phase space prevent a decisive
interpretation. The measurement is important for this review because it
demonstrates that the medium can distinguish conventional and exotic
hadronization paths once the comparison is made between states in the same
final state and kinematic region.

%% file: F_Summary.tex
\section{Summary and Outlook}
\label{sec:Summary}

Heavy quarkonium has served for five decades as a laboratory for both
perturbative and non-perturbative QCD. The measurements reviewed here span the
Tevatron, $e^+e^-$ colliders, and the LHC. Taken
together, they show that the inclusive production rates can be described by
NRQCD factorization over a broad kinematic range once LDMEs are fitted and
feed-down is modeled. The spin observables and hadronization inputs
distinguish otherwise comparable rate
descriptions~\cite{Brambilla:2004wh,Brambilla:2010cs,Bodwin:1994jh}.

At the Tevatron, CDF and D0 established the experimental failure of the
simplest color-singlet picture and supplied the first high-statistics tests of
the color-octet mechanism~\cite{CDF:1997psi2S,Abe:1997yz,Bodwin:1994jh,Braaten:1994vv}.
The $J/\psi$, $\psi(2S)$, and $\Upsilon(nS)$ cross sections can be described
after the appropriate long-distance inputs are fitted, but the polarization
measurements favor zero or longitudinal values where the leading
color-octet prediction tends toward transverse
polarization~\cite{Abulencia:2007us,Abazov:2008za,Leibovich:1996pa}. The CDF
and D0 bottomonium results are not fully consistent with one another, so part
of the apparent ``polarization puzzle'' is an experimental question that must
be settled before a unique theoretical explanation can be claimed. The
double-parton measurements also indicate a smaller effective transverse
overlap for gluon-dominated than for quark-dominated channels, under the
usual double-parton-factorization
assumption~\cite{Abazov:2014qba,Abazov:2015fbq}.
Finally, the CDF $X(3872)$ and $J/\psi\phi$ studies showed that precision
exclusive spectroscopy is necessary for identifying threshold and exotic
structures that an inclusive LDME fit cannot resolve~\cite{Abulencia:2005gc,Aaltonen:2009tz, Zhou:2025nyj}.

The $e^+e^-$ measurements supply the complementary normalization and
quantum-number information. Belle, BaBar, and BESIII determined the absolute
branching fractions needed to convert inclusive hadron-collider yields into
physical cross sections and established the prompt continuum and double-charmonium
benchmarks~\cite{Ablikim:2013wzq,Ablikim:2017pkk,Belle:2002doublecc,Belle:2004doublecc}.
The LEP measurement of prompt quarkonium in hadronic $Z$ decays tests gluon
fragmentation at a still larger virtuality, while the charged $Z_c$ and $Z_b$
states and the $X(3872)$ demonstrate that the spectrum cannot be reduced to
the conventional $Q\bar{Q}$ multiplets~\cite{OPAL:1996jpsi,Ablikim:2013mio,Bondar:2011xe}.
Spin-correlation measurements in charmonium decays provide additional
precision tests of the strong-interaction amplitudes and weak-decay
parameters, although they do not replace the direct hadroproduction
polarization measurements~\cite{BESIII:2019baryonpol}.

At the LHC, differential cross sections now extend to
$p_T=200~\mathrm{GeV}$ for the $\Upsilon(nS)$ states
\cite{CMS-BPH-24-004}, and forward measurements cover rapidities not accessible
to the central detectors. The polarization data remain frame- and
$p_T$-dependent and do not restore the simple transverse-polarization
expectation~\cite{ALICE:2012jpsipol}. Bottomonium suppression in PbPb shows the
expected ordering of ground and excited states, but its quantitative
interpretation must include cold-nuclear-matter effects, feed-down, breakup,
regeneration, and detector-dependent backgrounds~\cite{CMS:2024upsilon}.
The increasing $X(3872)$-to-$\psi(2S)$ ratio from $pp$ to $p$Pb to PbPb provides
a new, structure-sensitive probe of hadronization, but the present
uncertainties do not yet identify the internal state of the
$X(3872)$~\cite{LHCb2021mult,LHCb2024pPb,CMS2022x3872}.

Several fundamental questions therefore remain open. No single LDME set
simultaneously reproduces all rate and polarization data across
$e^+e^-$, $pp$, and nuclear collisions
\cite{Butenschoen:2010rq,Gong:2012ru,Brambilla:2010cs}. The origin of the
polarization pattern, the size of regeneration in the QGP, and the internal
structure of threshold states are not settled. Progress requires measurements
that reduce the present degeneracies: wider $p_T$ coverage at fixed rapidity
and frame, improved feed-down separation, and measurements of relative
quarkonium yields in the same final states.

\subsection{Outlook for Conventional Quarkonium}

The high-luminosity LHC, Belle~II, and BESIII will mainly improve the
statistics and control of systematic uncertainties rather than simply raise
the collision energy~\cite{Kou:2018nap}. The most valuable measurements will
be those that separate production mechanisms: polarization as a function of
$p_T$, rapidity, and frame; prompt versus non-prompt charmonium at low $p_T$;
absolute $\Upsilon(2S)$ and $\Upsilon(3S)$ cross sections; and
multi-quarkonium final states in $pp$, $p$Pb, and PbPb. Lattice QCD and
temperature-dependent pNRQCD calculations will be needed to turn the observed
suppression hierarchy into quantitative statements about the medium
\cite{Brambilla:2004jt}.

\subsection{Fully Charmed Tetraquarks: An Emerging Production Frontier}

The di-$J/\psi$ spectrum has become a particularly active frontier. LHCb
reported the $X(6900)$-like structure~\cite{LHCb:2020bwg}; CMS found three
interfering structures near $6.6$, $6.9$, and $7.1~\mathrm{GeV}$ and later
determined spin-parity properties~\cite{CMS:2024JJ,CMS:2025tetraquarkJP}; and
ATLAS independently confirmed an excess in the same
region~\cite{ATLAS:2023bft}. The most recent CMS data support a family of
overlapping, possibly radially related states, but the spectroscopic
interpretation remains model dependent~\cite{CMS:JJ2026,Zhu:2024tetra}.

Production can sharpen that interpretation. NRQCD and fragmentation-based
calculations predict different normalizations and $p_T$ shapes for
different spin, parity, color, and internal-structure
assumptions~\cite{Zhang:2020hoh,PhysRevD.108.L051501,Wang:2025hex,Celiberto:2025vra}.
A recent development in this direction is the TQ4Q2.0 framework, which
provides fragmentation functions for fully heavy $S$-wave tetraquarks
with scalar, axial-vector, and tensor quantum numbers. The framework
incorporates all partonic fragmentation channels at the initial scale,
including gluon-, constituent heavy-quark-, and nonconstituent
quark-initiated contributions, and evolves the fragmentation functions
using DGLAP equations with the threshold-aware HF-NRevo scheme.
It also provides a systematic treatment of theoretical uncertainties
from perturbative multiscale variations and long-distance matrix
elements, with the resulting fragmentation-function grids publicly
available in LHAPDF6 format~\cite{Celiberto:2026TQ4Q2}.
These developments provide a more systematic theoretical basis for
investigating the sensitivity of tetraquark production spectra to
different quantum-number assignments and for assessing whether their
differences can be distinguished experimentally.

Despite this progress, predictions for the absolute production rates
of specific states remain dependent on the underlying structural
assumptions and non-perturbative inputs, and have not yet been tested
against absolute differential measurements. The available LHCb result
is a ratio to inclusive di-$J/\psi$ production rather than an absolute
differential cross section~\cite{LHCb:2020bwg}. The decisive next step
is therefore a measurement of absolute $p_T$- and rapidity-differential
cross sections for each structure, with a consistent treatment of
interference and the di-$J/\psi$ continuum.

The coming decade should therefore improve not only the precision of known
quarkonium observables but also the consistency of the complete theoretical
picture. The key question is no longer whether individual measurements can be
described. It is whether one QCD-based framework can describe rates,
polarization, threshold structure, and medium modification with common
non-perturbative inputs. The data summarized in this review provide the
experimental map for that program.